\documentclass[11pt]{amsart}
\usepackage[usenames, dvipsnames]{xcolor}
\usepackage{amsmath, latexsym, amsfonts, amssymb, amsthm,MnSymbol,stmaryrd}
\usepackage{graphicx,hyperref,dsfont,epsfig,caption,wrapfig,subfig,tikz}
\usetikzlibrary{patterns}
\usetikzlibrary{arrows}
\usetikzlibrary{calc}
\usepackage{epigraph}
\usepackage{etoolbox}
\makeatletter
\patchcmd{\epigraph}{\@epitext{#1}}{\itshape\@epitext{#1}}{}{}
\makeatother
\usepackage{datetime}
\usepackage{movie15}
\hypersetup{
    linktoc=page,
    linkcolor=red,          
    citecolor=blue,        
    filecolor=blue,      
    urlcolor=cyan,
    colorlinks=true           
}

\newcommand{\nocontentsline}[3]{}
\newcommand{\tocless}[2]{\bgroup\let\addcontentsline=\nocontentsline#1{#2}\egroup}
\numberwithin{equation}{section}

\newtheorem{theorem}{Theorem}[section]
\newtheorem{lemma}[theorem]{Lemma}
\newtheorem{proposition}[theorem]{Proposition}

\theoremstyle{definition}

\pgfdeclareradialshading[tikz@ball]{ball}{\pgfqpoint{0bp}{0bp}}{%
 color(0bp)=(tikz@ball!0!white);
 color(7bp)=(tikz@ball!0!white);
 color(15bp)=(tikz@ball!70!black);
 color(20bp)=(black!70);
 color(30bp)=(black!90)}
 \tikzset{pics/.cd,
handle/.style={code={
\draw (-2.6,-1.5) coordinate (-left) to [out=320, in=70] (-2,-4) 
to [out=260, in=60] (-3,-6) 
to [out=240, in=110] (-3,-8) 
to [out=290,in=175] (0,-9) 
to [out=5,in=250] (3,-8) 
to [out=70,in=300] (3,-6) 
to [out=120,in=280] (2,-4) 
to [out=110,in=220] (2.6,-1.5)  coordinate (-right);
\draw (-1,-6.5) to[bend left] (1,-6.5);
\draw (-1.2,-6.4) to[bend right] (1.2,-6.4);
}}}

\renewcommand{\tilde}{\widetilde}          
\DeclareMathSymbol{\leqslant}{\mathalpha}{AMSa}{"36} 
\DeclareMathSymbol{\geqslant}{\mathalpha}{AMSa}{"3E} 
\DeclareMathSymbol{\eset}{\mathalpha}{AMSb}{"3F}     
\renewcommand{\leq}{\;\leqslant\;}                   
\renewcommand{\geq}{\;\geqslant\;}                   
\newcommand{\dd}{\text{\rm d}}             
\newcommand{\bs}{\boldsymbol}

\newcommand{\C}{\mathbb{C}}

\newcommand{\D}{\mathbb{D}}
\newcommand{\R}{\mathbb{R}}
\newcommand{\Z}{\mathbb{Z}}
\newcommand{\N}{\mathbb{N}}
\newcommand{\Q}{\mathbb{Q}}

\newcommand{\A}{\mathbb{A}}
\newcommand{\E}{\mathds{E}}

\newcommand{\cjd}{\rangle}
\newcommand{\cjg}{\langle}

\newcommand{\hf}{\frac{_1}{^2}}

\def\l{\mathbf{l}}
\def\k{\mathbf{k}}

\newcommand{\pl}{\partial}
\newcommand{\bbar}{\overline}
\newcommand{\mc}{\mathcal}
\newcommand{\la}{\lambda}

\usepackage{xparse}

\def\eps{\varepsilon}
\def\S{\mathbb{S}}
\def\T{\mathbb{T}}

\def\bi{\begin{itemize}}
\def\ei{\end{itemize}}

\def\bnum{\begin{enumerate}}
\def\enum{\end{enumerate}}
\def\<#1{\langle #1 \rangle}

\newcommand{\caA}{{\mathcal A}}

\newcommand{\caF}{{\mathcal F}}

\newcommand{\caH}{{\mathcal H}}

\newcommand{\caP}{{\mathcal P}}

\author{Colin Guillarmou}
\address{Universit\'e Paris-Saclay, CNRS,  Laboratoire de math\'ematiques d'Orsay, 91405, Orsay, France.}
\email{colin.guillarmou@universite-paris-saclay.fr}

\author{Antti Kupiainen}
\address{University of Helsinki, Department of Mathematics and Statistics}
\email{antti.kupiainen@helsinki.fi}

\author{R\'emi Rhodes}
\address{Aix Marseille Universit\'e, CNRS, I2M, Marseille, France
  and Institut Universitaire de France (IUF)}
\email{remi.rhodes@univ-amu.fr}

\title[Review on Probabilistic construction and conformal bootstrap for Liouville CFT]{Review on the probabilistic construction and Conformal bootstrap in Liouville Theory}
\begin{document}
 
 \begin{abstract}
In the paper, we review the recent construction of the Liouville conformal field theory (CFT) from probabilistic methods, and the formalization of the conformal bootstrap. This model has offered a fruitful playground to unify the probabilistic construction of the path integral, the geometric axiomatics of CFT by Segal and the representation theoretical content of the conformal bootstrap. We explain and extract the main steps and ideas behind the construction and resolution of this non-compact CFT.
  \end{abstract}
 
 \maketitle
 
 
\tableofcontents

\section{Introduction}
\subsection{Brief physics history of  Conformal Field Theory}
Quantum  Field Theory (QFT) provides the basic framework for  the study of physical systems with infinite number of degrees of freedom.  It was originally introduced  for the purpose  of extending the quantum formalism to the description of the electromagnetic field and its interaction with charged matter, also described by fields.  In QFT, quantum fields are viewed as 
a map $W_\alpha:\underline x\mapsto W_\alpha(\underline x)$ taking values in the space of  operators  acting on a Hilbert space $\caH$, and depending on space-time  $\underline x=(t,{\bf x})$ (here $t\in\R$ and  ${\bf x}\in\R^d$, $d=3$ being the physical case), and possibly other data (index $\alpha$). The basic objects of interest, with direct experimental connection, are the Wightman functions, which are matrix elements  $\cjg\Omega,W_{\alpha_1}(\underline x_1)\dots W_{\alpha_n}(\underline x_n)\Omega\cjd_{\mc{H}}$ with $\Omega\in\caH $ a special vector called the vacuum state. In the 50's, a set of axioms were postulated to these  functions that define a Wightman QFT. One assumes to have a representation of the  space time symmetry group, namely the Poincar\'e group, as operators acting on $\caH$. Then these axioms postulate  regularity properties of the Wightman functions in their arguments $\underline x_i$ and their behaviour under the action of the  Poincar\'e group. In particular,  two special operators are singled out in this representation: the generator of time translations defines the Hamiltonian operator, i.e. the energy  of the QFT, and the generator of spatial translations defines the momentum operator.   

An important observation going back to Julian Schwinger is that  the Wightman functions have an analytic continuation to the so-called Euclidean domain of imaginary time $\underline x=(it,{\bf x})$ with  $t$ real. Furthermore, in that domain, the field operators at different points  have a probabilistic interpretation and can be viewed as   random (generalized) functions $V_\alpha(x)$ defined on the Euclidean spacetime $x=(t,{\bf x})\in\R^{d+1}$. The analytically continued Wightman functions are then given by 
\begin{align}\label{corre}
(\Omega,W_{\alpha_1}(\underline x_1)\dots W_{\alpha_n}(\underline x_n)\Omega)=\langle V_{\alpha_1}(x_1)\dots V_{\alpha_n}(x_n)\rangle
\end{align}
where $\langle -\rangle$ denotes the expectation in the underlying probability space. These {\it correlation functions} of the fields are then the fundamental objects of interest in the Euclidean QFT. 
A concrete  formal expression for the expectation $\langle -\rangle$ is provided by Richard Feynman's path integral, which is a formal integration measure over some functional space $E(\Sigma)$ of maps $\phi: \Sigma\to M$, where $\Sigma$ is the Euclidean space-time ($\R^{d+1}$ or a manifold in general) and $M$ is  the manifold in which the fields take values.  In the simplest QFT, $M=\R$ and the maps $\phi:\Sigma\to\R$ are then called scalar fields. Feynman's path integral is based on  an action functional $S$, which has the same origins as the action functional in the Lagrangian description of a classical field theory where it is studied via the calculus of variations.
This action functional is  a map $S:
E(\Sigma)\to\R$, and the path integral takes the form
\begin{equation}\label{thePI}
F\mapsto \int_{E(\Sigma)} F(\phi) e^{-\frac{1}{\hbar}S(\phi)} \,D\phi 
\end{equation}
where $D\phi$ is the formal Lebesgue measure on $E(\Sigma)$, $\hbar$ the Planck constant and $F$ are  test functions on $E(\Sigma)$. The correlation functions are then formally given by
\begin{align}\label{action1}
\langle V_{\alpha_1}(x_1)\dots V_{\alpha_n}(x_n)\rangle=\int  V_{\alpha_1}(x_1,\phi)\dots V_{\alpha_n}(x_n,\phi)e^{-\frac{1}{\hbar}S(\phi)}D\phi
\end{align}
where  the functionals $V_{\alpha}(x,\phi)$ depend locally on $\phi$ (i.e. on $\phi(x)$ and possibly its derivatives) and their exact form depends on the model.

The formula \eqref{action1} can be viewed as a Gibbs measure describing classical statistical mechanics where $S(\phi)$ is the energy of the field configuration $\phi$ provided the Planck constant  $\hbar$ is replaced by the temperature variable. This analogy provided the basis for the grand synthesis of the 1970s, which allowed to extend QFT techniques to the study of phase transitions in condensed matter and to non-equilibrium systems such as turbulence in fluid flow.  

In mathematics, the precise definition of the path integral \eqref{thePI} posed a hard problem. The typical field configurations for the measure \eqref{thePI} are expected to be generalized functions on $\Sigma$, like distributions in the sense of Schwartz, and making sense of $S(\phi)$ for a nonlinear $S$ is then not obvious. When  $S$ is a quadratic functional in $\phi$, the formal measure $e^{-\frac{1}{\hbar}S(\phi)}D\phi$ can be interpreted as a Gaussian measure on a space of distributions. The physically relevant case is the Gaussian Free Field (GFF) where 
$$S(\phi)=\frac{1}{4\pi}\int_\Sigma |d\phi|^2_g \dd {\rm v}_g$$ is the Dirichlet energy (here $g$ is a Riemannian metric on $\Sigma$ with volume form ${\rm v}_g$) and, taking it as a starting point, one could try to study its perturbation by non-Gaussian terms in $S$, namely 
\begin{equation}\label{Smu}
S(\phi)=\frac{1}{4\pi}\int_\Sigma \big(|d\phi|^2_g  +\mu V(\phi)\big)\dd {\rm v}_g.
\end{equation}
In the 70's this program was carried out 
in several cases of so-called superrenormalisable QFT's\footnote{Roughly speaking, these are QFTs requiring to tame only a finite number of ultraviolet divergencies to be defined.} and in the 80's to some renormalisable asymptotically free cases\footnote{In such theories the effect of the non-Gaussian terms become  small in small spatial scales and the theory becomes Gaussian upon zooming into such scales.} as well.  

 In physics the path integral was originally viewed as a formal but convenient way to derive perturbation (in $\mu$ in \eqref{Smu}) theory around the free field and this approach   led to spectacular   experimental success in weakly coupled theories such as quantum electrodynamics.  However, in many cases, such as in the theory of strong nuclear force or in the theory of  second order phase transitions, this perturbative approach failed and   nonperturbative methods were called for. To tackle such problems, 
 a new picture of QFT based on the renormalisation group theory by Ken Wilson emerged. In that approach one probes the space of all QFTs by a flow generated by coarse graining and scaling. The fixed points of this flow are called conformal field theories (CFT) and  the small and/or large scale properties of other QFTs are expected to be described  by a  CFT.

 A CFT is a QFT with a larger space time symmetry group extending the Poincare invariance, namely it is invariant under the conformal group.
 In statistical mechanics, it is conjectured that a  physical system, tuned at the critical temperature where a second order phase transition occurs,  exhibits scale invariance and the QFT describing it becomes scale invariant. Also, it was first conjectured by Polyakov \cite{Pol}  that, under general conditions, a scale invariant QFT is also  invariant under conformal transformations.  
 Systems that undergo a second order phase
transition see their correlation length diverge as they approach the critical point.
Then, their continuum limit can be described by a conformal field theory. In this context there is a striking
phenomenon know as critical universality which refers to the fact that, near
the critical point, the continuum limit becomes independent of the microscopic
details of the underlying models, and many different systems are described by
the same conformal field theory. The canonical example of a system with this
behaviour is the Ising model. In physics, spectacular progress has   been made since then by classifying all possible CFTs and by using this classification to identify the CFT arising in the scaling limit of lattice statistical physics models. Such a philosophy has also spread over mathematics in the field of statistical physics, as exemplified by the proof of the  conformal invariance of the lattice Ising model   by Chelkak-Smirnov \cite{Sm10,CS10} or the conformal invariance of percolation \cite{Smirnov_2001}.

The simplest CFT is the (massless)  free field and it describes the small scale behaviour of the aforementioned superrenormalisable and asymptotically free theories. It also describes the large scale behaviour in the simplest ferromagnetic second order phase transitions in high dimensions $d\geq 4$. However other examples of CFTs are non-perturbative and finding concrete examples has been a challenge. In particular numerical approaches using the renormalisation group have proved to be challenging. The bootstrap approach, developed in physics since the 60's, is a non-perturbative method that has allowed to study CFTs both analytically and numerically.   
The origins of  the modern bootstrap approach lie in the Operator Product Expansion (OPE) introduced by Wilson in \cite{Wilson:1969}. He postulated that a QFT can be characterized by a family of local fields $(W_\alpha)_{\alpha\in\mathfrak{L}}$ (namely for each $\underline x$,  $W_\alpha(\underline x)$ is an endomorphism of $\mc{H}$) so that the field operators satisfy an  expansion
\begin{align}\label{wilson}
\cjg\Psi,W_\alpha(\underline x)W_\beta(\underline y)\Psi'\cjd_{\mc{H}}=\sum_{\gamma\in\mathfrak{L}} c_{\alpha\beta}^\gamma(\underline x-\underline y)\cjg\Psi,W_\gamma(\underline x)\Psi'\cjd_{\mc{H}}
\end{align}
for all $\Psi,\Psi'$ in some dense domain in $\caH$. The series involves in general infinitely many terms, among which  all but finitely many of the  functions $c_{\alpha\beta}^\gamma(\underline x-\underline y)$ are nonsingular. A similar expansion is expected to hold for the Schwinger functions \eqref{corre}.
If the QFT is a CFT,  it was argued in \cite{Ferrara}, \cite{Pol74} that  this series can be drastically simplified. First, there is a  set of special local fields $(W_\alpha)_{\alpha\in\mathfrak{S}}$, called \emph{primary fields}, so that all other local fields can be expressed as their (multiple) derivatives.  Second, using the conformal symmetry, if $W_{\alpha_1}$ and $W_{\alpha_2}$ are primary fields,  the OPE \eqref{wilson} can be written solely in terms of the primary fields so that  in the Euclidean setup it reads as
\begin{align}\label{boot}
\langle V_{\alpha_1}(x_1)V_{\alpha_2}(x_2)\dots V_{\alpha_n}(x_n)\rangle=\sum_{\gamma\in\mathfrak{S}} C_{\alpha_1\alpha_2}^\gamma(x_1-x_2,\partial_{x_1})\langle V_{\gamma}(x_1)V_{\alpha_3}(x_3)\dots V_{\alpha_n}(x_n)\rangle
\end{align}
where $V_\gamma $ are primary fields and  the differential operators $C_{\alpha\beta}^\gamma(x-y,\partial_x)$ are completely determined in terms of the three point correlation functions $\langle V_{\alpha_1}(y_1)V_{\alpha_2}(y_2)V_\gamma(y_3)\rangle $, whose  dependence on the $y_i$'s  in turn is explicit  due to the conformal symmetry, up to a constant $C_{\alpha_1,\alpha_2, \gamma}$. These constants are called the  {\it  structure constants} of the CFT. Obviously, by iterating \eqref{boot}, the correlation functions are completely determined in terms of the structure constants.  Last but not least, it was argued in \cite{Mack1} and proved in \cite{mack2} under some (axiomatic) assumptions that the sum in \eqref{boot} is convergent for $x_1-x_2$ small enough. 

Thus we are led to the picture that, in order to "solve a CFT", one needs to determine its  primary fields and determine the structure constants. Actually, only a subset of the primary fields enters the sum\footnote{This holds for a unitary QFT or a reflection positive statistical field theory, concept that we will soon explain.} in \eqref{boot} and these fields are said to be in the {\it spectrum}\footnote{The terminology is misleading in mathematics: as we will see, these states correspond to some eigenstates of some operator, whereas the term spectrum is rather used in maths for the eigenvalues.} of the CFT. Thus they form a family that is closed under the OPE. As we will see later, the spectrum is related to the eigenstates  of the Hamiltonian of the conformal group, as represented in the Hilbert space of the QFT. To find the spectrum and the structure constants,  physicists explored the consequences of the OPE as applied to four-point functions $\langle V_{\alpha_1}(x_1)V_{\alpha_2}(x_2)V_{\alpha_3}(x_3)V_{\alpha_4}(x_4)\rangle $. We can evaluate this by applying the OPE to $ V_{\alpha_1}$ and $V_{\alpha_2}$ or to $ V_{\alpha_1}$ and $V_{\alpha_3}$ both leading to quadratic expressions in the structure constants. Their equality is called the crossing symmetry equation and it turns out to pose strong constraints on the spectrum and the structure constants. Up to plugging some further a priori knowledge on the model in the crossing symmetry equations, they may determine completely the spectrum and the structure constants. The first successful application of this approach was carried out in the ground breaking work by Belavin, Polyakov and Zamolodchikov  in 1984 \cite{BPZ84} in the context of two dimensional CFTs:   under the assumption that the spectrum is finite, they classified the so-called minimal models. 

In two dimensions, the (Euclidean) conformal group is ${\rm PSL}(2,\C)$ consisting of M\"obius transformations of the extended complex plane, i.e. the Riemann sphere. However this symmetry is just a tip of the iceberg. The {\it local conformal} symmetry were used in \cite{BPZ84} to uncover a rich algebraic structure in 2d CFT. The crux of their approach was the postulate  of a local field, the stress energy-momentum  tensor  in CFT. Recall that, in a Wightman QFT, the generators of space-time translations are represented by the operators of  energy and momentum in the Hilbert space of the QFT. However in a classical field theory with a local action functional, the  energy and momentum are integrals over space of energy and momentum densities which can be collected to a tensor valued local  field called the energy-momentum tensor. Thus in the corresponding QFT one expects such a  local field to exist.  In a 2d CFT, this tensor reduces to one complex valued field $T(z)$ and it was observed by \cite{BPZ84} that correlation functions   should be {\it holomorphic}  in the arguments of $ T$-insertions and furthermore have an explicit OPE of the form
\begin{equation}
T(z)V_{\alpha}(u)  =   \frac{ \Delta_{\alpha} }{(z-u)^2} V_{\alpha}(u)  +\frac{1}{z-u} \partial_{u} V_{\alpha}(u)  +{\rm reg} \label{Wardidentityexp}  
\end{equation} 
\begin{align}
T(z) T(u)= \frac{{\rm c}}{2}\frac{1}{(z-u)^4} + \frac{2}{(z-u)^2}T(u)
 +  \frac{1}{z-u}\partial_{u} T(u)  +{\rm reg} \label{Ward} 
\end{align}
where $ {\rm reg} $ denote  regular terms as $z-u\to 0$. Here $\Delta_\alpha$ is called the {\it conformal weight} of the field $V_\alpha$ and the number ${\rm c}\in \C$ the {\it central charge} of the CFT.  Belavin-Polyakov-Zamolodchikov then showed that  these OPE have drastic  consequences to the structure of a CFT.  
Being holomorphic with eventual poles, the stress energy tensor can be expanded as a Laurent series and the coefficients of the series 
provides a family of operators $(\mathbf{L}_n)_{n\in\Z}$, as well as the the Laurent expansion of the complex conjugate $\bar{T}(z)$ provides another   family of operators $(\tilde{\mathbf{L}}_n)_{n\in\Z}$. These operators $\mathbf{L}_n, \tilde{\mathbf{L}}_m$ form two commuting representations of the Virasoro algebra, namely with the commutators obeying
\begin{align*}
[\mathbf{L}_n,\mathbf{L}_m]=(n-m)\mathbf{L}_{n+m}+\frac{{\rm c}}{12}\delta_{n,-m}{\rm Id}
\end{align*}
and the same relation for the family $(\tilde{\mathbf{L}}_n)_{n\in\Z}$. These families provide a representation of the local conformal symmetries.
Furthermore, these operators act on the primary fields $V_\alpha$ to produce new states, called descendant states, of the form $\tilde{\mathbf{L}}_{m_1}\dots \tilde{\mathbf{L}}_{m_l}\mathbf{L}_{n_1}\dots \mathbf{L}_{n_k}V_\alpha$ with $m_j,n_i\in \Z_-$. The correlation functions of such states can be inductively obtained from those of $V_\alpha,T(z),\bar{T}(z)$ by performing nested contour integrals.

Inserting these assumptions to the operator product expansion  \eqref{boot}, Belavin-Polyakov and Zamolodchikov drew two conclusions. First, the coefficient operators in \eqref{boot}   are determined algebraically and, due to the commutation of the two representations of the Virasoro algebra, they factor into holomorphic and antiholomorphic parts which could be studied separately. In particular, one can formulate the crossing symmetry equations for the holomorphic and antiholomorphic parts separately. Their solutions were dubbed (four point) {\it conformal blocks}. The original correlation functions were then obtained as sesquilinear combinations of the conformal blocks. In particular, an infinite class of solutions isolated by \cite{BPZ84} gives rise to the so-called {\it minimal models}, each of which has a finite set of primary fields. The result of these considerations is a beautiful conjectural structure for the  correlation functions\footnote{in the simplest  case of diagonal minimal models.}: if $\bs{\alpha}=(\alpha_1,\dots,\alpha_n)$,  
\begin{align}\label{final}
\langle V_{\alpha_1}(z_1)\dots V_{\alpha_n}(z_n)\rangle=\sum_a \rho(a,\bs{\alpha})|\mc{F}_{a,\bs{\alpha}}(z_1,\dots, z_n)|^2 
\end{align}
where $\{\mc{F}_{a,\bs{\alpha}}(z_1,\dots, z_n)\}_a$ is a finite family of conformal blocks, holomorphic in the region of non-coinciding points and  $ \rho(a,\bs{\alpha})$ is a product of structure constants. 
The structure constants are model dependent whereas the conformal blocks $\mc{F}_{a,\bs{\alpha}}$ are purely representation theoretical and a universal feature of CFT. After the paper \cite{BPZ84}, the formalism was extended to a large classes of rational CFT\footnote{Roughly speaking, rational CFT have the simplest possible algebraic structure, namely a finite spectrum.}  such as the Wess-Zumino-Witten models and their cosets. The reader may consult the pedagogical  review on CFT by Ribault \cite{Ribault14}. A numerical approach to solving the bootstrap equations also in dimensions greater than two was developed in  \cite{slava} with spectacular results on the 3d Ising model and other theories. However, to reconcile the conformal bootstrap with the path integral proved challenging, and this has some importance. Indeed, the conformal bootstrap is a powerful tool to construct and classify CFT but it does not make easily connections with statistical physics models, thus making the identification of the CFT arising in the scaling limit of lattice models sometimes tricky, whereas the path integral is the first accessible object from these scaling limits.

\subsection{Conformal Field Theory in Mathematics}
In math,  the paper of Belavin-Polyakov-Zamolodchikov  \cite{BPZ84} was both  a challenge and a deep source of  inspiration. There were several approaches developed to axiomatise what a conformal field theory should be in mathematical terms. 

Based on representation theory and algebraic geometry, the concept of 
 Vertex Operator Algebras  was introduced by Borcherds \cite{Borcherds} and Frenkel-Lepowsky-Meurman \cite{Frenkel:1988xz}  (see also \cite{Frenkel07} or \cite{Huang1997TwoDimensionalCG} for more recent developments) and this  led to a rigorous formalism for some aspects of  rational CFTs (say with finite spectrum, like the minimal models).  However, the mathematical connection between the  approach of Belavin-Polyakov-Zamolodchikov \cite{BPZ84} and the path integral is left open. 
 
 Friedan and Shenker \cite{FriedanShenker87} developped an abstract mathematical formulation of 2d CFT in terms of analytic geometry on moduli space of Riemann surfaces, by viewing correlation and partition functions as analytic functions on the moduli space (with marked points for the correlations).
They introduced a holomorphic vector bundle on the moduli space and a projectively flat Hermitian connection. The partition function is viewed as the norm squared, in the Hermitian metric, of a holomorphic section of the vector bundle. This approach has been the inspiration for further works in algebraic geometry.

Alternatively, Graeme Segal \cite{Segal87} designed a set of  axioms  to capture the conformal bootstrap approach to CFTs using a geometrical perspective (for a nice introduction to mathematicians see \cite{Gawedzki96_CFT}). They are based on heuristics derived from the path integral, if exists, and  are perhaps the most intuitive way to understand   the Operator Product Expansion. Also they are fundamental to obtain a consistent CFT on all closed surfaces.  But, as conceded by Segal himself \cite{Segal87}, they are rather demanding and finding non trivial examples of CFTs obeying these axioms turned out to be a difficult task: he wrote
\emph{``The manuscript that follows was written fifteen years ago...I just wanted to
justify my proposed definition by checking that all know examples of CFTs did  fit the definition. This task held me up...''}.

\begin{figure} 
\centering
\subfloat[Random triangulation]{\includegraphics[width=8cm]{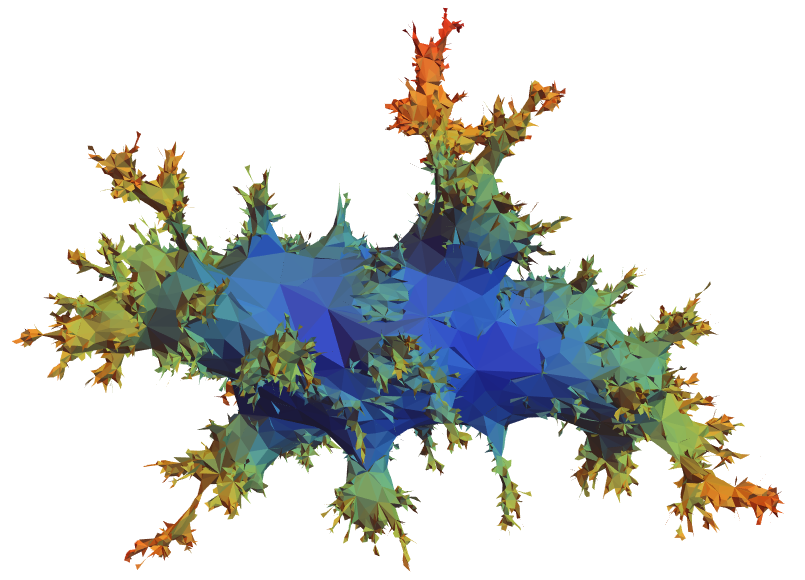}}%
\qquad
\subfloat[Local zoom in the triangulation once conformally embedded into the Riemann sphere]{\includegraphics[width=6.8cm,height=5.7cm]{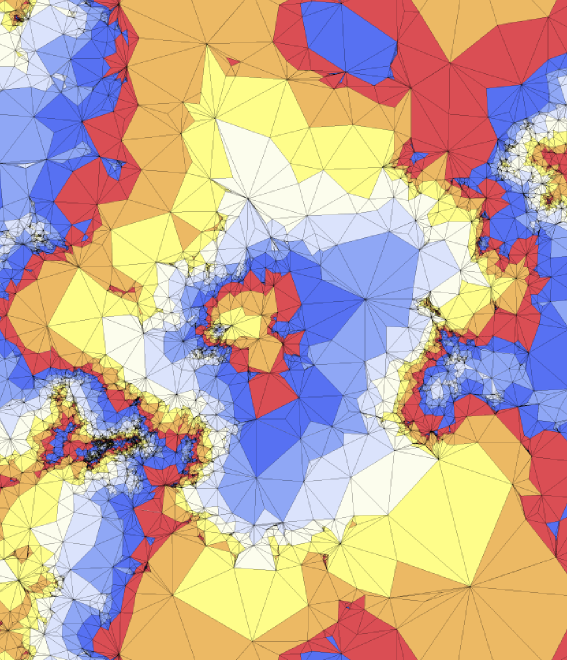}}
\caption{Triangulation of the Riemann sphere sampled from the uniform probability measure.@T.Budd}%
\label{triang}%
\end{figure}

The paper \cite{BPZ84} was originally motivated to solve the Liouville CFT, which was introduced by Polyakov as a model for random Riemannian metrics in 2 dimensions arising in string theory (see the reviews \cite{Klebanov,Nakayama,Teschner_revisited,SeibergRev}). 
In his article \cite{Pol08}, Polyakov explains about the paper \cite{BPZ84}: ``\emph{It started
with the attempt to build a conformal bootstrap for the Liouville theory. We worked on it with my friends Sasha Belavin and Sasha Zamolodchikov. We developed a general approach to conformal field theories, something like
complex analysis in the quantum domain. It worked very well in the various
problems of statistical mechanics but the Liouville theory remained unsolved.
I was disappointed and inclined not to publish our results.}''
The Liouville model later played a central role in models of 2d quantum gravity and random surface theory in the references \cite{Knizhnik:1988ak} and \cite{David,Distler_Kawai}, in particular as scaling limit of Random Planar Maps, sort of random probability measures on all the possible  triangulations of a Riemann surface (see Figure \ref{triang}). The reader may consult \cite{kostov} for a physics review on the topic, and \cite[Section 5.3]{DKRV16} or \cite[Section 5.5]{GRV2019} for mathematical conjectures relating random planar maps and Liouville CFT. Mathematical progress on this aspect has been achieved in \cite{XinCardy}.
 
If one thinks of the representation theoretical content of CFT as an infinite dimensional version  of the classification of representations of finite dimensional Lie groups, then Liouville CFT is the analog of ${\rm SL}_2(\R)$,  which involves continuous principal series (continuous spectrum), whereas rational CFTs, like minimal models, are the analog of compact Lie groups.
Liouville CFT also differs from many other theories in that it has a heuristic path integral description.  The study of the crossing symmetry equations was challenging and, only after 10 years, a conjectural solution was given in physics by Dorn-Otto \cite{DornOtto94} and independently by Zamolodchikov-Zamolodchikov \cite{Zamolodchikov96}, which was thus dubbed the DOZZ formula. This solution was later confirmed to be correct  in \cite{KRV_DOZZ} through a probabilistic construction of Liouville CFT \cite{DKRV16,GRV2019}.  This provides a test case for both providing a mathematical  construction of a CFT and deriving the conformal bootstrap solution rigorously from that setup. 
  
    The natural setup for the  Liouville theory (and in fact for 2d CFT in general)  is 2d surfaces with a complex structure, i.e. Riemann surfaces. These in turn can be viewed as Riemannian surfaces: a complex stucture determines a conformal class of Riemannian metrics. Thus, fixing a closed surface $\Sigma$ and a smooth Riemannian metric $g$ on $T\Sigma$, the Liouville action functional of a smooth $\phi:\Sigma\to\R$  is given by
\begin{equation}\label{AL}
S_{{\rm L}}(\phi ,g  )=       \frac{1}{4\pi} \int_{\Sigma}\big(|d\phi |^2_{{g }}+QK_{{g }} \phi  +4\pi \mu e^{ \gamma \phi  }\big)\, {\rm dv}_{{g }}  .
\end{equation}
Here $|\cdot|_g$ is the induced metric on $T^\ast\Sigma$,  ${\rm v}_g$ the Riemannian volume measure,
  $K_{{g }}$  is  the scalar curvature of the metric $g$, $\mu>0$ and $\gamma>0$ and $Q$ parameters. At the classical level and for $Q=\frac{2}{\gamma}$, the critical points $\phi_c$ of the Liouville action give rise to metrics $e^{\gamma \phi_c}g$ which have constant scalar curvature. If the surface has genus $\geq 2$, $\phi_c$ is a minimiser and the new metric has constant negative curvature\footnote{This picture is also valid in genus $0$ or $1$ if one considers further conical singularities on the Riemann surface.}. In this Riemannian setup, conformal invariance can be formulated as covariance under two actions on the space of smooth metrics: the action of the group of smooth diffeomorphisms $\psi:\Sigma\to\Sigma$ and the action of Weyl scalings $g\to e^\phi g$ with $\phi\in C^\infty(\Sigma)$. Indeed by a change of variables 
  \begin{align}\label{diffc}
  S_{{\rm L}}(\phi\circ\psi ,\psi^\ast g  )=S_{{\rm L}}(\phi ,g  )
\end{align} 
and using $K_{e^\varphi g}=e^{-\varphi}(K_g+\Delta_g\varphi)$ one readily checks
\begin{align}\label{weylc}
S_{{\rm L}}(\phi ,e^\varphi g  )=S_{{\rm L}}(\phi +\hf Q\varphi,g  )-\frac{Q^2}{16\pi}\int_{\Sigma}\big(|d\varphi |^2_{{g }}+2 K_{{g }} \varphi \big)\, {\rm dv}_{{g }}
\end{align}
provided we take $Q=2/\gamma$.  We will see later that in the quantum theory this value gets modified to $Q=2/\gamma+\gamma/2$.

To study the path integral \eqref{action1} with the action \eqref{AL}, we still need the local fields for Liouville CFT. It turns out these are formally given by $V_\alpha(z,\phi)=e^{\alpha\phi(z)}$ with $\alpha\in\C$ so that the correlation functions are formally given by
\begin{equation}\label{correl}
\langle\prod_{i=1}^nV_{\alpha_i}(z_i)\rangle=\int_{\Sigma}  \Big(\prod_{i=1}^ne^{\alpha_i\phi(z_i)}\Big)e^{- S_{{\rm L}}(\phi ,g  )}D\phi .
\end{equation}
The modification of the classical action functional \eqref{AL} by the term $\sum \alpha_i\phi(z_i)$ in \eqref{correl} has an interesting geometric interpretation. Indeed, for $\alpha_i\in\R$ the extremizers in its presence are constant curvature metrics with conical curvature singularities at the $z_i$'s, see \cite{Troyanov}.  
Hence the path integral \eqref{correl} can be  seen as the probabilistic theory of random Riemannian surfaces: it describes random fluctuations around such metrics. 

In the present manuscript, we review the probabilistic construction of this path integral \cite{DKRV16,GRV2019} using Gaussian multiplicative chaos theory (Gaussian Multiplicative Chaos) and how this led to  the derivation of the conformal bootstrap   in the series of works \cite{KRV_DOZZ,GKRV20_bootstrap,GKRV21_Segal} culminating to the Liouville CFT version of the expression \eqref{final}. 
Perhaps most interestingly, this series of works  offers a unified picture,  in the context of  a non trivial interacting CFT, of all the concepts relating probability, moduli space theory, representation theory and spectral/scattering theory,  presented in this introduction.

\subsection{Subsequent or related works and applications}

Many integrability formulas for Gaussian Multiplicative Chaos theory (in the one dimensional context) have been conjectured in statistical physics in the study  of  disordered systems. In particular an explicit formula for the moments of the  total mass of the Gaussian Multiplicative Chaos  measure on the circle (based on exponentiating the free boundary GFF) was proposed by Fyodorov-Bouchaud  \cite{Fyodorov_2008} (for generalizations to  other 1d geometries like the segment see the work by Fyodorov-Le Doussal-Rosso \cite{Fyodorov_2009} and Ostrovsky \cite{Ostrovsky_2009, Ostrovsky_2018}). It turns out that their formula is a particular case of the structure constants  for LCFT  with boundary. Building on \cite{KRV_DOZZ}, the structure constants of boundary Liouville CFT were determined in a series of works by Ang-R\'emy-Sun-Zhu \cite{AngRemySun21_FZZ,ang2023derivation}. The simplest and more pedagogical instance is the derivation of the Fyodorov-Bouchaud formula via CFT techniques by Remy in \cite{Remy20},  extended by Remy-Zhu  to the case of an interval in \cite{Remy_2020}.  Baojun Wu initiated in  \cite{Wu} the study of the conformal bootstrap for surfaces with a boundary in the case of the annulus. Building on this work, the law of the random modulus    for RPM with the topology of an annulus was obtained by Ang-Remy-Sun in \cite{Xin}.

The structure constants in Liouville CFT, i.e. the DOZZ formula, is expected to encode many other statistical physics models in some way. This has been highlighted via the powerful interplay between Liouville CFT and  Schramm Loewner Evolution (or Conformal Loop Ensembles), which can be coupled via the Mating of Tree formalism\footnote{A  framework developed by Duplantier-Miller-Sheffield  to study the coupling between SLE or CLE  and the GFF in terms of more classical probabilistic objects such as Brownian motion, Levy process, and Bessel process.}  \cite{MatingOfTrees,nina}. This has led to   unexpected results for various statistical physics models, like the computation of the three point correlation functions of Conformal Loop Ensembles  \cite{AngSun21_CLE}, shown to be given by the imaginary DOZZ formula,  or backbone exponent in critical percolation  \cite{nolin} .

Liouville CFT fits as a particular case of a more general CFT called Toda conformal field theory, where the field takes values in the Cartan algebra of a semi-simple Lie algebra. This model is also related to the AGT correspondence. The probabilistic construction was done recently by Cercl\'e-Rhodes-Vargas \cite{Cercle-Rhodes-Vargas} and Cercl\'e  \cite{cercle} has been able to prove the formula of 
Fateev-Litvinov for the $3$-point function for certain range of parameters. The general expression of the $3$-point function is not known in physics.

A  path integral based on the Liouville action functional with imaginary parameters was constructed in \cite{CILT}, based  on the compactified Gaussian Free Field  and imaginary Gaussian Multiplicative Chaos theory.  In physics this path integral is conjectured to describe the scaling limit of critical loop models such as Potts and O(n) models.   It is proved 
 that the  path integral  satisfies Segal's axioms of CFT and has conjecturally   the structure of a  logarithmic CFT.  
 
 In another  development, Liouville correlation functions show up in a seemingly completely different setup of four dimensional gauge theories ($N=2$ supersymmetric Yang-Mills) via the so-called AGT correspondence  \cite{Alday_2010} (see the work by Maulik-Okounkov \cite{Davesh_MAULIK_2019} and Schiffmann-Vasserot \cite{Schiffmann_2013} for the mathematical implications in quantum cohomology  of these ideas). The AGT correspondence   conjectures  that that the conformal blocks in Liouville CFT coincide  with special cases of Nekrasov's partition function \cite{Nek}. A probabilistic representation   of the 1-point toric Virasoro conformal block for central change greater than 25 based on Gaussian multiplicative chaos was obtained in  \cite{blocguillaume}.

\subsection{Organization of the manuscript}
 We first review Graeme Segal's definition of CFT in Section   \ref{sub:segal}. Then the path integral construction is  sketched out  in Section \ref{sec:PI} and we explain how it fits Segal's definition of CFT in Section \ref{sec:segalL}, in particular we construct the Hilbert space and the Liouville amplitudes. From Segal's axioms of CFT, we explain then how to construct the Hamiltonian of Liouville CFT in Section \ref{sec:hamiltonian} and how it can be diagonalized. In Section \ref{sec:LCS}, we   further exploit Segal's axioms in order to construct a representation of the full symmetry algebra, which acts on the primary states to form the descendant states. Then we explain how this representation connects to the stress energy tensor and how this leads to explicit computations for the geometrical building block amplitudes. Combining this elements, we state the conformal bootstrap in Section \ref{sec:CB} and describe the construction of the conformal blocks in Section \ref{sec:block}.

\section{Segal's approach of CFT}\label{sub:segal}
 
\subsection{Classical field theory}
A classical field theory  can be described roughly as follows. A field on a manifold $\Sigma$ 
is a function $\phi:\Sigma\to V$ where $V$ is another manifold, or more generally $\phi$ is a section of a  bundle on $\Sigma$ with fiber $V$. The simplest case which occupies us in this review is a scalar field where $V=\R$ and $\Sigma$ is a Riemannian manifold with metric $g$. 
One needs a topology on the space $E(\Sigma)$ of fields, for example in the scalar case, this can be 
the Sobolev space $H^s(\Sigma,V)$ for some $s\in \R$.
An action is then a continuous functional 
\[ S_\Sigma: E(\Sigma) \mapsto \C. \]
An important property of physical action functionals is {\bf locality}. This means that the action is given as $S_\Sigma(\phi)=\int_\Sigma s(x,\phi) {\rm dv}_g(x)$ where ${\rm v}_g$ is the Riemannian volume and $s(x,\phi)$ depends on $\phi$ only through a finite number of derivatives of $\phi$ at the point $x$. Thus if we decompose $\Sigma=\Sigma_1\cup \Sigma_2$ where $\Sigma_1\cap \Sigma_2=\partial\Sigma_1=\partial\Sigma_2$ is a codimension one submanifold then
\begin{align}\label{cutting}
S_\Sigma(\phi)=S_{\Sigma_1}(\phi)+S_{\Sigma_2}(\phi).
\end{align}
In classical field theory the typical problem is to find the critical points of $S_\Sigma$, which often leads to PDEs. For example, the Dirichlet energy on a closed Riemannian manifold $(\Sigma,g)$ with boundary, with $f\in H^{1/2}(\Sigma)$:  
\[ S_{\Sigma,f}: E_f:=\{\phi \in H^1(\Sigma)\,|\, \phi|_{\pl \Sigma}=f\} \to \R ,\quad S_{\Sigma,f}(\phi)=\int_{\Sigma}|d\phi|_g^2{\rm dv}_g\] 
admits a unique minimum $\phi_0$, which is a critical point of $S_{\Sigma,f}$, solving the Dirichlet problem
\[ \Delta_g\phi_0=0, \quad \phi|_{\pl \Sigma}=f.\]  

\subsection{Path integral and Segal's picture}
Given a classical field theory as above, in the Euclidean (or probabilistic)  Field Theory, we aim to construct a measure 
\begin{equation} \label{PIsegal}
e^{-S_\Sigma(\phi)}D\phi 
\end{equation}
on the space of fields $E(\Sigma)$ 
where $D\phi$ should be understood as the ``uniform measure'' on $E(\Sigma)$, i.e. some measure invariant under the symmetries of the action functional. 
When attempting to make sense of \eqref{PIsegal}, one encounters the problem that the support of the putative measure consists of  (generalized) functions for which the action is not defined and a rigorous definition consists of delicate regularisations, renormalizations and limiting procedures. Nevertheless the formal expression \eqref{PIsegal} provides a motivation for how the locality property of the classical field theory  \eqref{cutting} should be reflected in the probabilistic setup. Using naively  \eqref{cutting} and the corresponding formal  factorisation of $D\phi$,
 we can formally perform the path integral first over fields $\phi_i$ on $\Sigma_i$ with fixed boundary values $\phi_i|_{\Sigma_1\cap\Sigma_2}=\varphi$ and then integrate over $\varphi$. Given an observable $F$ that also factorises, i.e. $F(\phi)= F_1(\phi_1)F_2(\phi_2)$, we are led to 
 $$\int_{E(\Sigma)} F(\phi)e^{-S_\Sigma(\phi)}D\phi = \int_{E(\Sigma_1\cap\Sigma_2)} \mc{A}_{\Sigma_1,F_1}(\varphi)\mc{A}_{\Sigma_2,F_2}(\varphi)D\varphi$$
with
\begin{equation}\label{formamp}
 \mc{A}_{\Sigma_i,F_i}(\varphi):=
 \int_{E(\Sigma_i)}1_{\phi|_{\partial\Sigma_i}=\varphi}F_i(\phi)e^{-S_{\Sigma_i}(\phi)}D\phi.
 \end{equation}
 For a probabilist, such a factorisation can be interpreted as a  conditioning on $E(\Sigma_1\cap\Sigma_2)$ and this is indeed how we will approach this question later. 
The conditional path integrals $\mc{A}_i$ are called {\it amplitudes} of the manifolds $\Sigma_i$ with boundaries  $\Sigma_i$.
\begin{figure}
   \begin{tikzpicture}
    \node[inner sep=0pt] (pant) at (0,0)
{\includegraphics[scale=0.4]{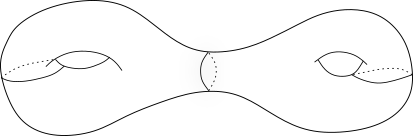}};
    \draw (-0.1,0.5) node[right,black]{$\mc{C}$} ;
     \draw (-1,0) node[right,black]{$\Sigma_1$} ;
       \draw (0.6,0) node[right,black]{$\Sigma_2$} ;
\end{tikzpicture}
\caption{A surface cut along a circle $\mc{C}$}
\label{picsegal}

\end{figure}
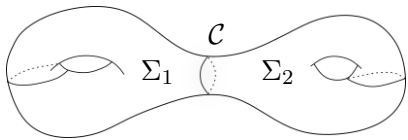

Applying this idea to the correlation functions, i.e. $F=\prod V_{\alpha_i}(x_i)$, we cut 
small balls $B_i$ around each insertion point $x_i$ and the result should be of the form
   \begin{align*}
\langle V_{\alpha_1}(x_1)\dots V_{\alpha_n}(x_n)\rangle=\int\mc{A}_{\Sigma\setminus\cup B_i,1}(\varphi_1,\dots,\varphi_n)\prod_{i=1}^n\Psi_{i}(\varphi_i)\prod_{i=1}^nD\varphi_i
\end{align*}
 where $\Psi_{i}$ is the amplitude of the ball $B_i$ with observable $F=V_{\alpha_i}(x_i)$. 

 In the 80's Graeme Segal gave an axiomatics for 2d CFT, based on these intuitions. The axioms deal with amplitudes $\mc{A}_\Sigma$ associated to Riemann surfaces with boundary given by a union of circles and obeying a composition property upon cutting the surface along  circles. This reduces the study of a CFT to the study of the amplitudes of  
 simple building blocks namely disks, annuli and pairs of pants. Before stating these axioms 
 we will make a detour to explain a bit the thread of ideas that led to obtaining the Hilbert space of a QFT from the path integral.

 \subsection{Hilbert space and reflection positivity}
 We explained in introduction how relativistic (Lorentzian) QFT leads to Euclidean QFT using a  Wick rotation, justified by the positivity of energy. An important  question is whether we can go the opposite direction, namely, starting from a Euclidean path integral, can we get a relativistic QFT out of it? In the 70s, Osterwalder and Schrader \cite{osterwalder-schrader} proposed a set of axioms  for a Euclidean field theory that allows to construct a  relativistic  Wightman QFT by analytic continuation. The key axiom for this is a positivity property called OS positivity or reflection positivity. Following this discovery, Euclidean fields became the fundamental tool to investigate relativistic field theory. The beautiful simplicity of this approach is that it
relies on a positivity condition, and in many cases this positivity is easy to preserve in approximating (cutoff) field theories. Their condition of OS positivity yields the existence of the Hilbert space $\mc{H}$, along with the existence of a  positive-energy Hamiltonian. Let us briefly summarize how this goes in order to connect it to the amplitudes of the previous subsection.
The reader may consult the review by Jaffe \cite{Jaffe} for more details.

 
Suppose the Euclidean space time is of the form $\Sigma=\R\times S$ where $\R$ is the Euclidean time. Suppose we have an expectation 
$
\langle F\rangle =\int F(\phi)d\mu(\phi)
$
defined by a measure $\mu$ on some space $E(\Sigma)$ of fields $\phi:\Sigma\to\R$. Let $\caF$ be the space of functions  $F:E(\Sigma)\to\C$ measurable w.r.t. $\phi|_{\Sigma_+}$ where $\Sigma_+=\R_+\times S$ i.e. functionals depending on the field variables at $t\geq 0$. Reflection positivity is the condition 
 $$ 
 \langle \bar F \Theta F\rangle\geq 0
 $$ 
 for all $F\in\caF$ where  $\Theta F$ is a time reflection  $(\Theta F)(\phi)=\overline{F(\theta\phi)}$ and $(\theta\phi)(t,{\bf x}))=\phi(-t,{\bf x})$. Thus  $\langle \bar F \Theta G\rangle
 $ defines an inner product in $\caF$ and the OS Hilbert space is obtained by modding out zero norm states and completing: $\caH$ is  the completion of  $\caF/\{F:\langle \bar F \theta F\rangle= 0\}$. The Hamiltonian $\mathbf{H}$ is then obtained from time translations in $\caF$: it is the generator of a contraction semigroup 
 $$e^{-t\mathbf{H}}[F]=[T_tF]$$
  where $T_tF(\phi)=F(\phi(\cdot+t))$ for $t\geq 0$.
 
 The existence of $\caH$ and $\mathbf{H}$ thus follow under quite general assumptions on a QFT. In many concrete cases like the Liouville CFT,  $\caH$ can indeed, as expected from the classical case, be realized as an $L^2$ space on some  space $E(S)$ of fields defined on the fixed time slice $S$: $\caH=L^2(E(S),d\nu)$ where $\nu$ is a positive measure on $E(S)$. 
%
 If we have a 2d CFT on the Riemann sphere we can take $\Sigma_+$ the unit disc (think of the radial coordinate as $e^{-t}$, $t\in\R_+$) and then the "zero time" slice is the  equator, i.e. the unit circle $\T$.  
   Thus, in such a setup, we would expect the Segal amplitudes  $A_{\Sigma\setminus\cup_i D_i}$ of the previous subsection naturally define elements in the tensor product $\caH^{\otimes n}$. We now turn to a more precise formulation of Segal's axioms.

\subsection{Conformal invariance}
 As explained in the introduction we will consider field theories where the underlying space is a  Riemannian manifold $(\Sigma,g)$. The action functional then is viewed as a function of the fields and of the metric so that we shall denote it $S_{\Sigma,g}(\phi)$. We assume given a collection of datas $\cjg V_{\alpha_1}(x_1)\dots V_{\alpha_m}(x_m)\cjd_{\Sigma,g}$ depending on the surface $(\Sigma,g)$ and some index $\alpha$, which are functions of the points $x_1,\dots,x_n$ on $\Sigma$. Of course, the bracket  $ \cjg - \cjd_{\Sigma,g}$ stands for expectation values with respect to the putative path integral \eqref{PIsegal} but it makes sense to be given these datas even though the path integral is ill-defined: these are just functions, called correlation functions. The first condition, usually assumed for all QFTs, is
 \vskip 1mm
\noindent (1) \textbf{Diffeomorphism covariance}: if $\psi: \Sigma\to \Sigma'$ is a diffeomorphism, then \footnote{We formulate this for scalar fields $V_\alpha$}:
 \begin{align}\label{diffeo_inv_correl}
& \cjg V_{\alpha_1}(\psi_1(x_1)),\dots V_{\alpha_m}(\psi_m(x_m))\cjd_{\Sigma',\psi_*g}= \cjg V_{\alpha_1}(x_1)\dots V_{\alpha_m}(x_m)\cjd_{\Sigma,g}
\end{align}
 
 \vskip 2mm
 
\noindent The second condition, imposed for CFTs is local scale invariance. In two dimensions it reads as
 \vskip 2mm

\noindent(2) \textbf{Conformal covariance }: Let $\omega\in C^\infty(\Sigma)$.  Then  
\begin{align}\label{conf_cov_cor}
& \cjg V_{\alpha_1}(x_1)\dots V_{\alpha_m}(x_m)\cjd_{\Sigma,e^\omega g}= e^{cS_{\rm L}^0(\Sigma,g,\omega)}e^{-\sum_{j=1}^m\Delta_{\alpha_j}\omega(x_j)}\cjg V_{\alpha_1}(x_1)\dots V_{\alpha_m}(x_m)\cjd_{\Sigma,g}\end{align}
 where
 \begin{equation}\label{SL0} 
 S_{\rm L}^0(\Sigma,g,\omega):=\frac{1}{96\pi}\int_{\Sigma}(|d\omega|^2_g+2K_g\omega){\rm dv}_g
 \end{equation}
 is called the {\bf Weyl anomaly}. 
 \vskip 1mm

The parameters $c$ and $\Delta_\alpha$ are the same as in \eqref{Wardidentityexp}: ${\rm c}\in \C$ is a parameter depending on the CFT, called the \textbf{central charge} and 
$ \Delta_\alpha\in \C$ is called the scaling dimension or \textbf{conformal weight} of the operator $V_\alpha$. 
As before $K_g$ is the scalar curvature of $g$ (twice the Gauss curvature).

Let us note also that in CFT the expectation $ \cjg - \cjd_{\Sigma,g}$ is not normalised. Its mass $ \cjg 1 \cjd_{\Sigma,g}:=Z_{\Sigma,g}$ is called the {\bf partition function}. It satisfies the same axioms as above with $m=0$. Notice that this means that the partition and correlation functions, viewed as function on the metric on a fixed conformal class $[g]$, is completely explicit and determined by its value at the minimizer, which is the constant curvature metric, and that it is also invariant by diffeomorphisms. All in all, it says that partition function can be viewed as a function on the moduli space of complex structures (or conformal classes) and the $n$-point 
correlation function as a function  on the moduli space of complex structures with $n$ marked points. The local conformal symmetries that will be discussed in Section \ref{sec:LCS} allow us to show, in certain cases, that these functions on moduli space satisfy certain differential equations, called BPZ or Ward identities, sometimes leading to exact formulas. This is the approach used in \cite{BPZ84} for the sphere, and more generally by \cite{FriedanShenker87} for higher genus surfaces.
 
\subsection{Structure constants}

An immediate consequence of these axioms is that the  spatial dependence of the three point function on the sphere is determined up to a constant. Indeed consider the sphere $\mathbb{S}^2$, which we can represent as the complex plane $\hat{\C}$ with the point $\infty$ added, and equip it with the round metric  $g_{\mathbb{S}^2}=g(z)|dz|^2$ with $g(z)=\frac{4}{(1+|z|^2)^2}$.   For $3$ points $z_1,z_2,z_3\in \hat{\C}$ let $\gamma\in {\rm PSL}_2(\C)$ be  a M\"obius transformation  mapping them to $(0,1,e^{i\pi/3})$. 
 Using first \eqref{diffeo_inv_correl} with $\psi=\gamma$ and then \eqref{conf_cov_cor} and the relation 
$|\gamma(z)-\gamma(z')|=|z-z'||\gamma'(z)|^{1/2}|\gamma'(z')|^{1/2}$ it is readily checked  that 
\begin{equation}\label{str_const}
\begin{split}
 \cjg V_{\alpha_1}(z_1)V_{\alpha_2}(z_2)V_{\alpha_3}(z_3)\cjd_{\mathbb{S}^2,g_{\mathbb{S}^2}}=&
C(\alpha_1,\alpha_2,\alpha_3)\prod_{j=1}^3g(z_j)^{-\Delta_{\alpha_j}}  |z_1-z_2|^{2(\Delta_{\alpha_3}-\Delta_{\alpha_2}-\Delta_{\alpha_2})} \\
& \times|z_1-z_3|^{2(\Delta_{\alpha_2}-\Delta_{\alpha_1}-\Delta_{\alpha_3})}|z_2-z_3|^{2(\Delta_{\alpha_1}-\Delta_{\alpha_2}-\Delta_{\alpha_3})}
 \end{split}\end{equation}
 where $C(\alpha_1,\alpha_2,\alpha_3)=\cjg V_{\alpha_1}(0)V_{\alpha_2}(1)V_{\alpha_3}(e^{i\pi/3})\cjd_{\mathbb{S}^2,g_{\mathbb{S}^2}}$. We used also that the conformal anomaly vanishes for the M\"obius map $\gamma$.
 \begin{figure}[h] \label{picsphere} 
 \begin{tikzpicture}
  \shade[ball color = gray!40, opacity = 0.4] (0,0) circle (2cm);
  \draw (0,0) circle (2cm);
  \draw (-2,0) arc (180:360:2 and 0.6);
  \draw[dashed] (2,0) arc (0:180:2 and 0.6);
   \node at (2,0) [circle,fill,inner sep=1.5pt]{};
    \node at (0,-2) [circle,fill,inner sep=1.5pt]{};
     \node at (0,2) [circle,fill,inner sep=1.5pt]{};
     \draw (2.1,0) node[right,black]{$z_1$} ;
        \draw (0,-2.2) node[right,black]{$z_2$} ;
           \draw (0,2.2) node[right,black]{$z_3$} ;
\end{tikzpicture}
\end{figure}
The constant $C(\alpha_1,\alpha_2,\alpha_3)$ is called the \textbf{structure constant} of the CFT and is a fundamental quantity 
of a CFT, as we shall see later.\\

\subsection{Segal's axioms for CFT}\label{sec:segalcft}

To define the amplitudes and their cutting and pasting we need to introduce surfaces with parametrised boundaries.
We call a Riemann surface with parametrized boundary a compact oriented surface $\Sigma$ with boundary $\pl \Sigma=\cup_{j=1}^b\pl_j \Sigma$ (where $\pl_j\Sigma$ are the connected components, thus circles), a complex structure $J$, and analytic parametrizations of the boundaries
\[ \zeta_j: \mathbb{T}\to \pl_j\Sigma.\]
This means in particular that $\zeta_j$ extends holomorphically on a neighborhood of $\mathbb{T}$ to a neighborhood of $\pl_j\Sigma$. Note that $\theta\mapsto \zeta_j(e^{i\theta})$ can be either positively oriented or negatively oriented inside 
$\Sigma$ (which has an orientation): for example, the circle $\theta \mapsto e^{i\theta}$ is positively oriented in $\mathbb{A}_T$ and 
$\theta \mapsto e^{-T+i\theta}$ is negatively oriented in the annulus $\mathbb{A}_T$.
 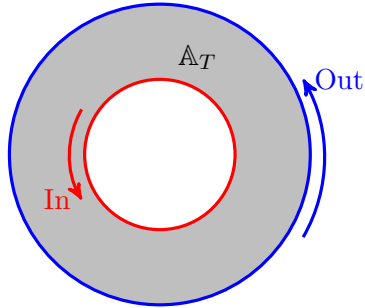
\begin{figure}[h] \label{pic2}
\begin{tikzpicture}[scale=1]
\draw[fill=lightgray,draw=blue,very thick] (0,0) circle (2) ;
\draw[fill=white,draw=red,very thick] (0,0) circle (1) ;
\draw[->,>=stealth',draw=red,very thick] (150:1.2cm) arc[radius=1.2, start angle=150, end angle=210] node[left,red]{In};
\draw[->,>=stealth',draw=blue,very thick] (-30:2.2cm) arc (-30:30:2.1cm) node[right,blue]{Out};
\coordinate (Z1) at (1.5,0) ;
\coordinate (Z2) at (-1,0) ;
\coordinate (Z3) at (-0.8,1.3) ;
\coordinate (Z4) at (-0.7,-0.9) ;
 \draw (0.5,1) node[above]{$ \mathbb{A}_T$} ;
\end{tikzpicture}
\caption{In/out boundaries on $\mathbb{A}_T$ }
\end{figure}
 In the case of positive orientation induced by $\zeta_j$, we say that $\pl_j\Sigma$ is \textbf{outgoing} while for negative orientation we say that  $\pl_j\Sigma$ is \textbf{incoming}.
A metric $g$ on $\Sigma$ is said \textbf{admissible} if it is compatible with the complex structure $J$ 
(i.e. the rotation of angle $\pi/2$ in the tangent space $T_x\Sigma$ according to $g_x$ is equal to $J_x$ for each $x\in \Sigma$) 
and 
\[\zeta_j^*g=\frac{|dz|^2}{|z|^2} \textrm{ near }\mathbb{T}.\]
We shall write $\bs{\zeta}=(\zeta_1,\dots,\zeta_b)$ for the parametrizations of $\pl \Sigma$, and we will also consider marked points 
${\bf x}=(x_1,\dots,x_m)$ and weights $\boldsymbol{\alpha}=(\alpha_1,\dots,\alpha_m)$ attached to the marked points.
Consider  two admissible surfaces $(\Sigma^1,g^1,{\bf x}^1,\bs{\alpha}^1,\boldsymbol{\zeta}^1)$ and $(\Sigma^2,g^2,{\bf x}^2,\bs{\alpha}^2,\boldsymbol{\zeta}^2)$ with respectively $b^1$ and $b^2$ boundary components; if $\pl_i\Sigma^1$ is incoming and  $\pl_j\Sigma^2$ is outgoing, we can form a new admissible surface $(\Sigma,g,{\bf x},\bs{\alpha},\boldsymbol{\zeta})$ with $b^1+b^2-2$ boundary components by gluing $\Sigma^1$ to $\Sigma^2$ along $\pl_i\Sigma^1$ and $\pl_j\Sigma^2$: this is done by setting $\Sigma=(\Sigma^1\sqcup \Sigma^2)/\sim$ where $\sim$ means that we identify $\zeta^1_i(e^{i\theta})= \zeta_j(e^{i\theta})$. We check that $g_1$ and $g_2$ glue smoothly to a metric $g$ on $\Sigma$ and we set ${\bf x}={\bf x}\cup {\bf x}^2$, $\bs{\alpha}=(\bs{\alpha}^1,\bs{\alpha}^2)$ and 
$\bs{\zeta}=(\hat{\bs{\zeta}}^1_i,\hat{\bs{\zeta}}^2_j)$ where $\hat{\bs{\zeta}}^1_i$ is $\bs{\zeta}^1$ with $\zeta^1_i$ removed \[ \hat{\bs{\zeta}}^1_i=(\zeta^1_1,\dots,\zeta^1_{i-1},\zeta_{i+1}^1,\dots,\zeta^1_{b^1})\]
and similarly for $\bs{\zeta}^2_j$.

In the sense of Segal, a CFT is a correspondence between disjoint copies of circles and tensor product of a (separable) Hilbert space 
$\mc{H}$ 
\[(1) \quad \bigsqcup_{j=1}^b \mathbb{T} \rightarrow \mc{H}^{\otimes b}\]
 and between Riemann surfaces with admissible metrics, marked points (with weights) and parametrized boundary
 and amplitudes, which are bounded operators on tensor products of $\mc{H}$: 
   \[(2) \quad (\Sigma,g,{\bf x},\bs{\alpha},\boldsymbol{\zeta}) \rightarrow \mc{A}_{\Sigma,g,{\bf x},\bs{\alpha},\boldsymbol{\zeta}}\in \mc{H}^{\otimes b^+}\otimes (\mc{H}^*)^{\otimes b^-}  \]
 where $b^+$ is the number of positively oriented boundary circles and $b^-$ is the number of negatively oriented ones. Each copy of $\mc{H}$ is attached to a boundary component in this correspondence.
 The element  $\mc{A}_{\Sigma,g,{\bf x},\bs{\alpha},\boldsymbol{\zeta}}$ can also be considered as an operator $\mc{H}^{\otimes b^-}\to \mc{H}^{\otimes b^+}$, which in turn is Hilbert-Schmidt. 
 If $\mc{H}=L^2(E(\mathbb{T}),\mu_0))$ represents the $L^2$ on the space of fields (for some measure $\mu_0$), 
one writes $\mc{H}^{\otimes b}=L^2((E(\mathbb{T})^b,\mu_0^{\otimes b})$ and the 
amplitude $\mc{A}_{\Sigma,g,{\bf x},\bs{\alpha},\boldsymbol{\zeta}}$ can be represented as an integral kernel 
\[ \mc{A}_{\Sigma,g,{\bf x},\bs{\alpha},\boldsymbol{\zeta}}F(\bs{\varphi}) 
=\int_{E(\mathbb{T})^{b^-}}\mc{A}_{\Sigma,g,{\bf x},\bs{\alpha},\boldsymbol{\zeta}}(\bs{\varphi},\bs{\varphi}')
F(\bs{\varphi}')d\mu_0^{\otimes b^-}(\bs{\varphi}')\]
which represents the path integral 
 \begin{equation}\label{ampsegal}
 \mc{A}_{\Sigma,g,{\bf x},\bs{\alpha},\boldsymbol{\zeta}}(\bs{\varphi}) = \int_{ E_{\bs{\varphi}}(\Sigma)}\prod_{j=1}^m V_{\alpha_j}(\phi,x_j) e^{-S_{\Sigma,g}(\phi)}D\phi
 \end{equation}
 where $E_{\bs{\varphi}}(\Sigma)= \{ \phi \in E(\Sigma)\,|\, \phi|_{\pl_j \Sigma}=\varphi_j\}$ is a subspace of fields on $\Sigma$ with fixed boundary values given by $\bs{\varphi}=(\varphi_1,\dots,\varphi_b)$.
 There are 3 fundamental properties required on these amplitudes to define a CFT. The first two are as in the case of the correlation functions\\
 (1) \textbf{Diffeomorphism invariance}: if $\psi: \Sigma\to \Sigma'$ is a diffeomorphism, then  
 \begin{equation}\label{inv_diff_amp} 
 \mc{A}_{\Sigma',\psi_*g,\psi({\bf x}),\bs{\alpha},\psi\circ \bs{\zeta}}=\mc{A}_{\Sigma,g,{\bf x},\bs{\alpha},\boldsymbol{\zeta}},
\end{equation}
where $\psi({\bf x}):=(\psi(x_1),\dots,\psi (x_m))$ and $\psi\circ \bs{\zeta}:=(\psi\circ \zeta_1,\dots,\psi\circ \zeta_b)$.\\
 (2) \textbf{Conformal covariance}: if $\omega\in C_0^\infty(\Sigma\setminus\partial\Sigma)$ (i.e. $\omega$ vanishes near $\pl \Sigma$), 
 then  
\begin{equation}\label{conf_cov_amp}
\mc{A}_{\Sigma,g',{\bf x},\bs{\alpha},\boldsymbol{\zeta}}=e^{cS_L^0(\Sigma,g,\omega)}e^{-\sum_{j=1}^m\Delta_{\alpha_j}\omega(x_j)}
\mc{A}_{\Sigma,g,{\bf x},\bs{\alpha},\boldsymbol{\zeta}},
\end{equation}
 where ${\rm c}\in \C$ is the central charge and $\alpha  \mapsto \Delta_\alpha$ the conformal weights for $\alpha\in \mc{S}$.\\
The final property allows us to compose amplitudes by gluing the underlying surfaces:
\vskip 1mm
\noindent(2) \textbf{Gluing property}: if $(\Sigma^1,g^1,{\bf x}^1,\bs{\alpha}^1,\boldsymbol{\zeta}^1)$ and $(\Sigma^2,g^2,{\bf x}^2,\bs{\alpha}^2,\boldsymbol{\zeta}^2)$ are two admissible surfaces and  $(\Sigma,g,{\bf x},\bs{\alpha},\boldsymbol{\zeta})$ the glued surface along $\pl_i\Sigma^1$ and $\pl_j\Sigma^2$. Then the amplitude of the glued surface is given by 
\begin{equation}\label{gluing_prop} 
\mc{A}_{\Sigma,g,{\bf x},\bs{\alpha},\boldsymbol{\zeta}}(\hat{\bs{\varphi}}^1_i,\hat{\bs{\varphi}}^2_j)=\int_{E(\mathbb{T})}
(\mc{A}_{\Sigma^1,g^1,{\bf x}^1,\bs{\alpha}^1,\boldsymbol{\zeta}^1}(\bs{\varphi}^1)\mc{A}_{\Sigma^2,g^2,{\bf x}^2,\bs{\alpha}^2,\boldsymbol{\zeta}^2}(\bs{\varphi}^2))|_{\varphi^1_i=\varphi^2_j=\varphi}d\mu_{0}(\varphi).
\end{equation}
In terms of elements in tensor products of $\mc{H}$, this can also be written as  
\begin{equation}\label{partial_trace}
\mc{A}_{\Sigma,g,{\bf x},\bs{\alpha},\boldsymbol{\zeta}}={\rm Tr}_{ij}(\underbrace{\mc{A}_{\Sigma^1,g^1,{\bf x}^1,\bs{\alpha}^1,\boldsymbol{\zeta}^1}}_{\in \mc{H}^{\otimes b^1_+}\otimes (\mc{H}^*)^{b^1_-}}\otimes \underbrace{\mc{A}_{\Sigma^2,g^2,{\bf x}^2,\bs{\alpha}^2,\boldsymbol{\zeta}^2}}_{\in \mc{H}^{\otimes b^2_+}\otimes (\mc{H}^*)^{b^2_-}}) \in \mc{H}^{\otimes b^1_++b^2_+-1}\otimes (\mc{H}^*)^{b^1_-+b^2_--1}
\end{equation}
where ${\rm Tr}_{ij}$ means the contraction (trace) of the i-th $\mc{H}^*$ component in $\mc{A}_{\Sigma^1,g^1,{\bf x}^1,\bs{\alpha}^1,\boldsymbol{\zeta}^1}$ with the j-th $\mc{H}$ component in $\mc{A}_{\Sigma^2,g^2,{\bf x}^2,\bs{\alpha}^2,\boldsymbol{\zeta}^2}$.

Technically, it may happen that  the amplitude of some surfaces can still be defined but do not belong to $\mc{H}^{\otimes b^+}\otimes (\mc{H}^*)^{b^-}$, but in some extended Hilbert spaces. The main property that is required in the end is that the amplitudes can be glued and that when gluing surfaces with boundary into a closed surface, the result converges as an element in $\C$ or $\R$, giving sense to the partition and correlation functions.

\subsection{Geometric building blocks and their amplitudes}\label{Sec_GBB}
The gluing property of amplitudes allows to express the correlation functions in terms of amplitudes of a few basic surfaces. 
In this section, we explain briefly the basic building blocks of a CFT. 
 \vskip 1mm
\noindent (1) \textbf{The disk with $1$ point.} Consider the disk $(\D,g)$ with a marked point $0$, attached weight $\alpha\in \C$, and parametrization $\zeta^0:e^{i\theta}\mapsto e^{i\theta}$ for the boundary. Here we have chosen a smooth  admissible metric
$g_{\D,{\rm ad}}=g(z)|dz|^2$ with $g(z)=g(|z|)$ a function of $|z|$ equal to  $1/|z|^2$ near $|z|=1$ and $g(z)=0$ near $z=0$. We extend the metric $g_{\D,{\rm ad}}$ on $\C$ by $g(z)=1/|z|^2$ for $|z|>1$. Then for $\la>1$, let $L_\la:\D\to \la \D$ the dilation $L_\la(z)=\la z$ and write $g_\la=(L_\la)_*g_{\D,{\rm ad}}=(\la^{-2}\frac{g(z/\la)}{g(z)}) g_{\D,{\rm ad}}$. Using \eqref{conf_cov_amp} for the first equality and \eqref{inv_diff_amp} for the second (with $\psi=L_\la$), we have the following identity of amplitudes
\[\mc{A}_{\la \D,g_1,0,\alpha,\la\zeta^0}=e^{{\rm c}S^0_L(\la \D,g_\la,g)-\Delta_{\alpha}\log(\la^2)}\mc{A}_{\la \D,g_\la,0,\alpha,\la\zeta^0}=e^{{\rm c}S^0_L(\la \D,g_\la,g_1)-\Delta_{\alpha}\log(\la^2)}\mc{A}_{\D,g_1,0,\alpha,\zeta^0}\]
where we recall \eqref{SL0}. A few lines computation gives that
\[ S_L^0(\la \D,g_\la,g_1)=\frac{\pi}{12}\log(\la)\]
showing the scaling relation 
\begin{equation}\label{scaling_disk} 
\mc{A}_{\la \D,g_1,0,\alpha,\la \zeta^0}=\la^{-2\Delta_\alpha+\frac{{\rm c}}{12}}\mc{A}_{\D,g_1,0,\alpha,\zeta^0}.
\end{equation}
 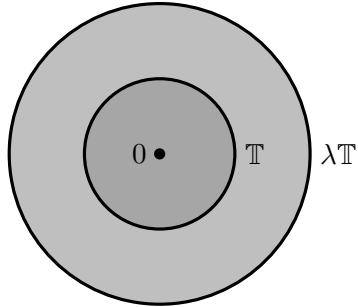
\begin{figure}[h] \label{pic1}
\begin{tikzpicture}[scale=1]
\draw[fill=gray!50,draw=black,very thick] (0,0) circle (2) ;
\draw[fill=gray!70,draw=black,very thick] (0,0) circle (1) ;
\coordinate (Z1) at (1.5,0) ;
\coordinate (Z2) at (-1,0) ;
\coordinate (Z3) at (-0.8,1.3) ;
\coordinate (Z4) at (-0.7,-0.9) ;
 \node at (0,0) [circle,fill,inner sep=1.5pt]{};
\draw (2,0) node[right,black]{$ \la\T$} ;
\draw (1,0) node[right,black]{$  \T$} ;
\draw (-0.5,0) node[right,black]{$0$} ;
\end{tikzpicture}
\caption{Disks $\D$ and $\la \D$, with marked point $z=0$}
\end{figure}
 \vskip 1mm
\noindent 
(2) \textbf{The annuli.}  For $q\in \D^\circ$, the annulus $\A_q=\D\setminus (|q|\D^\circ)$ is equipped with the flat 
metric $g_\A=|dz|^2/|z|^2$ and the boundary parametrizations $\bs{\zeta}_{\A_q}:=(\zeta_1,\zeta_2)$ defined by
\[ \zeta_1: \T\to \T, \quad \zeta(e^{i\theta}):=e^{i\theta} ,\quad  \zeta_2: \T\to |q|\T, 
\quad \zeta_2(e^{i\theta}):=qe^{i\theta}. \]
  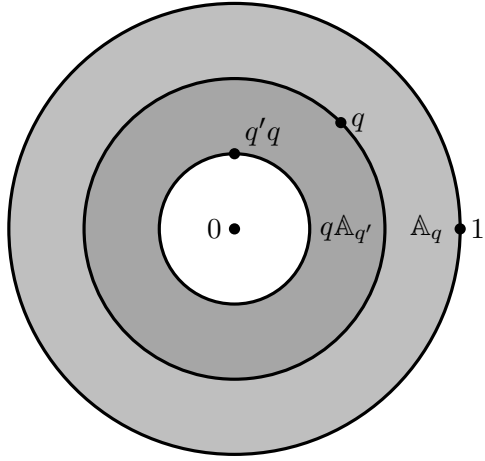
\begin{figure}[h] \label{picannulus}
\begin{tikzpicture}[scale=1]
\draw[fill=gray!50,draw=black,very thick] (0,0) circle (3) ;
\draw[fill=gray!70,draw=black,very thick] (0,0) circle (2) ;
\draw[fill=white,draw=black,very thick] (0,0) circle (1) ;
\coordinate (Z1) at (1.5,0) ;
\coordinate (Z2) at (-1,0) ;
\coordinate (Z3) at (-0.8,1.3) ;
\coordinate (Z4) at (-0.7,-0.9) ;
 \node at (0,0) [circle,fill,inner sep=1.5pt]{};
  \node at (3,0) [circle,fill,inner sep=1.5pt]{};
  \node at (1.414,1.414) [circle,fill,inner sep=1.5pt]{};
    \node at (0,1) [circle,fill,inner sep=1.5pt]{};
\draw (2.2,0) node[right,black]{$\A_q$} ;
\draw (1,0) node[right,black]{$q\A_{q'}$} ;
\draw (-0.5,0) node[right,black]{$0$} ;
\draw (3,0) node[right,black]{$1$} ;
\draw (1.414,1.414) node[right,black]{$q$} ;
\draw (0,1.3) node[right,black]{$q'q$} ;
\end{tikzpicture}
\caption{$\A_q$ glued to $q\A_{q'}$ is $\A_{qq'}$}
\end{figure}
For $q,q'\in \D$, we observe that the gluing $\A_q\# (q\A_{q'})$ of the annulus $\A_q$   with $q\A_{q'}$ by identifying the outgoing boundary of $q\A_{q'}$ with the incoming one of $\A_q$ leads to the annulus $\A_{qq'}$. Since the metric $g_\A$ is invariant by dilation $z\mapsto qz$, we deduce from \eqref{conf_cov_amp} and \eqref{inv_diff_amp} that, viewing the amplitude of the annulus as an operator $\mc{H}\to \mc{H}$,
 \[ \mc{A}_{\A_{qq'},g_\A,\bs{\zeta}_{\A_{qq'}}}=\mc{A}_{\A_q,g_\A,\bs{\zeta}_{\A_q}}\circ \mc{A}_{\A_{q'},g_\A,\bs{\zeta}_{\A_{q'}}}.\]
 This implies that the map 
 \[ q\in \D \mapsto {\bf T}_q:=\mc{A}_{\A_q,g_\A,\bs{\zeta}_{\A_q}}\in \mc{L}(\mc{H})\]
 is a complex semi-group. Restricting to $q=e^{-t}$ with $t>0$, this gives a real semi-group 
 ${\bf T}_{e^{-t}}=e^{-t({\bf H}-\frac{\rm c}{12})}$ where ${\bf H}-\frac{\rm c}{12}$ is the generator. Here we warn the reader that some care is needed due to the fact that often the generator is not bounded, and a priori not necessarily self-adjoint. For Liouville theory, we shall see later that ${\bf H}$ 
 is self-adjoint. This Hamiltonian plays a key role in the conformal bootstrap method, which will essentially consists in decomposing the pairing of building block amplitudes using a basis of eigenfunctions of ${\bf H}$. An amazing feature here is that this basis is constructed from Segal amplitudes of disks. In this direction, let us  make an important observation. If we glue the disk $(\D,g_{1},0,\alpha,\zeta^0)$ with marked point $0$, attached weight $\alpha$, admissible metric $g_{1}=g_{\D,{\rm ad}}$ defined above  and boundary parametrization $\zeta^0(e^{i\theta})=e^{i\theta}$, to  the annulus $(e^t\A_{e^{-t}},g_\A,e^{t}\bs{\zeta}_{\A_{e^{-t}}})$: we obtain the disk
 (note that $g_\A\# g_{1}|_{\D}=g_{1}|_{e^t\D}$ in view of our definitions of $g_\la$ above)
  \[( e^{t}\A_{e^{-t}}\# \D, g_\A\# g_{1}, 0,\alpha, e^t\zeta^0) = (e^{t}\D, g_{1}, 0,\alpha, e^{t}\zeta^0).\]
Using Segal's gluing axiom,  this tells us that 
 \[  \mc{A}_{e^{t}\D,g_{1},0,\alpha,e^{t}\zeta^0}={\bf T}_{e^{-t}} \mc{A}_{\D,g_1,0,\alpha,\zeta^0}\]
thus,  by  the scaling relation \eqref{scaling_disk},
 \[{\bf T}_{e^{-t}} \mc{A}_{\D,g_1,0,\alpha,\zeta^0} =e^{t(\frac{\rm c}{12}-2\Delta_\alpha)} \mc{A}_{\D,g_1,0,\alpha,\zeta^0}.\]
 This implies that 
 \begin{equation}\label{Psialpha} 
 \Psi_{\alpha}:= \mc{A}_{\D,g_1,0,\alpha,\zeta^0}  \textrm{ is an eigenstate of }{\bf H}: \quad {\bf H}\Psi_\alpha=2\Delta_\alpha \Psi_\alpha.\end{equation}
 
  \vskip 1mm
\noindent 
 (3) \textbf{Pairs of pants.} A pair of pants is a surface with genus $0$ and $3$ boundary components. 
 It can be represented in the Riemann sphere as 
 \[\mc{P}= \hat{\C}\setminus \cup_{j=1}^3\mc{D}_j\]
 with $\mc{D}_j$ three disjoint (topological) disks containing respectively $z_1=0$, $z_2=1$ and $z_3=e^{i\pi/3}$, with 
 $\zeta_j:\T\to \pl \mc{D}_j$ some analytic parametrization of $\pl \mc{D}_j$. We put an admissible metric $g_{\mc{P}}$ on $\mc{P}$.
   \begin{figure}[h] 
\begin{tikzpicture}[scale=1]
\draw[fill=gray!50,draw=black,very thick] (0,0) circle (0.7) ;
\draw[fill=gray!50,draw=black,very thick] (2,0) circle (0.7) ;
\draw[fill=gray!50,draw=black,very thick] (1,1.73) circle (0.7) ;
\coordinate (Z1) at (1.5,0) ;
\coordinate (Z2) at (-1,0) ;
\coordinate (Z3) at (-0.8,1.3) ;
\coordinate (Z4) at (-0.7,-0.9) ;
 \node at (0,0) [circle,fill,inner sep=1.5pt]{};
  \node at (1,1.73) [circle,fill,inner sep=1.5pt]{};
     \node at (2,0) [circle,fill,inner sep=1.5pt]{};
\draw (2.6,0.4) node[right,black]{$\mc{D}_2$} ;
\draw (-1.3,0.4) node[right,black]{$\mc{D}_1$} ;
\draw (-0.2,2.4) node[right,black]{$\mc{D}_3$} ;
\draw (-0.5,0) node[right,black]{$0$} ;
\draw (2,0) node[right,black]{$1$} ;
\draw (1,1.73) node[right,black]{$e^{i\frac{\pi}{3}}$} ;
\end{tikzpicture}
\caption{Pair of pants given by $\hat{\C}\setminus \cup_{j=1}^3\mc{D}_j$}
\end{figure}
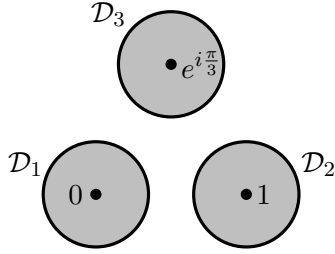
We can glue three disks $D_1,D_2,D_3=\D$ with admissible metric $g_1$ on the three boundary components of the pants, by using the parametrizations $\bs{\zeta}=(\zeta_1,\zeta_2,\zeta_3)$, this gives a sphere $\mathbb{S}^2$ with a metric $\hat{g}$ obtained from the gluing of $g_{\mc{P}}$ with 
$g_1$ on each glued disk $\D$, and there are three marked points $(x_1,x_2,x_3)=(0,1,e^{i\pi/3})$ given by the center of the three glued disks $D_j$. By the uniformization theorem on the sphere, there is a unique conformal, orientation preserving, diffeomorphism 
$\Phi: (\mathbb{S}^2,\hat{g})\to (\mathbb{S}^2,g_{\mathbb{S}^2})$ such that $\Phi(x_j)=x_j$.
Using Segal's gluing axiom and the properties \eqref{diffeo_inv_correl} and \eqref{conf_cov_cor} for the correlations, 
we get
 \[\begin{split} 
 \cjg \mc{A}_{\mc{P},g_{\mc{P},\bs{\zeta}}}, \Psi_{\alpha_1}\otimes \Psi_{\alpha_2},\Psi_{\alpha_3}\cjd_{\otimes ^3\mc{H}}=&
\prod_{j=1}^3g(x_j)^{\Delta_{\alpha_j}}|\Phi'(x_j)|^{-2\Delta_{\alpha_j}}e^{{\rm c}S_L^0(\mathbb{S}^2,g_{\mathbb{S}^2},\hat{g})} 
\cjg V_{\alpha_1}(0)V_{\alpha_2}(1)V_{\alpha_3}(e^{i\frac{\pi}{3}})\cjd_{\mathbb{S}^2,g_{\mathbb{S}^2}}\\
=& C(\alpha_1,\alpha_2,\alpha_3) \prod_{j=1}^3|\Phi'(x_j)|^{-2\Delta_{\alpha_j}}e^{{\rm c}S_L^0(\mathbb{S}^2,g_{\mathbb{S}^2},\hat{g})} 
\end{split} \]
where we recall that $g_{\mathbb{S}^2}=g(z)|dz|^2$ and we used \eqref{str_const} for the second line. This means that the pants amplitude evaluated against the eigenvectors $\Psi_\alpha$ produces the structure constant.

\section{The probabilistic construction of the path integral}\label{sec:PI}
We describe now the construction of the Liouville CFT on closed Riemann surfaces.
We use probabilistic methods to give a mathematical sense to the path integral \eqref{thePI} when the action functional $S$ is given by the Liouville functional      
\begin{equation}\label{AL2}
S_L(\phi,g):= \frac{1}{4\pi}\int_{\Sigma}\big(|d\phi|_g^2+QK_g \phi  + 4\pi \mu e^{\gamma \phi  }\big)\,{\rm dv}_g
\end{equation}
 on a given  two dimensional connected compact Riemannian manifold $(\Sigma,g)$ without boundary. 

First we stress that, at the classical level and when the value of $Q$ is set to $Q=\frac{2}{\gamma}$, finding the minimizer, call it  $u$, of this functional  allows one to uniformize $(\Sigma,g)$. Indeed, the metric $g'=e^{\gamma u}g$ has constant scalar curvature $K_{g'}=-2\pi\mu\gamma^2$ and it is the unique such metric in the conformal class of $g$. 

At the quantum level,  one wants to make sense of  the following measure on some appropriate functional space $\Sigma$ (to be defined later) made up of (generalized) functions $\phi:\Sigma\to \R$
\begin{equation}\label{pathintegral}
F\mapsto \langle F\rangle_{\Sigma,g}:=\int_\Sigma F(\phi)e^{-S_L(\phi,g)}\,D\phi
\end{equation}
where $D\phi$ stands for the ``formal uniform measure'' on $\Sigma$, with the following constraints on the parameters
\begin{equation}\label{param}
\gamma\in ]0,2],\quad Q=\frac{2}{\gamma}+\frac{\gamma}{2},\quad \mu>0.
\end{equation}
Up to renormalizing this measure by its total mass, this formalism describes the law of some random (generalized) function $\phi$ on $\Sigma$, called Liouville field, which stands for the (log-)conformal factor   of a random metric of the form $e^{\gamma \phi}g$ on $\Sigma$.

To make sense of this path integral, we first begin with the analysis of the quadratic part in the action, which gives rise to the Gaussian Free Field (GFF). The GFF is given by a random series: consider  an orthonormal basis $(e_j)_{j\geq 0}$  of eigenfunctions of $\Delta_g$ with eigenvalues $(\la_j)_{j\geq 0}$ (and with $\la_0=0$) and a sequence of independent and identically distributed random variables $(a_j)_j$ defined on some probability space $(\Omega,\mc{F},\mathbb{P})$, which are distributed as real Gaussians   $\mc{N}(0,1)$. The GFF on $(\Sigma,g)$ is then the random series
\begin{equation}\label{GFFclosed}
\forall x\in \Sigma,\quad X_g(x)=\sqrt{2\pi}\sum_{j\geq 1}a_j\frac{e_j(x)}{\sqrt{\la_j}} .
\end{equation}
This series converges in the Sobolev space   $H^s(\Sigma):=(1+\Delta_g)^{-s/2}(L^2(\Sigma))$, for any $s<0$ and with scalar product defined using the metric $g$, as can be seen with the computation of the expectation of his  $H^s(\Sigma)$-norm.  This random variable has vanishing expectation $\E[X_g(x)]=0$ and covariance $\E[X_g(x)X_g(y)]=G_g(x,y)$ (both equalities to be understood in the distributional sense), and $G_g$ is the Green function of the Laplacian on $\Sigma$ with boundary condition fixed by imposing the ${\rm v}_g$ mean over $\Sigma$ to vanish.

The interpretation of the Gaussian measure corresponding to the squared gradient in the Liouville action is then 
 \begin{equation}\label{mesureGFFint}
 \int F(\phi)e^{-\frac{1}{4\pi}\int_\Sigma|d\phi|_g^2{\rm dv}_g} D\phi:=
\big( \frac{{\rm Vol}_g(\Sigma)}{{\rm det}'(\Delta_g)}\big)^{1/2}\int_\R \E[F(c+X_g)]\,\dd c
\end{equation}
for any nonnegative measurable function $F$ on $H^{-s}(\Sigma) $, where ${\rm det}'(\Delta_g)$ is the regularized determinant of the Laplacian, defined as in Ray-Singer \cite{Ray-Singer}. 
 Note that the series \eqref{GFFclosed} is almost surely orthogonal to constant functions so that the $\dd c$ integral stands for integration over constant modes.  The formal equality \eqref{mesureGFFint} is an analogy with the finite dimensional setting for Gaussian integrals. 
To complete the Liouville action, the other terms will be seen as Radon-Nikodym derivatives with respect to this Gaussian measure. The curvature term will raise no difficulty as a linear perturbation of the Gaussian measure but the nonlinear (exponential) term in the Liouville action \eqref{AL2} is more problematic and requires the theory of Gaussian  Multiplicative Chaos theory introduced by Kahane \cite{Kahane85} in the eighties (see \cite{rhodes2014_gmcReview} for a review). Indeed, the GFF is a.s. a distribution so it cannot be understood as a pointwise defined function and cannot be exponentiated straightforwardly. One has to mollify the GFF, say with a mollification at scale $\epsilon$ in the metric $g$, to get a family of regularisations $(X_{g,\epsilon})_{\epsilon>0}$, whose covariance is of the form $G_{g,\epsilon}(x,y)=\ln\frac{1}{d_g(x,y)+\epsilon}+O(1)$ with $d_g$ the distance in the metric $g$. One can then define the random measure on $\Sigma$ by
\begin{equation}\label{GMC}
M^g_{\gamma}(\dd x):=\lim_{\epsilon\to 0}\epsilon^{\gamma^2/2}e^{\gamma X_{g,\epsilon}(x)}\,  {\rm v}_g(\dd x)
\end{equation}
where the limit holds in probability in the sense of weak convergence of measures. The limiting object is non trivial if and only if  $\gamma\in (0,2)$, and this is the reason for our constraint on $\gamma$. The reader may consult \cite[Section 3]{GRV2019} for a pedagogical introduction to the construction of these random measures for non-probabilists.

The mathematical definition of the Liouville path integral \eqref{pathintegral} on $(\Sigma,g)$ is then
\begin{align}\label{defLQFT}
\langle F\rangle_{\Sigma,g}:=& ({\det}'(\Delta_{g})/{\rm Vol}_{g}(\Sigma))^{-1/2}  \\
 &\times \int_\R  \E\Big[ F( c+  X_{g}) \exp\Big( -\frac{Q}{4\pi}\int_{\Sigma}K_{g}(c+ X_{g} )\,{\rm dv}_{g} - \mu  e^{\gamma c}M^g_{\gamma}(\Sigma)  \Big) \Big]\,\dd c \nonumber
\end{align}
for $F$ nonnegative and measurable on $H^{-s}(\Sigma)$. The total mass of this measure, obtained with $F=1$ which is also called the partition function,  is not necessarily finite. This can be understood at the level of the zero mode (also called minisuperspace approximation): this consists in forgetting any GFF contribution and get (up to the metric dependent overall factor)
$$\int_\R e^{-Q\chi (\Sigma)c-\mu e^{\gamma c}}\,\dd c$$
using the Gauss-Bonnet formula $\int_\Sigma K_G\,\dd {\rm v}_g=\chi(\Sigma)$, with $\chi(\Sigma)$ the Euler characteristics of $\Sigma$. On the Riemann sphere ($\chi(\Sigma)=2$)
or the torus ($\chi(\Sigma)=0$) the above integral is diverging, whereas on hyperbolic surfaces ($\chi(\Sigma)\leq -2$) the integral is finite.

Our first claim is that the  measure \eqref{defLQFT} exhibits a bunch of interesting symmetries
 \begin{theorem}\cite{DKRV16,DRV16_tori,GRV2019}\label{introweyl}
 Assume \eqref{param} and let   $g$ be a smooth metric on $\Sigma$. The path integral obeys:
 \begin{enumerate}
\item the total mass is finite iff  $\chi(\Sigma)\leq -2$
\item {\bf Weyl covariance:} for each  bounded continuous functional $F:  H^{-s}(\Sigma)\to\R$ (with $s>0$) and each $\omega\in C^\infty(\Sigma)$,
\[\langle F\rangle_{\Sigma,e^{\omega}g}=\langle F(\cdot\,-\tfrac{Q}{2}\omega)\rangle_{\Sigma,g}\exp\Big(\frac{c_{\rm L}}{96\pi}\int_{\Sigma}(|d\omega|_g^2+2K_g\omega) {\rm dv}_g\Big)\]
where $c_{\rm L}=1+6Q^2$ is a parameter called the central charge. 
\item {\bf Diffeomorphism invariance:} Let   $\psi:\Sigma'\to \Sigma$ be an orientation preserving diffeomorphism and let $\psi^*g$ be the pullback metric on $\Sigma'$. Then,  for   bounded measurable $F:H^{-s}(\Sigma)\to \R$ with $s>0$,
\[ \langle F\rangle_{\Sigma',\psi^*g}  =\langle F(\cdot \circ \psi)\rangle_{\Sigma, g}  .\]
\end{enumerate}
\end{theorem}
The key probabilistic component in the convergence of the path integral is the finiteness of negative moments of the Gaussian multiplicative chaos: 
$\E[M^g_\gamma(\Sigma)^{s}]<\infty$ for all $s\leq 0$.
The Weyl covariance expresses how the path integral reacts to local changes of scales, and this is a stronger statement than the standard global symmetries like scale invariance; this is a first manifestation of the local conformal symmetries which will be discussed further later in the paper. The key components for the proof of the conformal covariance are:
the conformal invariance of the GFF, i.e. if $g'=e^{\omega}g$ 
\[ \int_{\R} \E[ F(X_g+c)]dc=\int_{\R} \E[ F(X_{g'}+c)]dc,\]
the conformal change $e^{\omega}K_{g'}=K_g+\Delta_g\omega$ and the Girsanov transform.

\medskip
Now we turn to the construction of the  correlation functions of  Liouville CFT which  are basically expectation values of Laplace exponents of the Liouville field, meaning we want to plug arbitrary functionals of the form $F(\phi)=\prod_{j=1}^m e^{\alpha_j \phi(x_j)}$ in the path integral \eqref{defLQFT}. Yet, the GFF is not a well-defined function as it belongs to $H^{-s}(\Sigma)$ for $s>0$, so that the construction requires some care.  

We fix $m$ distinct points ${\bf x}=(x_1,\dots,x_n)\in\Sigma^m$ ($m\geq 0$)  with respective associated weights ${\boldsymbol \alpha} =(\alpha_1,\dots,\alpha_m)\in\R^m$.  We define the correlation functions via the regularisation procedure
\begin{align}\label{actioninsertion}
 \langle V_{\alpha_1}(x_1)\dots,V_{\alpha_m}(x_m)\rangle_{\Sigma,g}:=\lim_{\epsilon\to 0}\, \langle V_{\alpha_1,\epsilon}(\phi,x_1)\dots,V_{\alpha_m,\epsilon}(\phi, x_m)\rangle_{\Sigma,g}
\end{align}
where we have set, for fixed $\alpha\in\R$ and $x\in \Sigma$,
\begin{equation*}
V^{\alpha}_{g,\eps}(\phi,x)=\eps^{\alpha^2/2}  e^{\alpha  (\phi_\epsilon(x)) } .
\end{equation*}
The index $\epsilon$ means that we regularise the field with a mollification at scale $\epsilon$, like we did to regularise the GFF.  The operators $V_{\alpha}(x)$ are often called vertex operators. Under some conditions, the above limit makes sense  and is non trivial, as  summarized in the following statement

\begin{proposition}\label{limitcorel}  Consider the set  of conditions,  called Seiberg's bounds,
 \begin{align}\label{seiberg1}
 & \sum_{j=1}^m\alpha_j >\chi(\Sigma)Q,\qquad \text{and }\qquad\forall j=1,\dots,m,\quad \alpha_j<Q .
 \end{align}
The limit \eqref{actioninsertion} exists and is positive iff the Seiberg bounds are satisfied. 
 \end{proposition}

Note that, for the correlation functions to exist, one must have at least three vertex insertions on the Riemann sphere or at least one vertex insertion on the torus, whereas the partition function makes sense on hyperbolic surfaces. There is a geometric interpretation  if one thinks of vertex operators as quantum analogs of conical singularities on Riemann surfaces: recalling that minimizers of the Liouville functional give metrics with uniformized negative curvature then the Seiberg bounds echoes standard geometric conditions for Riemann surfaces to carry metric with negative curvature \cite{Troyanov} (for instance, the Riemann  sphere must then have at least three conical singularities).

The first condition on the Seiberg bound can be understood as  a probabilistic Gauss-Bonnet theorem. Again it can be understood if focusing on the minisuperspace approximation, in which case the zero mode contribution boils down, using the Gauss-Bonnet theorem to analyse the curvature term, to
\begin{equation}\label{defs}
\int_\R e^{sc-\mu e^{\gamma c}}\,\dd c,\qquad \text{ with }\quad s:=\sum_{j=1}^m\alpha_j-\chi(\Sigma)Q.
\end{equation}
This integral is finite provided $s>0$. The second condition is related to the  multifractal nature of the random measure $M_\gamma^g$. Indeed, in the expression \eqref{actioninsertion}, one can perform first the $c$ integration, and then the Cameron-Martin theorem to the exponential terms to land on the following probabilistic expression for the correlation functions,   interesting in its own right,
 \begin{equation*}
 \langle V_{\alpha_1}(x_1)\dots,V_{\alpha_m}(x_m)\rangle_{\Sigma,g}=C(g,\mathbf{x},\boldsymbol{\alpha})     \:   \mu^{-s}\Gamma(s)\:   \E\Big[     Z_\gamma^{g, {\bf  x},{\bf \alpha}}(\Sigma)^{-\frac{s}{\gamma}} \Big] 
 \end{equation*} 
 where $s$ is as above, $\Gamma$ is the usual Gamma function, $Z_\gamma^{g, {\bf  x},{\bf \alpha}}(\Sigma)$ is a random variable defined by 
 \begin{equation}\label{lamesure}
 Z_\gamma^{g, {\bf  x},{\bf \alpha}}(\Sigma):=\int_\Sigma e^{\gamma \sum_{j=1}^m \alpha_j 2\pi G_{g}(x_j,x)} M_\gamma^g(\dd x),  
\end{equation}
and $ C(g,\mathbf{x},\boldsymbol{\alpha}) $ is a less relevant totally explicit function of the parameters
 \begin{equation}\label{Z0Cx}
 C(g,\mathbf{x},\boldsymbol{\alpha}) :=\big(\frac{\det '(\Delta_{g})}{{\rm Vol}_{g}(M)}\big)^{-\frac{1}{2}}     \exp\Big(\sum_i\frac{\alpha_i^2}{2}2\pi m_{g}(x_i,x_i)+2\pi\sum_{i<j}\alpha_i\alpha_j G_{g}(x_i,x_j)\Big)
 \end{equation}
with $m_g$  the Robin mass of the Green function
\begin{equation}\label{greenfct2} 
m_g(x):=\lim_{x\to x'}G_g(x,x')+\frac{1}{2\pi}\log(d_g(x,x'))  .
\end{equation}
In the expression \eqref{lamesure}, the integrand presents a singularity of the form $|x-x_j|^{\gamma\alpha_j}$ around the point $x_j$ and the question is then whether this singularity is integrable with respect to $M_\gamma^g(\dd x)$. Classically one obtains $\alpha_j\gamma<2$ but the regularity properties of $M_\gamma^g$ are more subtle than the Lebesgue measure: this measure has random local H\"older exponents but if one fixes a point deterministically then it has H\"older exponent around this point given by $\gamma Q-\delta$ for any $\delta>0$, so that  one ends up with the condition $\alpha_j<Q$ in order for $ Z_\gamma^{g, {\bf  x},{\bf \alpha}}(\Sigma)$ to be well defined.  The expectation is then finite due to existence of negative moments for Gaussian Multiplicative Chaos measures (see \cite{rhodes2014_gmcReview}).

\medskip
The symmetries of the path integral (Theorem \ref{introweyl}) then translates to the correlation  functions as 

\begin{proposition}\cite{DKRV16,DRV16_tori,GRV2019}
\label{covconf2} Assume the Seiberg bounds are satisfied.  
\begin{enumerate}
\item {\bf Weyl covariance:} For  each $\omega\in C^\infty(\Sigma)$,
\[ \frac{\langle V_{\alpha_1}(x_1)\dots,V_{\alpha_m}(x_m)\rangle_{\Sigma,e^\omega g}}{ \langle V_{\alpha_1}(x_1)\dots,V_{\alpha_m}(x_m)\rangle_{\Sigma,g} }=\exp\Big(\frac{c_{\rm L}}{96\pi}\int_{\Sigma}(|d\omega|_g^2+2K_g\omega) {\rm dv}_g\Big)-\sum_{j=1}^{m}\Delta_{\alpha_j}\omega(x_j)\]
where $c_{\rm L}=1+6Q^2$ is the central charge  and the real numbers $\Delta_{\alpha_i}$, called {\it conformal weights}, are defined by the relation $\Delta_{\alpha}:=\frac{ \alpha}{2}(Q-\frac{\alpha}{2}) $ for $\alpha\in\R$.
\item {\bf diffeomorphism invariance:} Let   $\psi:\Sigma'\to \Sigma$ be an orientation preserving diffeomorphism and let $\psi^*g$ be the pullback metric on $\Sigma'$. Then,  
\[ \langle V_{\alpha_1}(x_1)\dots,V_{\alpha_m}(x_m)\rangle_{\Sigma',\psi^*g}  =\langle V_{\alpha_1}(\psi(x_1))\dots,V_{\alpha_m}(\psi(x_m))\rangle_{\Sigma, g}  .\]
\end{enumerate}
\end{proposition}

\subsection{Structure constants and DOZZ formula}\label{Sec:DOZZ}
In the conformal bootstrap description of CFTs, an important role is played by the three point functions on the Riemann sphere, which can be  identified     with the extended complex plane $\hat\C$ by stereographic projection. Apart from the dependence on the weights, all other parameters of the three point functions  can  be fixed using symmetries as explained in \eqref{str_const}: for $g_{\mathbb{S}^2}=g(z)|dz|^2$ with $g(z)=4/(1+|z|^2)^2$ the round metric, 
\begin{align}
\label{3pointDOZZ}
 \langle  &V_{\alpha_1}(z_1)  V_{\alpha_2}(z_2) V_{\alpha_3}(z_3)  \rangle_{\hat\C,g_{\mathbb{S}^2}}\\
  =&|z_1-z_3|^{2(\Delta_{\alpha_2}-\Delta_{\alpha_1}-\Delta_{\alpha_3})}|z_2-z_3|^{2(\Delta_{\alpha_1}-\Delta_{\alpha_2}-\Delta_{\alpha_3})}|z_1-z_2|^{2(\Delta_{\alpha_3}-\Delta_{\alpha_1}-\Delta_{\alpha_2})}
\Big(\prod_{i=1}^3g(z_i)^{-\Delta_{\alpha_i}}\Big)\nonumber \\
&\times C_0 C_{\gamma,\mu} (\alpha_1,\alpha_2,\alpha_3 ) \nonumber
 \end{align} 
 where $C_0$ is  a universal normalization constant ($\zeta_R$ is Riemann zeta function)
\begin{equation}\label{C_0def}
C_0=\sqrt{\pi} e^{-\frac{1}{4}+2\zeta_R'(-1)-Q^2(1-2\log(2))}
\end{equation}
and $C_{\gamma,\mu} (\alpha_1,\alpha_2,\alpha_3 )$ is called the \textbf{structure constants}.
  
   Dorn and Otto \cite{DornOtto94} (1994) and independently Zamolodchikov and Zamolodchikov \cite{Zamolodchikov96} (1996) proposed a remarkable explicit expression, the so-called DOZZ formula,  for the   structure constants. The original proposal, considered as a guess even in physics standards, was later given strong support  by Teschner \cite{Tesc}. A mathematical proof only appeared   more than twenty years later in \cite{KRV_DOZZ}.
   
 We introduce now the DOZZ formula.   We set $l(z)=\frac{\Gamma (z)}{\Gamma (1-z)}$ where $\Gamma$ denotes the standard Gamma function. We consider Zamolodchikov's special holomorphic function $\Upsilon_{\frac{\gamma}{2}}(z)$ by the following expression for $0<\Re (z)< Q$
\begin{equation}\label{def:upsilon}
\ln \Upsilon_{\frac{\gamma}{2}} (z)  = \int_{0}^\infty  \left ( \Big (\frac{Q}{2}-z \Big )^2  e^{-t}-  \frac{( \sinh( (\frac{Q}{2}-z )\frac{t}{2}  )   )^2}{\sinh (\frac{t \gamma}{4}) \sinh( \frac{t}{\gamma} )}    \right ) \frac{dt}{t}.
\end{equation}
The function $\Upsilon_{\frac{\gamma}{2}}$  is then defined on all $\C$ by analytic continuation of the expression \eqref{def:upsilon} as expression \eqref{def:upsilon}  satisfies the following remarkable functional relations: 
\begin{equation}\label{shiftUpsilon}
\Upsilon_{\frac{\gamma}{2}} (z+\frac{\gamma}{2}) = l( \frac{\gamma}{2}z) (\frac{\gamma}{2})^{1-\gamma z}\Upsilon_{\frac{\gamma}{2}} (z), \quad
\Upsilon_{\frac{\gamma}{2}} (z+\frac{2}{\gamma}) = l(\frac{2}{\gamma}z) (\frac{\gamma}{2})^{\frac{4}{\gamma} z-1} \Upsilon_{\frac{\gamma}{2}} (z).
\end{equation}
The function $\Upsilon_{\frac{\gamma}{2}}$ has no poles in $\C$ and the zeros of $\Upsilon_{\frac{\gamma}{2}}$ are simple (if $\gamma^2 \not \in \Q$) and given by the discrete set $(-\frac{\gamma}{2} \N-\frac{2}{\gamma} \N) \cup (Q+\frac{\gamma}{2} \N+\frac{2}{\gamma} \N )$. 
With these notations, the DOZZ formula is defined for $\alpha_1,\alpha_2,\alpha_3 \in \C$ by the following formula where we set $\bar{\alpha}=\alpha_1+\alpha_2+\alpha_3$ 
\begin{equation}\label{theDOZZformula}
C_{\gamma,\mu}^{{\rm DOZZ}} (\alpha_1,\alpha_2,\alpha_3 )  = (\pi \: \mu \:  l(\frac{\gamma^2}{4})  \: (\frac{\gamma}{2})^{2 -\gamma^2/2} )^{\frac{2 Q -\bar{\alpha}}{\gamma}}   \frac{\Upsilon_{\frac{\gamma}{2}}'(0)\Upsilon_{\frac{\gamma}{2}}(\alpha_1) \Upsilon_{\frac{\gamma}{2}}(\alpha_2) \Upsilon_{\frac{\gamma}{2}}(\alpha_3)}{\Upsilon_{\frac{\gamma}{2}}(\frac{\bar{\alpha}}{2}-Q) 
\Upsilon_{\frac{\gamma}{2}}(\frac{\bar{\alpha}}{2}-\alpha_1) \Upsilon_{\frac{\gamma}{2}}(\frac{\bar{\alpha}}{2}-\alpha_2) \Upsilon_{\frac{\gamma}{2}}(\frac{\bar{\alpha}}{2} -\alpha_3)   }.
\end{equation}      
 The DOZZ formula is meromorphic with poles corresponding to the zeroes of the denominator of expression \eqref{theDOZZformula}. The next theorem is the first result connecting the probabilistic Liouville theory to special functions

\begin{theorem}\cite{KRV_DOZZ}\label{DOZZth}
 Assume the Seiberg bounds are satisfied. Then the structure constant for the Liouville theory is $C_{\gamma,\mu}(\alpha_1,\alpha_2,\alpha_3 )=C_{\gamma,\mu}^{{\rm DOZZ}} (\alpha_1,\alpha_2,\alpha_3 )$, i.e.
 \begin{equation*}
 \langle  V_{\alpha_1}(0)  V_{\alpha_2}(1) V_{\alpha_3}(\infty)  \rangle_{\mathbb{S}^2,g_{\mathbb{S}^2}} =\frac{C_04^{-\Delta_{\alpha_3}-\Delta_{\alpha_1}}}{2}C_{\gamma,\mu}^{{\rm DOZZ}} (\alpha_1,\alpha_2,\alpha_3 ).
  \end{equation*} 
\end{theorem} 
The reader can consult the lecture notes \cite{Vargas} for more on the DOZZ formula,  at the level of both  its heuristic derivation or its proof.

\section{Segal's axioms in Liouville CFT}\label{sec:segalL}
Since the Liouville action \eqref{AL} is local, one can expect the Segal axioms to be valid for the Liouville path integral.  In the case when the surface $\Sigma$ is closed, this amplitudes are nothing but the partition function and correlation functions defined in the previous section, with $E(\Sigma)=H^1(\Sigma)$. When $\Sigma$ has a non empty boundary (and we stick to the notations in Subsection \ref{sec:segalcft}), then the Liouville amplitude \eqref{ampsegal} involves, like in the case of the Dirichlet energy, the space of maps $E_{\bs{\varphi}}(\Sigma)= \{ \phi \in H^1(\Sigma)\,|\, \phi|_{\pl_j \Sigma}=\varphi_j\}$, which can be itself decomposed as $E_{\bs{\varphi}}(\Sigma)=H^1_0(\Sigma)+P\bs{\varphi}$, where $H^1_0(\Sigma)$ is the Sobolev space of functions vanishing on the boundary and  $P\bs{\varphi}$ stands for the harmonic function on $\Sigma$ that coincides in local coordinates with the boundary values collected in the vector $\bs{\varphi}$, meaning $P\bs{\varphi} =\varphi_j\circ \zeta_j^{-1}$ on $\partial_j\Sigma$. Note that if we decompose $ \phi\in E_{\bs{\varphi}}(\Sigma)$ as $\phi=\phi_0+P\bs{\varphi}$ with $\phi_0\in H^1_0(\Sigma)$ then the Dirichlet energy obeys
$$ 
\int_{\Sigma}|d\phi|_g^2{\rm dv}_g=\int_{\Sigma}|d\phi_0|_g^2{\rm dv}_g+\int_{\Sigma}|dP\bs{\varphi}|_g^2{\rm dv}_g.
$$
The second term in the rhs can be interpreted in terms of the Dirichlet-to-Neumann operator  (DN map for short) on $\Sigma$. Equip the unit circle $ \T$ with the   probability measure $\dd \theta/(2\pi)$ and let  $(\cdot,\cdot)_2$ stands for the canonical inner product on $L^2( \T)^{b}$. The definition of   the DN map  $\mathbf{D}_\Sigma:C^\infty( \T)^{b}\to C^\infty( \T)^{b}$ reads
\[\mathbf{D}_\Sigma \boldsymbol{\varphi} =(-\partial_{\nu } P \boldsymbol{\varphi}_{|\partial_j\Sigma}\circ\zeta_j)_{j=1,\dots,b} \]
where $\nu$ is the inward unit normal vector fields to $\partial_j\Sigma$. Then the Green formula connects the Dirichlet energy of the harmonic extension to the DN map
\begin{equation}\label{Greenformula}
\int_{\Sigma} |dP \boldsymbol{\varphi}|_g^2{\rm dv}_g = 2 \pi (  \boldsymbol{\varphi},\mathbf{D}_\Sigma \boldsymbol{\varphi})_2.
\end{equation}
We can thus write  formally the amplitude \eqref{ampsegal} as
\begin{equation}\label{amP}
e^{-\frac{1}{2}(  \boldsymbol{\varphi},\mathbf{D}_\Sigma \boldsymbol{\varphi})_2} \int_{ H^1_0(\Sigma)}\prod_{j=1}^m V_{\alpha_j}(\phi_0+P\bs{\varphi},x_j) e^{-\frac{1}{4\pi} \int_{\Sigma}\big( QK_{{g }} (\phi_0+P\bs{\varphi})  +4\pi \mu e^{ \gamma( \phi+P\bs{\varphi})  }\big)\, {\rm dv}_{{g }}  }e^{-\frac{1}{4\pi}\int_{\Sigma}|d\phi_0|_g^2{\rm dv}_g}D\phi_0.
\end{equation}
As we will soon explain, one can completely makes sense of this expression. There is however a big issue with this formulation: the gluing of amplitudes requires the Hilbert space to be the space $L^2( \T)$ equipped with the "uniform measure", which can be understood as follows. Write any $\varphi\in L^2( \T)$ as
$$\forall\theta\in \R,\quad\varphi(\theta)=c+\sum_{n\not=0}\varphi_ne^{in\theta}$$
where the Fourier coefficients $\varphi_n\in\C$ obey $\varphi_{-n}=\bar{\varphi}_n$. The uniform measure should be $\dd c\otimes_{n\geq 1}(\dd {\rm Re}(\varphi_n)\otimes \dd {\rm Im}(\varphi_n))$ but there is, of course, no mathematical meaning of such a measure. On infinite dimensional spaces, we usually work instead with Gaussian measures, which are well defined. So we want our measure to be
\begin{equation}\label{measuremu0}
e^{- (\mathbf{D}\varphi,\varphi)_2}\dd c\bigotimes_{n\geq 1}(\frac{n}{8\pi}\dd {\rm Re}(\varphi_n)\otimes \dd {\rm Im}(\varphi_n))
\end{equation}
where $\mathbf{D}$ is some operator defined by 
\[(\mathbf{D}\varphi,\varphi)_2:=2\sum_{n>0}n|\varphi_n|^2;\]
the $n/8\pi$ factor comes from the fact that for each $n>0$, $e^{-2n|\varphi_n|^2}\dd {\rm Re}(\varphi_n)
\dd {\rm Im}(\varphi_n)$ has total mass $8\pi/n$ on $\R^2$, and we want a probbility measure.
A rigorous way to define this measure is reparametrize the Fourier coefficients $\varphi_n$, for $n>0$, as $\varphi_n=\frac{x_n+iy_n}{2\sqrt{n}}$, then the measure defining the formal expression \eqref{measuremu0} is 
\[\mu_0=d\varphi_0\bigotimes \prod_{n=1}^\infty \frac{e^{-\frac{x_n^2}{2}-\frac{y_n^2}{2}}}{2\pi}dx_ndy_n\]
and the real coordinates $(x_n,y_n)_{n\geq 0}$ behave under $\mu_0$ as i.i.d. standard Gaussian on a probability space $\Omega_\T=(\R^2)^{\N}$.
Note that $\mu_0$ also produces a Gaussian measure (through the random variable $\varphi$) supported on the Sobolev type space $H^{-\epsilon}(\S^1)$, for any $\epsilon>0$. 
Under $\mu_0$, the random trigonometric series $\varphi$ obtained this way turns out to coincide with the law of the whole plane GFF restricted to the unit circle.
The \textbf{Hilbert space} for the Liouville CFT will be
\[\mc{H}:=L^2(H^{-\epsilon}(\S^1),\mu_0)=L^2\Big(\R\times \Omega_\T, dc\otimes \prod_{n=1}^\infty \frac{e^{-\frac{x_n^2}{2}-\frac{y_n^2}{2}}}{2\pi}dx_ndy_n\Big).\]

The amplitudes need to be corrected  to cancel out  the extra factor $e^{- (\mathbf{D}\varphi,\varphi)_2}$ that we put in the measure $\mu_0$. So we need to multiply the amplitudes by 
 $e^{\frac{1}{2}(\mathbf{D}\varphi,\varphi)_2}$ for each boundary component so that this compensates exactly for the term  $e^{- (\mathbf{D}\varphi,\varphi)_2}$ in $\mu_0$ when we glue two boundary components in $\mc{H}$. If the boundary fields on the surface $\Sigma$ are collected in $\boldsymbol{\varphi}$ in $(H^{-\epsilon}(\S^1))^b$, we still denote by $\mathbf{D}$ the operator
 $$(\mathbf{D}\boldsymbol{\varphi},\boldsymbol{\varphi})_2=\sum_{j=1}^b(\mathbf{D}\varphi_j,\varphi_j)_2.$$
 So we need to multiply \eqref{amP} by $e^{\frac{1}{2}(\mathbf{D}\boldsymbol{\varphi},\boldsymbol{\varphi})_2}$. The integral over $H^1_0(\Sigma)$ in \eqref{amP} then makes sense with the help of the GFF with Dirichlet boundary conditions $X_{g,D}$. It has a  series expansion of the type \eqref{GFFclosed} but now with $(e_j)_{j\geq 0}$ an orthonormal basis of eigenfunctions of the Laplacian with Dirichlet boundary conditions, with eigenvalues $(\la_j)_{j\geq 0}$ and the sum runs over $j\geq 0$.  The random series converges in $H^{-s}(\Sigma)$ ($-1/2<s<0$) and has now the Green function for the Laplacian with Dirichlet boundary condition has covariance and we have the interpretation
 \begin{equation}\label{mesureGFFintdir}
 \int_{H^1_0(\Sigma)} F(\phi_0)e^{-\frac{1}{4\pi}\int_\Sigma|d\phi_0|_g^2{\rm dv}_g} D\phi_0:=
\big({\rm det}'(\Delta_g)\big)^{-1/2}  \E[F( X_{g,D})] 
\end{equation}
for any nonnegative measurable function $F$ on $H^{-s}(\Sigma) $, where ${\rm det}(\Delta_{g,D})$ is the regularized determinant of the Laplacian with Dirichlet b.c., defined as in Ray-Singer \cite{Ray-Singer}.   Starting from this observation, the construction is then similar to the path integral, dealing with the Dirichlet GFF instead of the GFF on closed surfaces. The definition is
\begin{multline}\label{ampL}
 \mc{A}_{\Sigma,g,{\bf x},\bs{\alpha},\boldsymbol{\zeta}}(\bs{\varphi}) =\big({\rm det}(\Delta_{g,D})\big)^{-1/2} e^{-\frac{1}{2}(  \boldsymbol{\varphi},(\mathbf{D}_\Sigma-\mathbf{D}) \boldsymbol{\varphi})_2}
 \\
  \lim_{\epsilon\to 0}\E_{\bs{\varphi}}\Big[  \prod_{j=1}^mV^{\alpha_j}_{g,\eps}(x_j,X_{g,D}+P\boldsymbol{\varphi})e^{-\frac{Q}{4\pi}\int_\Sigma K_g(X_{g,D}+P\boldsymbol{\varphi})\dd {\rm v}_g-\mu M^g_\gamma(X_{g,D}+P\boldsymbol{\varphi},\Sigma)}]
\end{multline}
 where $\E_{\bs{\varphi}}$ means expectation conditional on $\bs{\varphi}$ (we integrate out $X_{g,D}$), 
  the Gaussian Multiplicative Chaos measure is again defined as the limit
 \begin{equation}\label{GMCdir}
M^g_{\gamma}(\phi,\dd x):=\lim_{\epsilon\to 0}\epsilon^{\gamma^2/2}e^{\gamma \phi_\epsilon(x)}\,  {\rm v}_g(\dd x),
\end{equation}
 with $\phi_\epsilon$ stands for a mollification at scale $\epsilon$ of the field $\phi$, and where
 we have set, for fixed $\alpha\in\R$ and $x\in \Sigma$,
\begin{equation*}
V^{\alpha}_{g,\eps}(x,\phi)=\eps^{\alpha^2/2}  e^{\alpha  (\phi_\epsilon(x)) } .
\end{equation*}
 Using Gaussian Multiplicative Chaos theory, one can justify the existence of the limits for $\gamma\in (0,2)$ and $\alpha_j<Q$.  The amplitudes are in $\mc{H}^{\otimes b}$ if the Seiberg bound $\sum_j\alpha_j>\chi(\Sigma)Q$, and  the Euler characteristic is equal to $\chi(\Sigma)=2-2\mathbf{g}-b$. 
 \begin{theorem}\cite{GKRV21_Segal}\label{Th:Segal_gluing}
 If $\alpha_i<Q$ for each $i$, the amplitudes $\mc{A}_{\Sigma,g,{\bf x},\bs{\alpha},\bs{\zeta}}\in e^{ac_-}L^2(H^{-s}(\Sigma)^{b})$ where $a>-\sum_{i=1}^m \alpha_i+Q\chi(\Sigma)$, they satisfy the conformal equivariance and diffeomorphism invariance of \eqref{inv_diff_amp} and \eqref{conf_cov_amp} and the gluing property \eqref{gluing_prop}. When considered as integral kernels of  operators, they are bounded as maps 
 \[\mc{A}_{\Sigma,g,{\bf x},\bs{\alpha},\bs{\zeta}}: \mc{H}^{\otimes b^-}\to \mc{H}^{b^+}\]
 if $b^{+/-}$ is the number of outgoing/incoming boundary components of $\Sigma$.
 \end{theorem}

\section{Hamiltonian}\label{sec:hamiltonian}
As explained in Section \ref{Sec_GBB}, the Segal amplitudes of the annuli $\A_{e^{-t}}$ generates a semi-group ${\bf T}_t$ and its generator is called the Hamiltonian of the Liouville theory. Its spectral resolution plays a key role in the conformal bootstrap.
Below we discuss its probabilistic construction, following \cite[Section 4 \& 5]{GKRV20_bootstrap} and \cite[Section 6]{GKRV21_Segal}.

When viewed as integral kernel of an operator on $L^2(H^{-s}(\T))$, the Segal amplitude of the annulus $\A_{e^{-t}}$ 
\begin{equation}\label{amplitude_Liouville_prop}
\mc{A}_{\A_{e^{-t}},g_\A,\bs{\zeta}_{\A_{e^{-t}}}}(\varphi_1,\varphi_2)=\frac{e^{-\frac{1}{2}(({\bf D}_{\A_{e^{-t}}}-{\bf D})(\varphi_1,\varphi_2),\varphi_1,\varphi_2)_2}}{{\rm det}(\Delta_{\A_{e^{-t}},D})^{1/2}}\mathbb{E}_{\varphi_1,\varphi_2}[e^{-\mu M^{g_\A}_\gamma(X_{g_\A,D}+P(\varphi_1,\varphi_2),\A_{e^{-t}})}]
\end{equation}
where $X_{g_\A,D}$ is the Dirichlet GFF on $\A_{e^{-t}}$.

\subsection{The Free Field Hamiltonian}
The Free Field part corresponds to taking $\mu=0$. Using the Fourier decomposition in the angular  variable $\theta$ for functions on the annulus $\A_{e^{-t}}$, a direct computation leads to the following expression for 
$\mc{A}^0_{\A_{e^{-t}},g_\A,\bs{\zeta}_{\A_{e^{-t}}}}(\varphi_1,\varphi_2)=e^{-\frac{1}{2}(({\bf D}_{\A_{e^{-t}}}-{\bf D})(\varphi_1,\varphi_2),\varphi_1,\varphi_2)_2}$ the free-field amplitude 
\[ \mc{A}^0_{\A_{e^{-t}},g_\A,\bs{\zeta}_{\A_{e^{-t}}}}(\varphi_1,\varphi_2)=\exp\Big(-\frac{(c-c')^2}{2t}-\sum_{n\geq 1}\frac{(x_n'-e^{-nt}x_n)^2}{2(1-e^{-2tn})}-\frac{x_n^2}{2}+\frac{(y_n'-e^{-nt}y_n)^2}{2(1-e^{-2tn})}-\frac{y_n^2}{2}\Big)\]
where $\varphi_1(\theta)=c+\sum_{n=1}^\infty \frac{x_n+iy_n}{2\sqrt{n}}e^{in\theta}+\sum_{n=1}^\infty \frac{x_n-iy_n}{2\sqrt{n}}e^{-in\theta}$ and $\varphi_2(\theta)=c'+\sum_{n=1}^\infty \frac{x_n'+iy_n'}{2\sqrt{n}}e^{in\theta}+\sum_{n=1}^\infty \frac{x_n'-iy_n'}{2\sqrt{n}}e^{-in\theta}$.
The determinant of Laplacian can also be computed explictly using separation of variables:
\[{\rm det}(\Delta_{\A_{e^{-t}},D})^{-1/2}= (\frac{\pi}{t})^{1/2}e^{\frac{t}{12}} \prod_{n\geq 1}(1-e^{-2nt})^{-1}.\]
The action of $\mc{A}^0_{\A_{e^{-t}},g_\A,\bs{\zeta}_{\A_{e^{-t}}}}$  is diagonal with respect of the variables $x_n,y_n,c$: we recognize 
 the integral kernel  $(2\pi t)^{-1/2}e^{-(c-c')^2/(2t)}$ of the heat operator $e^{\frac{t}{2}\pl_c^2}$ in the $0$-mode variable $c$. The semi-group $e^{-t(\pl_{x_n}^*\pl_{x_n})}$ associated to the $1$-dimensional Ornstein-Uhlenbeck 
 operator (the measure being $(2\pi)^{-1/2}e^{-\frac{x_n^2}{2}}dx_n$ and the adjoint $\pl_{x_n}^*$ taken with respect to this measure) 
 has for integral kernel (called Mehler kernel)
 \[e^{-t\pl_{x_n}^*\pl_{x_n}}(x_n,x_n')= \frac{1}{\sqrt{1-e^{-2t}}}\exp\Big(\frac{(x_n'-e^{-t}x_n)^2}{2(1-e^{-2t})}-\frac{x_n^2}{2}\Big).\]
These considerations show that 
\[   \mc{A}^0_{\A_{e^{-t}},g_\A,\bs{\zeta}_{\A_{e^{-t}}}}=\sqrt{2}\pi e^{t\frac{c_{\rm L}}{12}} e^{-t{\bf H}^0}\] 
where ${\bf H}^0$ is the \textbf{Free Field Hamiltonian}
\begin{equation}\label{HamGFF}
{\bf H}^0:= -\frac{1}{2}\pl_c^2+\sum_{n=1}^\infty n(\pl_{x_n}^*\pl_{x_n}+\pl_{y_n}^*\pl_{y_n})+\frac{Q^2}{2}. 
\end{equation}
The Laplacian $-\frac{1}{2}\pl_c^2$ is the generator of the Brownian motion, and $\pl_{x_n}^*\pl_{x_n}$ and $\pl_{y_n}^*\pl_{y_n}$ generate Ornstein-Uhlenbeck processes. This aspect can also be read in the decomposition of the Gaussian Free Field $X=X_{\D,D}+P\varphi$ on the disk $\D$ using the radial coordinates $(t,\theta)\mapsto e^{-t+i\theta}$
\[ X(e^{-t+i\theta})= X_0(t)+c +\sum_{n\geq 1}e^{in\theta}X_n(t)+\sum_{n\geq 1}e^{-in\theta}\bbar{X_n}(t)\]
with $X_0(t)$ a Brownian with covariance $\min(t,t')$, $X_n(t)=\frac{x_n(t)+iy_n(y))}{2\sqrt{n}}$ with $x_n(t),y_n(t)$ independent Ornstein-Uhlenbeck processes. The propagator $e^{-t{\bf H}^0}$ can then be represented as the following \textbf{Markov process}: for $F\in L^2(H^{-s}(\T))$ with $s>0$
\begin{equation}\label{FKGFF} 
e^{-t{\bf H}^0}F(\varphi)=e^{-\frac{Q^2}{2}t}\E_{\varphi}[ F((X\circ e^{-t})|_{\T})].
\end{equation}
The operator ${\bf H}_0$ is associated to the quadratic form 
\[\mc{Q}_0(F,F):=\frac{1}{2}\|\pl_c F\|^2_{\mc{H}}+\sum_nn(\|\pl_{x_n}F\|_{\mc{H}}^2+\|\pl_{y_n}F\|_{\mc{H}}^2)+\frac{Q^2}{2}\|F\|^2_{\mc{H}}\] and is self-adjoint. It can be diagonalized by using the family  $\Psi_{Q+ip,{\bf k},{\bs \ell}}^0$ for $p\in \R,{\bf k},\bs{\ell}\in \mc{N}$ where 
\[ \Psi_{\alpha,{\bf k},{\bs \ell}}^0:= e^{(\alpha-Q) c}\prod_{n\geq 1}{\rm He}_{k_n}(x_n){\rm He}_{\ell_n}(y_n), \quad p\in \R, \, {\bf k}=(k_n)_{n\in\N},\, \bs{\ell}=(\ell_n)_{n\in\N}\]
where ${\rm He}_k(x)=(-1)^ke^{x^2/2}\pl_x^k(e^{-x^2/2})$ is the Hermite polynomial and $\mc{N}$ is the set of finite sequences of integers. If $|{\bf k}|:=\sum_nnk_n$, we have $({\bf H}_0-2\Delta_{\alpha}+|{\bf k}|+|\bs{\ell}|) \Psi_{\alpha,{\bf k},{\bs \ell}}^0=0$ for all $\alpha \in \C$. 
\subsection{The Liouville Hamiltonian}\label{Liouville_Hamiltonian}
Using the expression \eqref{amplitude_Liouville_prop} for the integral kernel of the annulus propagator ${\bf T}_{e^{-t}}$ and the Markov properties of the GFF, one can write this operator under the Feynman-Kac representation (just as \eqref{FKGFF}),  with $X=X_{\D,D}+P\varphi$ the GFF on the disk, 
\begin{equation}\label{FK}
{\bf T}_{e^{-t}}F(\varphi)=e^{-\frac{Q^2t}{2}}\E_{\varphi}[ F((X\circ e^{-t})|_{\T})e^{-\mu\int_{\A_{e^{-t}}}|z|^{-\gamma Q}M_\gamma(X,dx)}].
\end{equation} 
Since ${\bf T}_{e^{-t}}$ is a self-adjoint bounded semigroup, there is a self-adjoint generator 
${\bf H}$, i.e. ${\bf T}_{e^{-t}}=e^{-t{\bf H}}$, which is then computed by differentiating at $t=0$:
\begin{equation}\label{HamiltonianL}
{\bf H}={\bf H}^0+\mu V, \quad V(\varphi):=\int_{\T}e^{\gamma \varphi(\theta)}d\theta
\end{equation}
where the potential $V$ can be represented as a multiplication by a positive function if $\gamma<\sqrt{2}$, 
of the form $V=e^{\gamma c}V_0$ for some $V_0$ independent ot $c$ such that $V_0 \in L^p(\Omega_\T)$ for all $p<2/\gamma^2$ (if $\gamma\in [\sqrt{2},2)$, $V$ has to be understood via a quadratic form and is a non-negative operator). The domain of ${\bf H}$ is defined from the semi-group, and it is a non-tivial result \cite[Section 5.2]{GKRV20_bootstrap} that ${\bf H}$ is the operator coming from the Friedrichs extension of the quadratic form 
\[ \mc{Q}(F,F)=\mc{Q}_0(F,F)+\mu\cjg VF,F\cjd_{\mc{H}} \textrm{ with domain }\mc{D}(\mc{Q}).\]
The potential, as function of $c$, acts as a barrier for $c\to \infty$ while at $c\to -\infty$ it becomes negligible, which roughly means that in this regime ${\bf H}$ behaves like the free field Hamiltonian.  
 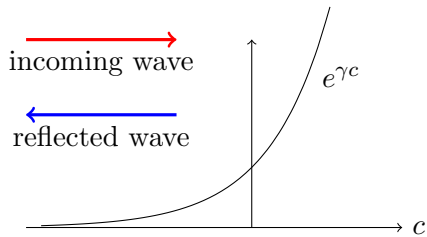
\begin{figure}[h]
\begin{tikzpicture}
  \draw[->] (-3, 0) -- (2, 0) node[right] {$c$};
  \draw[->] (0,0) -- (0, 2.5) node[above] {$$};
  \draw[scale=0.8, domain=-3.5:1.3, smooth, variable=\x, black] plot ({\x}, {exp(\x)});
  \draw (0.8,2) node[right,black]{$e^{\gamma c}$} ;
   \draw[->,draw=red,very thick] (-3,2.5) --node[midway,below]{$\textrm{incoming wave}$} (-1,2.5) ;
    \draw[->,draw=blue,very thick] (-1,1.5) --node[midway,below]{$\textrm{reflected wave}$} (-3,1.5) ;
\end{tikzpicture}
\caption{The toy model potential $V=e^{\gamma c}$}
\end{figure}
For the toy model operator ${\bf H}_{\rm mod}:=-\frac{1}{2}\pl^2_c+e^{\gamma c}$ on $L^2(\R)$, we can easily show that the spectrum is only made of continuous spectrum $[0,\infty)$ and for each $p\in \R$ there is a unique solution $\psi_p$ of
$({\bf H}_{\rm mod}-\frac{p^2}{2})\psi_{p}(c)=0$ which satisfies for some $S(p)\in \C$
\[ \psi_p|_{\R^+}\in L^2(\R^+), \quad  \psi_p|_{\R^-}=e^{ipc}+S(p)e^{-ipc}+\tilde{\psi}_p, \,\textrm{with } \tilde{\psi}_p\in L^2(\R^-).\]
Here, $e^{ipc}$ is an incoming plane wave (and an eigenfunction of the free operator $-\frac{1}{2}\pl_c^2$)
while $S(p)e^{-ipc}$ is an outgoing reflected plane wave as $c\to -\infty$. Moreover, $\psi_p$ can be written in terms of modified Bessel function and thus extends to $p\in \C$ meromorphically as a $C^\infty(\R)$ function.
For the true Hamiltonian ${\bf H}$, a similar result is proved: 
\begin{theorem}\cite{GKRV20_bootstrap}\label{Theorem_spectral}
For each ${\bf k},\bs{\ell}\in \mc{N}$ with $|{\bf k}|+|\bs{\ell}|=N$, there is a connected open set $\Omega_{N}\subset \C$ containing
$(Q+i\R)\cup (-\infty,Q-C_N)$ for some $C_N>0$, and an 
analytic family\footnote{Here $c_-$ denote a smooth function on $\R$ equal to $c$ on $(-\infty,0)$ and $0$ in $(1,\infty)$} $\alpha \in \Omega_{{\bf k},\bs{\ell}} \mapsto \Psi_{\alpha,{\bf k},\bs{\ell}}\in e^{(-|Q-{\rm Re}(\alpha)|-\eps)c_-}\mc{H}$ such that
$({\bf H}-2\Delta_{\alpha}-|{\bf k}|+|\bs{\ell}|)\Psi_{\alpha,{\bf k},\bs{\ell}}=0$ and satisfying the following properties:
\begin{align*}
& ({\bf Intertwinning})  \quad\quad\quad   \forall \alpha < Q-C_N, \quad \Psi_{\alpha,{\bf k},\bs{\ell}}=\lim_{t\to +\infty}e^{t(2\Delta_{\alpha}+|{\bf k}|+|\bs{\ell}|)}e^{-t{\bf H}}\Psi^0_{\alpha,{\bf k},\bs{\ell}},\\ 
& ({\bf  spectral} \,\,{\bf resolution})\quad \forall u,v\in \mc{H},\,\,   \cjg u,v\cjd_{\mc{H}}=\frac{1}{2\pi}\int_0^\infty \sum_{{\bf k},\bs{\ell}\in \mc{N}} \cjg u,\Psi_{Q+ip,{\bf k},\bs{\ell}}\cjd_2 \cjg \Psi_{Q+ip,{\bf k},\bs{\ell}},v\cjd_{\mc{H}}dp.
\end{align*} 
If $\alpha<Q$, the interwining formula leads to 
\[\Psi_{\alpha}(\varphi)=e^{(\alpha-Q)c}\E_\varphi[ e^{-\mu \int_{\D}|x|^{-\gamma \alpha}M_\gamma(X,dx)}]=\mc{A}_{\D,|dx|^2,0,\alpha,\zeta^0}(\varphi)\]
with $X=X_{\D,D}+P\varphi$, i.e. the eigenstate for $\alpha<Q$ is the disc amplitude as expected from \eqref{Psialpha}.
\end{theorem}
The difficult part is to extend analytically the eigenstate to the non-probabilistic region, in particular to $Q+i\R$. The proof is done using scattering theory.
 \begin{figure}[h]
 \begin{tikzpicture}[xscale=1,yscale=1]  
\newcommand{\B}{(0,-0.1) rectangle (4.4,3.3)};
\newcommand{\Ci}{(4,-0.2) arc (30:150:1.1) -- (3,-0.2)};
\fill[pattern=north east lines, pattern color=gray] \B;
\fill[white] \Ci;
\node[below] at (4.2,-0.2) {$Q$};
\draw[line width=1pt] (2.1,-0.1) -- (2.1,0.1) ;
\draw [line width=2pt,color=blue](4.2,-0.1) --   (4.2,3);
\draw[line width=1pt,<-] (4.2,2) -- (4.6,2)node[right]{Spectrum line } ;
\node[below] at (5.5,1.8) {$Q+i\R$};
\shade[left color=gray!10,right color=gray,opacity=0.7] (2,0) arc (0:25:2) -- (0,3) -- (0,0) -- (2,0);
\shade[left color=gray!10,right color=gray] (2,0) arc (0:-5:2) -- (0,-0.2) -- (0,0) -- (2,0);
\draw[style=dashed,line width=1pt,->] (3,-0.1) node[below]{$0$} -- (3,3) node[left]{{\rm Im} $\alpha$};
\draw[style=dashed,line width=1pt,->] (0,0) -- (6,0)node[below]{{\rm Re} $\alpha$};
\node[below] at (0,1) {{\small Probabilistic region}};
\node[below] at (2.4,2.5) {{\small Analyticity region}};
\node[below] at (2.1,-0.2) {$Q-C_N$};
\end{tikzpicture}
\caption{Region of analyticity $\Omega_N\cap {\rm Im}(\la)\geq 0$ is the gray region. The probabilistic region is the region where the interwtining formula still converge.}
\label{art:intro}
\end{figure}
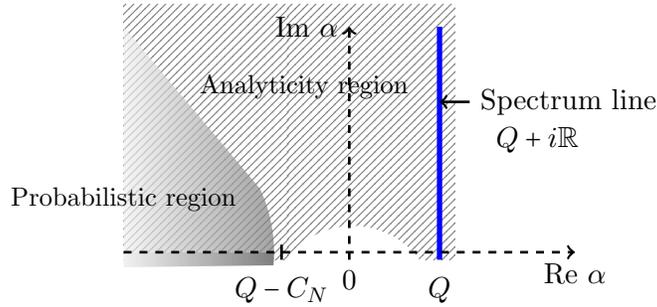

\section{Local conformal symmetries}\label{sec:LCS}

The Hamiltonian of Liouville CFT encodes the effect of dilations on the path integral. Yet the theory has way more symmetries, the local conformal symmetries, that still can be captured by Segal's framework. Recall that the Hamiltonian was defined as the generator associated to the semigroup of annulus amplitudes. Local conformal symmetries emerge when considering more general annuli, when one boundary is given by the image of the unit circle by a conformal map.

\subsection{Flow of deformations}
More precisely, let ${\rm Hol}(\D)$ be the complex vector space of holomorphic maps on the unit disk $f:\D\to\C$ which are smooth up to the boundary $\S^1=\pl\D$. Segal's semigroup of holomorphic annuli is the open subset of ${\rm Hol}(\D)$ defined by
\begin{equation}\label{defS}
\mc{S}:=\{f\in {\rm Hol}(\D)\,\,\, |\,\, f(0)=0,\,\,f_{|_\T}\text{ injective and }f(\D)\subset\D^\circ\}
\end{equation}
  \begin{figure}[h] 
\begin{tikzpicture}[scale=0.8]
\draw[fill=lightgray,draw=blue,very thick] (0,0) circle (2) ;
\coordinate (Z1) at (1.5,0) ;
\coordinate (Z2) at (0,0) ;
\coordinate (Z3) at (-0.8,1.3) ;
\coordinate (Z4) at (-0.7,-0.9) ;
\draw[>=latex,draw=red,very thick,fill=white] (Z1) to[out=120,in=40] (Z2) to [out=220,in=40](Z3) to [out=220,in=70](Z4) to [out=250,in=300](Z1);
  \draw (0,-2) node[above,red]{$ f(\T)$} ;
\draw (2,0) node[right,blue]{$ \T$} ;
 \draw (0.5,0.5) node[above]{$\mathbb{A}_f$} ;
 \draw (0.3,-0.9) node[above]{$f(\D)$} ;
\end{tikzpicture}
\end{figure}  
Such a $f\in \mc{S}$ defines an amplitude $\mc{A}_{\mathbb{A}_f,g,\bs{\zeta}_{f}}$ on the annulus $\mathbb{A}_f:=\D\setminus f(\D)$, with boundary parametrization  $\bs{\zeta}_{f}:=(\zeta_1,\zeta_2)$ defined by
\[ \zeta_1: \T\to \T, \quad \zeta(e^{i\theta}):=e^{i\theta} ,\quad  \zeta_2: \T\to f(\T), 
\quad \zeta_2(e^{i\theta}):=f(e^{i\theta}). \]
In fact, the choice of the amplitude associated to $f$ is unique up to the choice of the admissible metric $g$ on $\A_f$. There is a choice of  specific admissible metric  for which the operator $\mc{H}\to\mc{H}$ with integral kernel given by the amplitude $\mc{A}_{\mathbb{A}_f,g,\bs{\zeta}_{f}}$ has a nice Feynman-Kac representation. For this, consider the operator  $\mc{H}\to\mc{H}$ defined by
\begin{align*}
& \mathbf{T}_fF( \varphi):=|f'(0)|^\frac{Q^2}{2}\E \left[F\Big((X \circ f +Q\ln\Big|\frac{f'}{f}\Big|)|_\T\Big)e^{-\mu\int_{\A_f}|x|^{-\gamma Q}M^{g=|\dd z|^2}_\gamma (X, \dd x) }\right], 
\end{align*}
where $X:=X_{\D}+P\varphi$, $X_{\D}$ is the Dirichlet GFF on the unit disk $\D$ and expectation above is with respect to $X_\D$.
\begin{proposition}\label{cor:annulus_propagator}\cite{BGKR1}
For $f\in \mc{S}$ and $g=e^{\omega}|dz|^2$ conformal to. $|dz|^2$ and admissible, then   
\[ e^{W(f,g)} {\bf T}_fF( \varphi_1)= 
\int \mc{A}_{\mathbb{A}_f,g,\boldsymbol{\zeta}}(\tilde{\varphi}_1,\tilde{\varphi}_2)F(\tilde{\varphi}_2) \,\dd\mu_0(\tilde{\varphi}_2)
\]
for some constant $W(f,g)$ which is is explicit but whose value will not serve in this review. 
\end{proposition}

In fact, $\mc{S}$ has a natural structure of complex manifold, as an open subset  of ${\rm Hol}^\bullet(\D)$, made up of those $f\in {\rm Hol}(\D)$ such that $f(0)=0$. Furthermore, the operators $\mathbf{T}_f$ compose nicely in the sense that $\mathbf{T}_{f_1}\mathbf{T}_{f_2}=\mathbf{T}_{f_1\circ f_2}$ for $f_1,f_2\in \mc{S}$ so that we have a representation of the semigroup $\mc{S}$ with the $\mathbf{T}_f$'s. A natural further step is thus to identify the representation of the Lie algebra associated to $\mc{S}$.
For this, note that the tangent space to  $\mc{S}$ is made up of holomorphic vector fields $v(z)\partial_z$ with $v\in {\rm Hol}^\bullet(\D)$. Such a  holomorphic vector field   is called \emph{Markovian} if  
\begin{equation}\label{eq:def_markovian}
{\rm Re}(\bar{z}v(z))<0,\qquad \forall z\in\T. 
\end{equation} 
This condition ensures that its flow, defined by the ODE $\pl_tf_t(z)=v(f_t(z))$ with $f_0(z)=z$, is a family of holomorphic univalent maps with 
$f_{t+s}(\D)\subset f_t(\D)$ for all $t,s\geq 0$ and so that   $f_t(\D)\to \{0\}$ as $t\to \infty$. 
\begin{proposition}\label{genvectorfield}
\cite{BGKRV} Let $v(z)\partial_z$ be a Markovian vector field, and $f_t$ its flow. Then the family $(\mathbf{T}_{f_t})_{t\geq 0}$ is a continuous contraction semigroup on $\mc{H}$, with a generator    ${\bf H}_{v}$   which can be decomposed as 
\begin{equation}\label{defopHv}
{\bf H}_{v}=\sum_{n=0}^\infty v_n\mathbf{L}_n+\sum_{n=0}^\infty \bar{v}_n\tilde{\mathbf{L}}_n, \quad \textrm{ if } v(z)=-\sum_{n=0}^\infty v_nz^{n+1}  
\end{equation}
where 
\begin{equation}\label{formofLn} 
\mathbf{L}_n =\mathbf{L}_n^0+ \frac{\mu}{2}\int_0^{2\pi}e^{in\theta}e^{\gamma \varphi(\theta)}\dd \theta, \quad \tilde{\mathbf{L}}_n =\tilde{\mathbf{L}}_n^0+ \frac{\mu}{2}\int_0^{2\pi}e^{-in\theta}e^{\gamma \varphi(\theta)}\dd \theta. 
\end{equation}
Here $\mathbf{L}_n,\tilde{\mathbf{L}}_n$ and ${\bf H}_{v}$ are unbounded operators on $\mc{H}$, mapping continuously $\mc{D}(\mc{Q})\to \mc{D}'(\mc{Q})$.  
\end{proposition}

We should comment on the statement of this proposition. First, note that the operators $\mathbf{T}_f$ make sense in the case $\mu=0$, which we call the case of $Q$-GFF, and then the expansion \eqref{defopHv} is in terms of the operators $(\mathbf{L}^0_n)_n$ and $(\tilde{\mathbf{L}}^0_n)_n$ since $\mathbf{L}_n =\mathbf{L}_n^0$ and $\tilde{\mathbf{L}}_n =\tilde{\mathbf{L}}_n^0$ when $\mu=0$.  The operators $\mathbf{L}^0_n$ and $\tilde{\mathbf{L}}^0_n$  are thus the generators of the Q-GFF. It turns out that we can have a completely  explicit description of these operators on $\mc{H}$, which is called the Sugawara construction. We will not expand this any further here because it will not be important for our discussion, we refer to \cite[Section 4.4.]{GKRV20_bootstrap}.
When $\mu\not=0$, 
 the potential term $\int_0^{2\pi}e^{in\theta}e^{\gamma \varphi(\theta)}\dd \theta$ appearing in the expression of $\mathbf{L}_n$ (and similarly for  $\tilde{\mathbf{L}}_n $) is a Gaussian Multiplicative Chaos on the unit circle and must then be defined as before with a regularization procedure. It makes sense $\mu_0$-almost surely when $\gamma$ is small enough, meaning $\gamma^2<2$ as predicted by Gaussian Multiplicative Chaos theory (the circle has dimension 1). When $2\leq \gamma^2<4$ this random variable cannot be anymore defined by Gaussian Multiplicative Chaos theory but still makes sense as a multiplication operator. In any case, we obtain the operators $\mathbf{L}_n$ and $\tilde{\mathbf{L}}_n$ as explicit perturbations of the Q-GFF generators. Finally we stress that the Hamiltonian of the Liouville theory corresponds to  $v_0(z)=-z$, in which case $f_t(z)=e^{-t}z$ is the flow of dilations, and then we get $\mathbf{H}=\mathbf{L}_0+\tilde{\mathbf{L}}_0$. Let us explain where the commutation relation between the ${\bf L}_n$ comes from: observe that for Markovian $v,w\in {\rm Hol}^\bullet(\D)$
$$e^{-t\mathbf{H}_{ v}}e^{-s\mathbf{H}_{ w}}= e^{-s\mathbf{H}_{{  w}_t}}e^{-t\mathbf{H}_{  v}}$$ where $w_t(z):=\frac{w(f_t^{-1}(z))}{(f_t^{-1})'(z)}$ and $f_t$ is the flow $\partial_tf_t(z)=v(f_t(z))$ and $f_0(z)=z$. This relation uses the fact that, if $h_s$ is  the solution  of the flow $\partial_sh_s(z)=w(h_s(z))$ and $h_0(z)=z$ then $\bar h_s:=f_t\circ h_s\circ f_t^{-1}$ is the solution of the flow $\partial_s\bar h_s(z)=w_t(\bar h_s(z))$ and $\bar h_0(z)=z$.  Differentiating this relation at $t=0$ we obtain
$$[\mathbf{H}_{  v},\mathbf{H}_{ w}]=\mathbf{H}_{[{  v},{  w}]}$$
with $[{ v},{ w}]$ the commutator of the two vector fields. We deduce that the families $(\mathbf{L}_n)_{n\geq 0}$ and $(\tilde{\mathbf{L}}_n)_{n\geq 0}$ commute, i.e. $[\mathbf{L}_n,\tilde{\mathbf{L}}_m]=0$,  and, specializing  to $v=z^n,w=z^m$, that they obey the commutation relations of the Witt algebra $[\mathbf{L}_n,\mathbf{L}_m]=(n-m)\mathbf{L}_{n+m}$ (and similarly for the $\tilde{\mathbf{L}}_n$'s). In fact, this argument is not quite correct due to the fact that ${\bf L}_n{\bf L}_m$ can not be easily defined due 
to domain issues, but it can be made rigourous for $\mu=0$, i.e. for $[{\bf L}_n^0,{\bf L}^0_m]$.

\subsection{Representation of the Virasoro algebra}\label{rep_virasoro}
Actually, restricting to positive modes ($n\geq 0$) does not give the full picture. The  definition of the operators $\mathbf{L}_n,\tilde{\mathbf{L}}_n:\mc{D}(\mc{Q})\to \mc{D}'(\mc{Q})$ can be extended  to $n\in\Z$ by the relation: $\forall F,G\in \mc{D}(\mc{Q})$
\[ \langle \mathbf{L}_{-n}F,G\rangle_{\mc{H}}:=\langle F,\mathbf{L}_n G\rangle_{\mc{H}}\qquad\text{ and }\qquad  \langle  \tilde{\mathbf{L}}_{-n}F,G\rangle_{\mc{H}}:=\langle  F,\tilde{\mathbf{L}}_n G\rangle_{\mc{H}}.\]
The same definition applies to the $\mu=0$ case, with $\mc{D}(\mc{Q})$ replaced by $\mc{D}(\mc{Q}_0)$, 
and the operators $\mathbf{L}^0_n$ and $\tilde{\mathbf{L}}^0_n$ produce two commuting  unitary representations of the Virasoro algebra, which is the central extension of the Witt algebra.  
The fact that the two families  $(\mathbf{L}^0_n)_{n\in\mathbb{Z}}$ and $(\tilde{\mathbf{L}}^0_n)_{n\in\mathbb{Z}}$ commute is expressed by $[\mathbf{L}^0_n,\tilde{\mathbf{L}}^0_m]=0$. The representations are unitary in the sense that $(\mathbf{L}_n^0)^*=\mathbf{L}^0_{-n}$ and $(\tilde{\mathbf{L}}_n^0)^*=\tilde{\mathbf{L}}^0_{-n}$. And finally they obey the commutation relations of the Virasoro algebra, namely  for all $n,m\in\Z$ 
\begin{align}
&[\mathbf{L}^0_n,\mathbf{L}^0_m]=(n-m)\mathbf{L}^0_{n+m}+\frac{c_L}{12}(n^3-n)\delta_{n,-m},\label{vircom1}\\
&[\tilde{\mathbf{L}}^0_n,\tilde{\mathbf{L}}^0_m]=(n-m)\tilde{\mathbf{L}}^0_{n+m}+\frac{c_L}{12}(n^3-n)\delta_{n,-m}\label{vircom2}
\end{align}
 where $c_{\rm L}=1+6Q^2$ is the central charge. To understand the case $\mu\not=0$ and give a sense to the commutation relations of the ${\bf L}_n$ and to capture the origins of the probabilistic representation,
 we need to come back to the eigenstates $\Psi_{Q+ip,{\bf k},\bs{\ell}}^0$ of the free Hamiltonian \eqref{HamGFF}. These eigenstates are generated by the Heisenberg algebra (whose creation operators increase the degrees of the Hermite polynomials) but are not so much adapted to the symmetries in the Liouville theory. We shall therefore use a more adapted basis. For this, let us start with the state $ \Psi_{\alpha}^0:=\Psi_{\alpha,0,0}^0$, which is an eigenstate $\mathbf{H}^0\Psi_{\alpha}^0=2\Delta_\alpha \Psi_{\alpha}^0$. Actually, it is not hard to see, using basic GFF computations, that it is even an eigenstate of both $\mathbf{L}_0^0 $ and $\tilde{\mathbf{L}}_0^0$, that is $\mathbf{L}_0^0  \Psi_{\alpha}^0=\Delta_\alpha \Psi_{\alpha}^0$ and $\tilde{\mathbf{L}}_0^0  \Psi_{\alpha}^0=\Delta_\alpha \Psi_{\alpha}^0$, and that it is annihilated by the $\mathbf{L}_n^0 $ and $\tilde{\mathbf{L}}_n^0$ for $n>0$, meaning  $\mathbf{L}_n^0  \Psi_{\alpha}^0=0$ and $\tilde{\mathbf{L}}_n^0  \Psi_{\alpha}^0=0$. Then, using the commutation relations \eqref{vircom1}, one can see that, for $m>0$, $\mathbf{L}_{-m}^0 \Psi_{\alpha}^0$ is another eigenstate of $\mathbf{L}_0^0 $ (and similarly for $\tilde{\mathbf{L}}_{-m}^0$)
 $$\mathbf{L}_0^0(\mathbf{L}_{-m}^0 \Psi_{\alpha}^0)=\mathbf{L}_{-m}^0\mathbf{L}_0^0\Psi_{\alpha}^0+[\mathbf{L}_0^0,\mathbf{L}_{-m}^0]  \Psi_{\alpha}^0=(\Delta_\alpha+m)\mathbf{L}_{-m}^0 \Psi_{\alpha}^0.$$
Iterating this procedure, we construct a  large family of eigenstates $\mathbf{L}_{-m_1}^0\dots \mathbf{L}_{-m_k}^0\tilde{\mathbf{L}}_{-\tilde{m}_1}^0\dots \tilde{\mathbf{L}}_{-\tilde{m}_{\tilde{k}}}^0 \Psi_{\alpha}^0$, but there are some redundancies: indeed, using the commutation relations and the fact that $ \Psi_{\alpha}^0$ is annihilated by the positive modes, one can restrict to sequences of integers with $m_1\geq \dots \geq m_k$ and $\tilde{m}_1\geq \dots \geq \tilde{m}_{\tilde{k}}$. Such finite sequences of decreasing integers are called \textbf{Young diagrams}.  We will denote by    $\mc{T}$   the set of Young diagrams. An element $\nu\in\mc{T}$ is thus  a  sequence  of integers   with the further requirements that $\nu(k)\geq \nu(k+1)$ and $\nu(k)=0$ for $k$ large enough.  
For a Young diagram $\nu$ we denote its length as $|\nu|=\sum_k\nu(k)$ and its size as $s(\nu)=\max\{k\,|\,\nu(k)\not=0\}$. Given two Young diagrams $\nu= (\nu(i))_{i \in [1,k]}$, $\tilde{\nu}= (\tilde{\nu}(i))_{i \in [1,j]}$ with size $k$ and $j$, we define the operators
\begin{equation*}
\mathbf{L}^0_{-\nu}=\mathbf{L}^0_{-\nu(k)} \cdots \, \mathbf{L}^0_{-\nu(1)}, \quad \quad \quad   \tilde{\mathbf{L}}^0_{-\tilde \nu}=\tilde{\mathbf{L}}^0_{-\tilde\nu(j)} \cdots\, \tilde{\mathbf{L}}^0_{-\tilde\nu(1)}.
\end{equation*}
We set
\begin{align}\label{psibasis}
\Psi^0_{\alpha,\nu, \tilde\nu}=\mathbf{L}_{-\nu}^0\tilde{\mathbf{L}}_{-\tilde\nu}^0  \: \Psi^0_\alpha,
\end{align}
with the convention that $\Psi^0_{\alpha,0,0}:=\Psi^0_{\alpha}$.
The states $\Psi^0_{\alpha,\nu, \tilde\nu}$ are called the \emph{descendant} states of $\Psi^0_\alpha$. 

 \begin{proposition}\label{prop:mainvir0}\cite{GKRV20_bootstrap}
For all $\alpha \in \C$ and each pair of Young diagrams $\nu,\tilde{\nu}\in \mathcal{T}$, 
\begin{equation*}
\mathbf{L}_0^0\Psi^0_{\alpha,\nu, \tilde\nu} = (\Delta_{\alpha}
+|\nu|)\Psi^0_{\alpha,\nu ,\tilde\nu},\ \ \  \tilde{\mathbf{L}}_0^0\Psi^0_{\alpha,\nu, \tilde\nu} = (\Delta_{\alpha}+|\tilde\nu|)\Psi^0_{\alpha,\nu ,\tilde\nu}
\end{equation*}
and thus, since $\mathbf{H}^0=\mathbf{L}_0^0+\tilde{\mathbf{L}}_0^0$,
\begin{equation*}
\mathbf{H}^0\Psi^0_{\alpha,\nu,\tilde{\nu}} = (2 \Delta_\alpha+|\nu|+|\tilde{\nu}| )\Psi^0_{\alpha,\nu,\tilde{\nu}}  .
\end{equation*}
 \end{proposition}
The next step is to construct a basis of $\mc{H}$ out of these eigenstates and this is a non trivial task carried out in \cite{FeiginFuchs} using the Kac determinant. First one has the change of basis formula (with $N=|\nu|+|\tilde\nu|$)
 \begin{equation}\label{defqalpha}
\begin{gathered}
\Psi^0_{\alpha,\nu,\tilde{\nu}} =   \sum_{\k,\l, |{\bf k}|+|{\bf l}|=N}M^{N}_{\alpha,\k\l,\nu\tilde\nu}\Psi^0_{\alpha,\k,\l},
\end{gathered}
\end{equation}  
for some coefficients $M^{N}_{\alpha,\k\l,\nu\tilde\nu}$ polynomial in $\alpha\in \C$. The matrix $M^{N}_{\alpha}$ is invertible for $\alpha\in \C$, outside of a set of discrete values located on the real line. In particular, when $\alpha$ is restricted to the spectrum line $Q+i\R$, the family $(\Psi^0_{Q+ip,\nu,\tilde{\nu}})_{p\in\R,\nu,\tilde{\nu}\in\mc{T}}$ is a complete family of eigenstates of $\mc{H}$, but they are not orthogonal. An important input is then to identify the Gram-Schmidt coefficients. The inner products of the descendant states obey (as a measure in the $p$ variable)
\begin{equation}\label{scapo}  
\cjg \Psi^0_{Q+ip,\nu,\tilde{\nu}},\Psi^0_{Q+ip',\nu',\tilde{\nu}'} \cjd_{\mc{H}}=\delta_{p=p'}\delta_{|\nu| ,|\nu'|}\delta_{|\tilde\nu| ,|\tilde\nu'|}F_{Q+ip}(\nu,\nu')F_{Q+ip}(\tilde\nu,\tilde\nu')
\end{equation} 
where the matrices  $(F_{Q+ip}(\nu,\nu'))_{|\nu|=|\nu'|=N}$ are positive definite and  called the \textbf{Schapovalov forms}; we denote by   $F_{Q+ip}^{-1} $ its inverse. The following {\bf spectral decomposition} thus holds: for $u,v\in\mc{H}$  
\begin{align}\label{fcompletevir}
\langle u,v\rangle_{\mc{H}}=\frac{1}{2 \pi}\sum \int_\R \langle u,
\Psi^0_{Q+ip,\nu',\tilde{\nu}'} \rangle_{\mc{H}} \langle \Psi^0_{Q+ip,\nu,\tilde{\nu}}, v\rangle _{\mc{H}}F^{-1}_{Q+ip}(\nu,\nu')F^{-1}_{Q+ip}(\tilde\nu,\tilde\nu')\, \dd p 
\end{align}
where the sum runs over Young diagrams $\nu,\tilde\nu,\nu',\tilde\nu'\in \mc{T}$ such that $|\nu|=|\nu'|$ and $|\tilde\nu|=|\tilde\nu'|$.

\medskip  
 
One can then define the states
\begin{equation}\label{defqalphaL}
\begin{gathered}
\Psi_{\alpha,\nu,\tilde{\nu}} :=   \sum_{\k,\l, |{\bf k}|+|{\bf l}|=N}M^{N}_{\alpha,\k\l,\nu\tilde\nu}\Psi_{\alpha,\k,\l}.
\end{gathered}
\end{equation}  
As linear combination of eigenstates of the Liouville Hamiltonian associated to the same eigenvalue, the properties of the eigenstates $\Psi_{\alpha,{\bf k},\bs{\ell}}$ transfer to the $\Psi_{\alpha,\nu,\tilde{\nu}} $'s   
 
 \begin{theorem}\cite{GKRV20_bootstrap}\label{th:desc:liouv}
For each $\nu,\tilde{\nu}\in\mc{T}$ with $|\nu|+|\tilde\nu|=N$, there is a connected open set $\Omega_{N}\subset \C$ containing
$(Q+i\R)\cup (-\infty,Q-C_N)$ for some $C_N>0$, and an 
analytic family $\alpha \in \Omega_{\nu,\tilde{\nu}} \mapsto \Psi_{\alpha,\nu,\tilde{\nu}}\in e^{(-|Q-{\rm Re}(\alpha)|+\eps)c_-}\mc{H}$ such that
$({\bf H}-2\Delta_{\alpha}-|\nu|-|\tilde\nu|)\Psi_{\alpha,\nu,\tilde{\nu}}=0$ and satisfying the following properties:
{\bf Intertwining:}
\[    \forall \alpha < Q-C_N, \quad \Psi_{\alpha,\nu,\tilde{\nu}}=\lim_{t\to +\infty}e^{t(2\Delta_{\alpha}+|\nu|+|\tilde\nu|)}e^{-t{\bf H}}\Psi^0_{\alpha,\nu,\tilde{\nu}}.\] 
{\bf Spectral decomposition:} for $u,v\in\mc{H}$  
\begin{align*}
\langle u,v\rangle_{\mc{H}}=\frac{1}{2 \pi}\sum_{\nu,\nu',\tilde{\nu},\tilde{\nu}'} \int_{\R_+} \langle u,,
\Psi_{Q+ip,\nu',\tilde{\nu}'} \rangle_{\mc{H}} \langle \Psi_{Q+ip,\nu,\tilde{\nu}}, v\rangle _{\mc{H}}F^{-1}_{Q+ip}(\nu,\nu')F^{-1}_{Q+ip}(\tilde\nu,\tilde\nu')\, \dd p 
\end{align*}
where the sum runs over Young diagrams $\nu,\tilde\nu,\nu',\tilde\nu'\in \mc{T}$ such that $|\nu|=|\nu'|$ and $|\tilde\nu|=|\tilde\nu'|$.
\end{theorem}

The spectral decomposition above is the cornerstone of the conformal bootstrap: it will serve as  a Plancherel type formula on the Hilbert space. Yet, we need more information on the eigenstates $\Psi_{Q+ip,\nu,\tilde{\nu}}$ to be in position to evaluate their correlation functions.
Using the intertwining property on the real line, some scattering theory agument and the form \eqref{formofLn} of ${\bf L}_n$, one can  show that the eigenstates  on the spectrum line $\Psi_{Q+ip,\nu,\tilde{\nu}}$ can be constructed through the action of the Virasoro generators on the primary states, like in the GFF case. The operators ${\bf L}_{n}$ and $\tilde{{\bf L}}_{-n}$ can be easily defined on the weighted space
$e^{-Ac_-}\mc{D}(\mc{Q})$ (for $A>0$) where the $\Psi_{Q+ip,\nu,\tilde{\nu}}$'s live in, and we show that
 ${\bf L}_{-n}\Psi_{Q+ip,\nu,\tilde{\nu}}=\Psi_{Q+ip,\nu',\tilde{\nu}}$ for some $\nu'\in \mc{T}$ 
 and  $\tilde{{\bf L}}_{-n}\Psi_{Q+ip,\nu,\tilde{\nu}}=\Psi_{Q+ip,\nu,\tilde{\nu}'}$ for some $\tilde{\nu}'\in \mc{T}$, 
 and furthermore  that these quantities belong to $e^{-\eps c_-}\mc{D}(\mc{Q})$. 
 This implies that ${\bf L}_{\nu}=\mathbf{L}_{-\nu(k)} \cdots \, \mathbf{L}_{-\nu(1)}$ and $\tilde{\mathbf{L}}_{-\tilde \nu}=\tilde{\mathbf{L}}_{-\tilde\nu(j)} \cdots\, \tilde{\mathbf{L}}_{-\tilde\nu(1)}$ are well defined as operators acting on the vector space $\mc{W}_{Q+ip}:={\rm span}\{\Psi_{Q+ip,\nu,\tilde{\nu}} | \, \nu,\tilde{\nu}\in \mc{T} \}$, and one has:
 \begin{proposition}\label{prop:mainvir1} \cite{BGKRV} 
 For $p\in\R_+$, $\nu,\tilde{\nu}\in\mc{T}$ one has 
 \[\Psi_{Q+ip,\nu,\tilde{\nu}} =\mathbf{L}_{-\nu}\tilde{\mathbf{L}}_{-\tilde\nu}  \: \Psi_{Q+ip}.\]
 Moreover, $\Psi_{\alpha,\nu,\tilde{\nu}}$ admit an analytic extension to $\alpha \in \C$ in weighted $\mc{D}(\mc{Q})$ spaces and the relation above extends as well.
 \end{proposition}
 When acting on $\mc{W}_{Q+ip}$, the ${\bf L}_{n}$ and $\tilde{{\bf L}}_n$ satisfy the Virasoro commuting relations introduced in \eqref{vircom1} and \eqref{vircom2}.
 
The relation of Proposition \ref{cor:annulus_propagator} between annuli amplitude and the generator of ${\bf L}_n+\tilde{{\bf L}}_{n}$ can be promoted to a differentiability result on amplitudes, which will be crucial to understand the conformal blocks and Ward identity.
\begin{proposition}\cite{BGKR1, BGKR2}\label{diff_amplitudes}
Let $(\Sigma^t,g^t,{\bf x},\bs{\zeta}^t)$ be a $1$-parameter $C^1$ family of surfaces with parametrized boundary and admissible metrics, with $\Sigma^t$ sitting inside a closed surface $(\hat{\Sigma},\hat{g})$ and $g^t$ conformal to $\hat{g}$ on $\Sigma^t$. 
If $\bs{\alpha}$ are weights attached to ${\bf x}$ satisfying Seiberg bounds,  $\zeta_1,\dots,\zeta_{b^+}$ 
are the outgoing boundary parametrizations and $\zeta_{b^++1},\dots,\zeta_b$ the incoming ones, then
$t\mapsto \mc{A}_{\Sigma^t,g^t,{\bf x},\bs{\alpha},\bs{\zeta}^t}$
is $C^1$ as map $\mc{D}(\mc{Q})^{\otimes b^-}\to \mc{D}'(\mc{Q})^{\otimes b^+}$, and 
\[\begin{split}
 \pl_t \mc{A}_{\Sigma^t,g^t,{\bf x},\bs{\alpha},\bs{\zeta}^t}\big|_{t=0}=&- \mc{A}_{\Sigma^0,g^0,{\bf x},\bs{\alpha},\bs{\zeta}^0}\circ \sum_{j=1}^{b^-}
{\bf H}^{(j)}_{v_{b^++j}}+\sum_{j=1}^{b^+}{\bf H}^{(j)}_{v_j}\circ \mc{A}_{\Sigma^0,g^0,{\bf x},\bs{\alpha},\bs{\zeta}^0}\\
& +c_{\rm L}(\sum_{j=1}^b\frac{\sigma_j}{12} {\rm Re}(v_{j0})+\pl_t S_{\rm L}^0(\Sigma^t,g^0,g^t)|_{t=0})\mc{A}_{\Sigma^0,g^0,{\bf x},\bs{\alpha},\bs{\zeta}^0}
\end{split}\]
where ${\bf H}^{(j)}_{v_j}$ is the action of ${\bf H}_{v_j}=\sum_{n\in \Z}v_{jn}{\bf L}_n+\bar{v}_{jn}\tilde{{\bf L}}_n$ 
on the $j$-th copy of $\mc{D}(\mc{Q})^{\otimes b^\pm}$ and 
$v_j:=\pl_t ((\zeta_j^{0})^{-1}\circ \zeta^t_j)|_{t=0}=-\sum_{n\in \Z}v_{jn}z^{n+1}$, with $\sigma_j=1$ if $j>b^+$ and $\sigma_j=-1$ if $j<b^+$. 
\end{proposition} 
Let us consider a particular case where $b^+=0$ and $\Sigma\subset \hat{\C}$ with metric $g_0$ conformal to the flat metric 
$|dz|^2$, and for simplicity we assume that $\zeta_j(z)=x_j+z$ for $z\in\T$.
Let $v$ be a holomorphic vector field $v$ near $\Sigma$, then its flow $f_t$ generates conformal transformations 
and we can consider the image surface $\Sigma^t=f_t(\Sigma)$ with parametrization $\bs{\zeta}^t=(f_t\circ \zeta_1,\dots,f_t\circ \zeta_b)$ and  metric $(f_t)_*g_0$, and we set
 ${\bf x}^t=(f_{-t}(x_1),\dots,f_{-t}(x_m))$ for the marked points.
The diffeomorphism invariance of amplitudes gives $\mc{A}_{\Sigma^t,g^t,{\bf x},\bs{\alpha},\bs{\zeta}^t}= \mc{A}_{\Sigma,g_0,{\bf x}^t,\bs{\alpha},\bs{\zeta}}$ and, differentiating at $t=0$, one obtains from Proposition \ref{diff_amplitudes} (see \cite{BGKR2})
 \begin{lemma}[\textbf{Ward identity}]\label{lemma:ward}
If $v$ is a holomorphic vector field defined on a neighborhood of $\Sigma$ in $\hat{\C}$, then with $v_j=\zeta_j^*v=-\sum_{n\in \Z}v_{jn}z^{n+1}\pl_z$
 \[
  \pl_t \mc{A}_{\Sigma,g_0,{\bf x}^t,\bs{\alpha},\bs{\zeta}}|_{t=0}=- \mc{A}_{\Sigma,g_0,{\bf x},\bs{\alpha},\bs{\zeta}}
 \sum_{j=1}^{b}{\bf H}^{(j)}_{v_{j}}.
\]
 \end{lemma}
 This is the form of the Ward identities appearing for instance in the Vertex operator algebra language. We will see soon another probabilistic formulation in terms of the stress energy tensor.
 
 \subsection{Matrix coefficients of the disk with $2$ points on the basis $\Psi_{Q+ip,\nu,\tilde{\nu}}$}\label{Matrix_coeff}
 In this section, we explain in simple terms how the Ward identity of Lemma \ref{lemma:ward} allows to compute the matrix elements of the disk amplitudes on the basis $\Psi_{Q+ip,\nu,\tilde{\nu}}$, i.e. the values of 
 \[ \mc{A}_{\D^*,g_0,{\bf x}, \bs{\alpha},\zeta}(\Psi_{Q+ip,\nu,\tilde{\nu}})\]
 where $\D^*=\{z\in \hat{\C}\,|\, |z|\geq 1\}$, ${\bf x}=(x_1,x_2)$, $\bs{\alpha}=(\alpha_1,\alpha_2)$, $\zeta(e^{i\theta})=e^{i\theta}$ and $g_0$ is an admissible metric. This will be essential in the 
conformal bootstrap method to calculate the $4$-points correlation functions on $\mathbb{S}^2$. Another approach will be explained below using the stress energy tensor, which for computational purpose is somehow more efficient, in particular when there are several descendants like for computing the matrix coefficients of pairs of pants. Consider the vector field 
$v=-z^{-n+1}\pl _z$ for $n\geq 1$. Then, if $f_t^v$ is the flow of $v$ and $f^{iv}_t$ the flow of $iv$, 
by Lemma \ref{lemma:ward} we have 
\[\begin{split}
 2\mc{A}_{\D^*,g_0,{\bf x}, \bs{\alpha},\zeta}({\bf L}_{-n}\Psi_{Q+ip,\nu,\tilde{\nu}})=& \pl_t\mc{A}_{\D^*,g_0,f_{-t}^v({\bf x}),\bs{\alpha}, \zeta}(\Psi_{Q+ip,\nu,\tilde{\nu}})|_{t=0}\\
& -i\pl_t\mc{A}_{\D^*,g_0,f_{-t}^{iv}({\bf x}),\bs{\alpha},\zeta}(\Psi_{Q+ip,\nu,\tilde{\nu}})|_{t=0}.
\end{split} \]
Writing $x_j(t):=x_j+tx_j^{-n+1}$ for $j=1,2$, we see by taking $\nu=\tilde{\nu}=\emptyset$ that
\begin{multline}\label{eval_descendant}
\mc{A}_{\D^*,g_0,{\bf x}, \bs{\alpha},\zeta}(\Psi_{Q+ip,n,\emptyset})=
\\
\frac{1}{2}\Big(\pl_t \cjg V_{Q+ip}(0)V_{\alpha_1}(x_1(t))V_{\alpha_2}(x_2(t))\cjd_{\hat{\C},\hat{g}_0}-i\pl_t\cjg V_{Q+ip}(0)V_{\alpha_1}(x_1(it))V_{\alpha_2}(x_2(it))\cjd_{\hat{\C},\hat{g}_0}\Big)
\end{multline}
where $\hat{g}_0$ is the metric\footnote{$\hat{g}_0$ is only piecewise smooth and continuous at $\T$, but this can be made smooth by modifying conformally the metric $|dz|^2$ in $\D$ using the Weyl covariance formula.} obtained by gluing $g_0$ on $\D^*$ with $|dz|^2$ on $\D$. Using the expression \eqref{3pointDOZZ} for the $3$-point function, one obtains an expression of the form
\[ \mc{A}_{\D^*,g_0,{\bf x}, \bs{\alpha},\zeta}(\Psi_{Q+ip,n,\emptyset})=w(\Delta_{\alpha_1},\Delta_{\alpha_2},\Delta_{Q+ip},n,{\bf x}) 
\mc{A}_{\D^*,g_0,{\bf x}, \bs{\alpha},\zeta}(\Psi_{Q+ip})\]
 for some explicit coefficient $w(\Delta_{\alpha_1},\Delta_{\alpha_2},\Delta_{Q+ip},n,{\bf x}) $ holomorphic in $x_1,x_2$. We can do the same with $\tilde{\bf L}_{-n}$ instead of ${\bf L}_{-n}$ (using $+i\pl_t$ in \eqref{eval_descendant}), and we observe that 
\[ \mc{A}_{\D^*,g_0,{\bf x}, \bs{\alpha},\zeta}(\Psi_{Q+ip,\emptyset,n})=\overline{w(\Delta_{\alpha_1},\Delta_{\alpha_2},\Delta_{Q+ip},n,{\bf x})}\mc{A}_{\D^*,g_0,{\bf x}, \bs{\alpha},\zeta}(\Psi_{Q+ip}).\]
Now this relation can be iterated using  \eqref{eval_descendant} and we can compute  $\mc{A}_{\D^*,g_0,{\bf x}, \bs{\alpha},\zeta}(\Psi_{Q+ip,\nu,\tilde{\nu}})$ from $\mc{A}_{\D^*,g_0,{\bf x}, \bs{\alpha},\zeta}(\Psi_{Q+ip,\nu',\tilde{\nu}'})$ with 
$\nu=(\nu',n)$ for some $n\geq 1$ and $\tilde{\nu}=(\tilde{\nu}',k)$ for some $k\geq 1$. Ultimately, this shows that 
\begin{equation}\label{matrixcoefampl}
  \mc{A}_{\D^*,g_0,{\bf x}, \bs{\alpha},\zeta}(\Psi_{Q+ip,\nu,\tilde{\nu}})=w(\Delta_{\alpha_1},\Delta_{\alpha_2},\Delta_{Q+ip},\nu,{\bf x})
\overline{w(\Delta_{\alpha_1},\Delta_{\alpha_2},\Delta_{Q+ip},\tilde{\nu},{\bf x})}
\mc{A}_{\D^*,g_0,{\bf x}, \bs{\alpha},\zeta}(\Psi_{Q+ip})
\end{equation}
where $w(\Delta_{\alpha_1},\Delta_{\alpha_2},\Delta_{Q+ip},\nu,{\bf x})$ are coefficients that are holomorphic in $x_1,x_2$ and can be computed iteratively by application of linear differential operators in the variables $x_1,x_2$, with holomorphic coefficients, acting
 on the function \eqref{3pointDOZZ}. This will discussed further below, in particular in Theorem \ref{pantDOZZ}, from the Stress Energy tensor.

 \subsection{Stress energy tensor}
As we have seen above, the descendant states  can be understood as the effect of infinitesimal deformations of the complex structure of the disk amplitude $\Psi_{Q+ip}$. There is actually another way to deform the complex structure: varying the metric in the amplitudes. And this observation is at the core of the probabilistic representation of the descendant states, which we explain now.  The idea is to differentiate correlation functions or amplitudes, which depend on the metric $g$, with respect to the inverse of the metric. This has the effect of inserting a new random field called the Stress Energy Tensor (SET for short) in the amplitudes. This is a general expected feature of CFT but it takes an explicit form when specializing to Liouville CFT (see \cite{Oikarinen} for more details). To be more explicit, we keep on the discussion when the underlying Riemann surface is a subset of the Riemann sphere, which we identify with the extended complex plane, equipped with a conformal metric $g:=e^{\omega}|dz|^2$. The holomorphic part of the SET then formally reads for $z\in \C$
\begin{equation}\label{SET}
\forall z\in\C,\quad T(\phi,z):=Q\partial_{zz}^2(\phi+\tfrac{Q}{2}\omega)(z)+:(\partial_z(\phi+\tfrac{Q}{2}\omega)(z))^2:
\end{equation}
and its anti-holomorphic part is the complex conjugate. The SET does not make sense as a random field but can be given sense at the level of correlation functions/amplitudes as the limit of a regularized SET and the notation $:-:$ above means Wick ordering. If $u_1,\dots,u_k$ are distinct point in $\C$, we will collect them in the vector ${\bf u}=(u_1,\dots,u_k)\in\C^k$ (and similarly for ${\bf v}\in \C^{\tilde{k}}$) and we will use the notation
$$T(\phi,{\bf u}):=\prod_{j=1}^kT(\phi,u_j)\quad \text{ and }\quad \bar{T}(\phi,{\bf v}):=\prod_{j=1}^{\tilde{k}}\bar{T}(\phi,v_j).$$
Correlation functions involving the SET are quantities  of the type
\begin{align}\label{actioninsertion2}
 \langle T(\phi,{\bf u}) \bar{T}(\phi,{\bf v})V_{\alpha_1}(x_1)\dots,V_{\alpha_m}(x_m)\rangle_{\Sigma,g},
\end{align}
which are defined similarly to \eqref{actioninsertion} with a regularisation procedure and the Wick ordering is a way to control the divergencies coming from the squared gradient. Amplitudes with SET insertions are defined similarly by plugging the products $T(\phi,{\bf u}) \bar{T}(\phi,{\bf v}) $ in the expression  \eqref{ampL}, again using a regularisation. One important feature of the SET is that these amplitudes are   holomorphic in ${\bf u}$ and antiholomorphic in ${\bf v}$ with eventual poles at the vertex insertions ${\bf x}$. From now on, we will skip the dependence on $\phi$ of the SET in the notations, thus writing $T({\bf u})$ and $\bar{T}({\bf v}) $,  and we will use the notation $ \mc{A}_{\Sigma,g,{\bf x},\bs{\alpha},\boldsymbol{\zeta}}(T({\bf u}) \bar{T}({\bf v}),\bs{\varphi}) $ for amplitudes with SET insertions.    

One can then provide  a probabilistic representation of the descendants in terms of contour integrals of disk amplitudes with SET insertions. For this we need to introduce a few notations. Given two Young diagrams $\nu$, $\tilde\nu$  with respective size $k,\tilde k$ and ${\bf u}\in\C^k$, ${\bf v}\in \C^{\tilde{k}}$, we denote
\begin{equation}\label{notation}
\mathbf{u}^{1-\nu}:=\prod u_i^{1-\nu(i)},\ \ \ \bar{\mathbf{v}}^{1-\tilde{\nu}}:=\prod \bar {v}_i^{1-\tilde{\nu}(i)}.
\end{equation}
Let  $f:\C^k\times\C^{\tilde k}\to\C$ and  ${\bf a} \in \C^k$,  ${\bf b} \in \C^{\tilde k}$.       We will denote multiple   nested contour integrals of $f$ as follows:
\begin{align*}
& \oint_{|\mathbf{u}-\boldsymbol{a}|=\boldsymbol{\delta} }\oint_{|\mathbf{v} -\boldsymbol{\tilde a}|=\tilde{\boldsymbol{\delta}}}f ({\bf u},{\bf v})\dd {\bf \bar v}\dd {\bf u}\\
& :=
\oint_{|u_k-a_k|=\delta_k}\dots \oint_{|u_1-a_1|=\delta_1} \oint_{|v_{\tilde k}-\tilde a_{\tilde k}|=\tilde{\delta}_{\tilde k}}\dots  \oint_{|v_1-\tilde a_1|=\tilde{\delta}_1}  f({\bf u},{\bf v}) \dd \bar v_1\dots \dd \bar v_{\tilde{k}}  \dd u_1\dots \dd u_k\nonumber
\end{align*}
where $\boldsymbol{\delta} :=(\delta_1,\dots, \delta_k)$ with $0<\delta_1<\dots<\delta_k<1$ and similarly for $\tilde{\boldsymbol{\delta}}$.  We always suppose $\delta_i\neq \tilde\delta_j$ for all $i,j$. Finally, for  ${\bf u}\in\C^k$ as ${\bf u}=(u_1,\dots,u_k)$ we will write ${\bf u}^{(l)}$ for the vector  in $\C^{k-1}$ with the same entries as  ${\bf u}$ with $l$-th entry removed and we will sometimes iterate this procedure to remove further entries and write   ${\bf u}^{(l,l')}$ and so on. The same notations are used for the vector ${\bf v}$.  We claim 
\begin{lemma} \label{ampdiskSET}
 Let $\alpha\in\R$ and $\nu,\tilde{\nu}\in \mc{T}$ with $|\nu|+|\tilde{\nu}|=N$ and  $\alpha<Q-C_N$ (given by Theorem \ref{th:desc:liouv}). Then 
\begin{align*}
\Psi_{\alpha,\nu,\tilde\nu}(\tilde\varphi)= &\frac{C}{(2\pi i)^{s(\nu)+s(\tilde \nu)}} 
 \oint_{|\mathbf{u}|=\boldsymbol{\delta}}   \oint_{|\mathbf{v}|=\boldsymbol{\tilde\delta}}  
\mathbf{u}^{1-\nu}\bbar{\mathbf{v}}^{1-\tilde\nu} \mathcal{A}_{\D,|dz|^2,0,\alpha ,\zeta_\D}(T({\bf u})   \bbar T({\bf v}), \varphi)    \dd   \bbar{\mathbf{v}}\dd   \mathbf{u}.
 \end{align*}
\end{lemma} 
 
This formula follows from basic Gaussian computations in the Q-GFF case (i.e. $\mu=0$)  and then can be transferred to the Liouville case via the intertwining property of Theorem \ref{th:desc:liouv} and the Feynman-Kac formula  \eqref{FK}. This formula will be crucial for the next step on the one hand because it identifies descendants  as disk amplitudes (with SET insertions) so that they fit into the Segal picture and on the other hand because correlation functions with SET insertions have a clear structure of poles in the variables ${\bf u}$ and   $\bbar {\bf v}$ which can be determined by the Ward identities
 
\begin{proposition}[{\bf Ward Identity}]\label{propward}
 Equip the extended complex plane $\hat\C$ with a conformal metric $g=e^{\omega}|dz|^2$. Then 
 \begin{align*} 
 \langle T( \mathbf{u})\bbar T( \mathbf{v})\prod_{i=1}^mV_{\alpha_i,g}(z_i)\rangle_{\hat\C,g} = & \frac{1}{2}\sum_{l=1}^{k-1}\frac{c_{\rm L}}{(u_k-u_l)^4} \langle  T(\mathbf{u}^{(k,l)})\bbar T({ \mathbf{v}})\prod_{i=1}^mV_{\alpha_i,g}(z_i)\rangle_{\hat\C,g}
 \\
 &+ \sum_{l=1}^{k-1}\Big(\frac{2}{(u_k-u_l)^2} +\frac{1}{(u_k-u_l) }\partial_{u_l} \Big)
  \langle  T(\mathbf{u}^{(k)})\bbar T({ \mathbf{v}})\prod_{i=1}^mV_{\alpha_i,g}(z_i)\rangle_{\hat\C,g} 
   \\
 &+\sum_{i=1}^{m}\Big(\frac{\Delta_{\alpha_i}}{(u_k-z_i)^2} +\frac{\partial_{z_i}+\Delta_{\alpha_i} \partial_{z_i}\omega }{(u_k-z_i) }\Big)
  \langle  T(\mathbf{u}^{(k)})\bbar T({ \mathbf{v}})\prod_{i=1}^mV_{\alpha_i,g}(z_i)\rangle_{\hat\C,g} 
 \end{align*}
 and  
 \begin{align}\nonumber
 \langle T( \mathbf{u})\bbar T( \mathbf{v})\prod_{i=1}^mV_{\alpha_i,g}(z_i)\rangle_{\hat\C,g} =& \frac{1}{2}\sum_{l=1}^{\tilde k-1}\frac{c_{\rm L}}{(\bar v_{\tilde k}-\bar v_l)^4} \langle  T(\mathbf{u})\bbar T({ \mathbf{v}^{(\tilde k,l)} })\prod_{i=1}^mV_{\alpha_i,g}(z_i)\rangle_{\hat\C,g}
 \\
 &+ \sum_{l=1}^{\tilde k-1}\Big(\frac{2}{(\bar v_{\tilde k}-\bar v_l)^2} +\frac{1}{(\bar v_{\tilde k}-\bar v_l) }\partial_{\bar v_l} \Big)
  \langle  T(\mathbf{u})\bbar T({ \mathbf{v}^{(\tilde k)}})\prod_{i=1}^mV_{\alpha_i,g}(z_i)\rangle_{\hat\C,g}\nonumber
   \\
 &+\sum_{i=1}^{m}\Big(\frac{\Delta_{\alpha_i}}{(\bar v_{\tilde k}-\bar z_i)^2} +\frac{\partial_{\bar z_i}+\Delta_{\alpha_i}\partial_{\bar z_i}\omega }{(\bar v_{\tilde k}-\bar z_i) } \Big)
  \langle  T(\mathbf{u})\bar T({ \mathbf{v}}^{(\tilde k)})\prod_{i=1}^mV_{\alpha_i,g}(z_i)\rangle_{\hat\C,g}\nonumber.
 \end{align} 
  \end{proposition}

 By applying recursively these formula we can obtain all the poles of the correlation functions with SET insertions. The idea of the proof is conceptually simple even though computationally painful: it consists in applying Gaussian integration by parts to the SET expression and then in reorganising all the terms in a clever way.

\subsection{Computation of the building block amplitude and holomorphic factorization}\label{holo_fact}
We can now coordinate Segal's gluing with the Ward identities to compute the amplitudes of building blocks evaluated at the eigenstates of the Liouville Hamiltonian. Any genus-$\mathfrak{g}$ surface $(\Sigma,g,{\bf x})$ with $m$ marked points can be cut along $3\mathfrak{g}-3+m$ curves, producing $2\mathfrak{g}-2+m$ geometric building blocks; The building blocks considered here are:
\begin{enumerate} 
\item the surfaces diffeomorphic to a disk with two marked points and one parametrized boundary circle,
\item the surfaces diffeomorphic to an annulus with one marked point and two parametrized boundary circles,
\item the surfaces diffeomorphic to a pair of pants with three parametrized boundary circles.
\end{enumerate}
We call   $(\caP,{\bf x}^{\rm mp},\boldsymbol{\zeta})$ such a complex building block.
Here the boundary is $\partial\caP=\cup_{i=1}^b\pl_i\mc{P}$ with parametrization $\boldsymbol{\zeta}=(\zeta_1,\dots,\zeta_b)$, $b\in \{1,2,3\}$, and $x_i$, $i\in \{b+1,\dots, 3\}$, are  points in the interior of $\caP$ collected in the vector ${\bf x}^{\rm mp}$ (the superscript ${\rm mp}$ refers to ``marked point'').  We have also set $\sigma_i=-1$ if the boundary component $\pl_i\mc{P}$ is outgoing and we set $\sigma_i=1$ if $\pl_i\mc{P}$ is incoming. 
We collect in the vector $\boldsymbol{\alpha}^{\rm mp}$ the weights $\alpha_i<Q$, $i=b+1,\dots, 3$, attached to the marked points in ${\bf x}^{\rm mp}$,  satisfying the condition $\sum_{j=b+1}^3\alpha_j>\chi(\Sigma)Q$ so that the amplitude associated to $\caP$ is in $\mc{H}^{\otimes b}$.
 We fix also an admissible metric $g_\caP$ on $\caP$.  
  
  The amplitude  $\caA_{\caP, g_\caP, {\bf x}^{\rm mp},\boldsymbol{\alpha}^{\rm mp},\boldsymbol{\zeta}}$ is a function $\caA_{\caP, g_\caP,{\bf x}^{\rm mp},\boldsymbol{\alpha}^{\rm mp},\boldsymbol{\zeta}}:H^s(\T)^b\to \C$. Let us use the notation
 \begin{align}\label{ampf}
 \big\cjg \caA_{\caP, g_\caP,{\bf x}^{\rm mp},\boldsymbol{\alpha}^{\rm mp},\boldsymbol{\zeta}},\otimes_{j=1}^{b}f_j\big\cjd_{\mc{H}^{\otimes b}}:=\int \caA_{\caP, g_\caP,{\bf x}^{\rm mp},\boldsymbol{\alpha}^{\rm mp},\boldsymbol{\zeta}}( \boldsymbol{\varphi})\big(\prod_{j=1}^b \bbar{f_j(\varphi_j)}\big) \dd \mu_0^{\otimes_b}(   \boldsymbol{\varphi}) .
\end{align}
 As explained above for the disk with $2$ marked points, we can evaluate the matrix coefficients of the amplitude on the eigenbasis $\Psi_{Q+ip,\nu,\tilde{\nu}}$,  that is when $f_j={\bf C}^{(1+\sigma_j)/2}\Psi_{Q+ip_j,\nu_j,\tilde\nu_j}$  with ${\bf C}$  the complex conjugation.
 
We will represent $\mc{P}$ under a \textbf{model form} as a subset of the Riemann sphere $\hat{\C}$ as follows: 
 we glue $b$ discs $\mc{D}_j= \D$  to each boundary curve $\pl_j\mc{P}$ using the parametrization 
 $\zeta_j:\T\to \pl_j\mc{P}$ to obtain a closed Riemann surface $\hat\caP$ that is biholomorphic conformal  to $\hat\C$ with its standard complex structure. Denote by $x_j\in \hat\caP$ the center of $\mc{D}_j$, $j=1,\dots,b$ and collect these entries in ${\bf x}^{\rm di}=(x_j)_{j=1,\dots,b}$ (the superscript ${\rm di}$ refers to ``disc insertion'', i.e. to emphasize that it collects the added $x_j$'s lying in the disks glued to the pant to obtain $\hat{\C}$). 
 We will also write ${\bf x}:=(x_1,x_2,x_3)=({\bf x}^{\rm di}, {\bf x}^{\rm mp})$ for the set of $3$ punctures of the punctured sphere $\hat{\mc{P}}$. The biholomorphism $\Phi: \hat{\Sigma}\to \hat{\C}$ is unique if we ask that $\Phi(x_j)=z_j$ where  ${\bf z}=(z_{1},z_2,z_3)$ is any fixed collection of $3$ disjoint points on the sphere.
Let us define, for $j=1,\dots,b$, the maps $\psi_j:=\Phi\circ \zeta_j$, 
which extend holomorphically from $\T$ to (a neighborhood) of $\D$ and  satisfying $\psi_j(0)=z_j$ (resp. of $\D^*:=\hat{\C}\setminus \D^\circ$ and satisfying $\psi_j(\infty)=z_j$) 
if $\zeta_j$ is incoming  (resp. outgoing) by construction of $\Phi$. 
If $o(z):=1/z$ is the inversion, one has that $\hat{\C}\setminus \cup_{j=1}^b\psi_j(o^{(1-\sigma_j)/2}(\D))$ with the parametrizations $\psi_j|_{\T}$ of the boundary is biholomorphic to $(\mc{P},\bs{\zeta})$ and $\psi_j(\D)$ are topological disks in ${\C}$ with analytic boundaries. Define the conformal radius of $\psi_j(\D)$ by
\begin{equation}\label{defrad} 
{\rm Rad}_{z_j}(\psi_j(\D)):=|\psi_j'(0)|.
\end{equation}
The main result concerning the computations of building block amplitudes is the following:

\begin{theorem}\label{pantDOZZ}\cite{GKRV21_Segal}
Let $(\caP,{\bf x}^{\rm mp},\boldsymbol{\zeta})$ be a building block. Let $\alpha_j= Q+ip_j$, $p_j\in\R$, $j=1,\dots, b$, let $\alpha_j<Q$ for $j=b+1,..,3$ and denote $\boldsymbol{\alpha}=(\alpha_1,\alpha_2,\alpha_3)$.  Assume (recall that $\chi(\mc{P})$ is the Euler characteristic of $\mc{P}$)
\begin{equation}\label{ass:starward}
   \sum_{j=b+1}^{3}\alpha_{j }-\chi(\mc{P})Q>0 .
\end{equation}
Then, if $\boldsymbol{\nu}=(\nu_1,\dots,\nu_{b})\in \mc{T}^b$ and $\boldsymbol{\tilde{\nu}}=(\tilde{\nu}_1,\dots,\tilde{\nu}_{b})\in \mc{T}^b$, 
the following formula holds

 \begin{align}\label{pantclaim}
\big\cjg  \caA_{\caP, g_\caP,{\bf x}^{\rm mp},\boldsymbol{\alpha}^{\rm mp},\boldsymbol{\zeta}},  \otimes _{j=1}^b{\bf C}^{(1+\sigma_j)/2}\Psi_{\alpha_j,\nu_j,\tilde\nu_j}  \big\cjd_{\mc{H}^{\otimes b}}
&=  C(\mc{P},g_{\mc{P}},\boldsymbol{\Delta}_{\boldsymbol{\alpha}})
C_{\gamma,\mu}^{{\rm DOZZ}} (\boldsymbol{\tilde\alpha}) \nonumber \\&\times 
  w_{\caP}(\boldsymbol{\Delta}_{\boldsymbol{\alpha}},
 \boldsymbol{\nu},{\bf z})
 \overline{ w_{\caP}(\boldsymbol{\Delta}_{\boldsymbol{\alpha}},
 \boldsymbol{\tilde\nu},{\bf z})}P({\bf z}) 
 \end{align}
 
where   $\boldsymbol{\tilde \alpha}=(\tilde\alpha_1,\tilde\alpha_2,\tilde\alpha_3)$ with  $\tilde\alpha_j=Q+i \sigma_j p_j$ for $j=1,..,b$ and $\tilde\alpha_j=\alpha_j$  for  $j=b+1,..,3$. Furthermore:
\begin{itemize}
\item the function $w_{\caP}$ is a polynomial in the conformal weights $\boldsymbol{\Delta}_{\boldsymbol{\alpha}}=(\Delta_{\alpha_1},\Delta_{\alpha_2},\Delta_{\alpha_3})$ with coefficients depending only on, and in a holomorphic way on, the uniformizing maps $\psi_j$, and on the Young diagram  
\begin{align*}
 w_{\caP}(\boldsymbol{\Delta}_{\boldsymbol{\alpha}},
 \boldsymbol{\nu},{\bf z})=\sum_{{\bf n}=(n_1,n_2,n_3)} a_{\bf n}(\caP, \boldsymbol{\nu},{\bf z})\boldsymbol{\Delta}_{\boldsymbol{\alpha}}^{\bf n},
 \end{align*}
where   $w_{\caP}(\Delta_{\alpha_1} ,\Delta_{\alpha_2},\Delta_{\alpha_3},\emptyset,{\bf z})=1$ and the sum runs over ${\bf n}\in\N^3$ with finitely many coefficients $a_{\bf n}(\caP, \boldsymbol{\nu},{\bf z})$ that are non zero, $\boldsymbol{\Delta}_{\boldsymbol{\alpha}}^{\bf n}:=\prod_{j=1}^3\Delta_{\alpha_j}^{n_j}$.
\item $P$ is given by
\begin{equation}\label{defp0}
 P({\bf z}):= |z_1-z_2|^{2(\Delta_{\alpha_3}-\Delta_{\alpha_2} -\Delta_{\alpha_1})}|z_3-z_2|^{2(\Delta_{\alpha_1}-\Delta_{\alpha_2} -\Delta_{\alpha_3})}|z_1-z_3|^{2(\Delta_{\alpha_2}-\Delta_{\alpha_1} -\Delta_{\alpha_3})}
\end{equation}
\item the  constant $C(\mc{P},g_{\mc{P}},\boldsymbol{\Delta}_{\boldsymbol{\alpha}})$ has the form
\begin{equation}\label{constantC(P,g)}
C(\mc{P},g_{\mc{P}},\boldsymbol{\Delta}_{\boldsymbol{\alpha}})=\prod_{j=1}^b|\psi'_j(0)|^{\Delta_{Q+ip_j}}C'(\mc{P},g_\mc{P},\Delta_{\alpha_{b+1}},\dots,\Delta_{\alpha_3})
\end{equation}
where $C'(\mc{P},g_\mc{P},\Delta_{\alpha_{b+1}},\dots,\Delta_{\alpha_3})$ is a Liouville anomaly which does not depend on $p_j$ and 
is obtained by comparing \eqref{pantclaim} with \eqref{3pointDOZZ} when $\bs{\nu}=\tilde{\bs{\nu}}=\emptyset$, and the Weyl covariance formula of Proposition \ref{covconf2}.
\end{itemize}
\end{theorem}

The proof of this statement is a long story that we   try to make short now. First, we analytically continue the eigenstates $\Psi_{\alpha_j,\nu_j,\tilde\nu_j} $ from the spectrum line $Q+i\R$ to the probabilistic region, namely the real line with  $\alpha_j<Q-C_N$, with Theorem \ref{th:desc:liouv}. In this region, the descendants have a probabilistic representation in terms of contour integrals of amplitudes   of a disk with flat metric, which is not admissible, and SET insertions, see Lemma \ref{ampdiskSET}. We can trade the flat metric on the disk for another admissible one with the Weyl invariance of amplitudes  \eqref{inv_diff_amp} and  then map the disk to $\psi_j(\D)$ using the diffeomorphism invariance of amplitudes \eqref{conf_cov_amp}. Now everything is set to use Segal's gluing: the pairing of the building block amplitudes with the amplitudes of the $\psi_j(\D)$'s and SET insertions produces, via Segal, a correlation function on the Riemann sphere of the type 
$$ \langle T(\phi,{\bf u}) \bar{T}(\phi,{\bf v})V_{\alpha_1}(x_1)\dots,V_{\alpha_m}(x_m)\rangle_{\Sigma,g}$$
and we have to evaluate contour integrals of such quantities. It turns out that these contour integrals can be evaluated if we know the structure of the poles, and this is where the Ward identities (Prop \ref{propward}) come into play. Based on this, a lengthy and  tricky computation shows that we are then left with applying partial differential operators to the three point function
$$    P(\boldsymbol{\Delta}_{\boldsymbol{\alpha}}, \boldsymbol{\nu},\partial_{{\bf z}})P(\boldsymbol{\Delta}_{\boldsymbol{\alpha}}, \boldsymbol{\tilde\nu},\partial_{\bar{{\bf z}}}) \langle  V_{\alpha_1}(x_1)\dots,V_{\alpha_3}(x_3)\rangle_{\Sigma,g}.$$
It remains to plug the exact expression \eqref {3pointDOZZ} for the 3 point function in order to get the claim.

\section{Conformal Bootstrap} \label{sec:CB}

The \textbf{conformal bootstrap} is a method for computing the correlation functions of a CFT on any surface from the $3$-point correlation functions on $\mathbb{S}^2$ (structure constant) and an algebraic function called conformal block. The ingredients are:\\ 
(1) Segal axioms to cut the path integral into path integrals (amplitudes) over building blocks (pants, annuli, disks) -- Section \ref{sec:segalL}.\\  
(2) The resolution of the identity using the eigenfunctions of the Hamiltonian ${\bf H}$ -- Section \ref{Liouville_Hamiltonian}.\\
(3) The computation of the matrix coefficients of the building block amplitude on the eigenbasis, based on Ward identity --   Sections  \ref{Matrix_coeff} and \ref{holo_fact}.\\
(4) The expression of the structure constant (DOZZ formula for Liouville) -- Section \ref{Sec:DOZZ}.

Instead of writing the general case, where one needs to introduce a heavier amount of notations,  
let us explain two examples: the $4$-point correlation function on $\mathbb{S}^2$ and a genus $2$ surface. For the general result we refer to \cite[Theorem 8.5]{GKRV21_Segal}.

\subsection{Bootstrap for $4$-point correlation on $\mathbb{S}^2$}\label{sphere_boot}

\begin{figure}[h] 
 \begin{tikzpicture}
  \shade[ball color = gray!40, opacity = 0.4] (0,0) circle (2cm);
  \draw (0,0) circle (2cm);
  \draw (-2,0) arc (180:360:2 and 0.6);
  \draw[dashed] (2,0) arc (0:180:2 and 0.6);
   \node at (1.2,-0.9) [circle,fill,inner sep=1.5pt]{};
    \node at (0,-2) [circle,fill,inner sep=1.5pt]{};
     \node at (0,2) [circle,fill,inner sep=1.5pt]{};
       \node at (0,0.2) [circle,fill,inner sep=1.5pt]{};
     \draw (1.2,-0.9) node[right,black]{$z$} ;
      \draw (0,0.2) node[right,black]{$2$} ;
        \draw (0,-2.2) node[right,black]{$0$} ;
           \draw (0,2.2) node[right,black]{$\infty$} ;
\end{tikzpicture} \caption{The sphere with $4$ points, cut into $2$ disks along the equator.}\label{picsphere2} 
\end{figure}
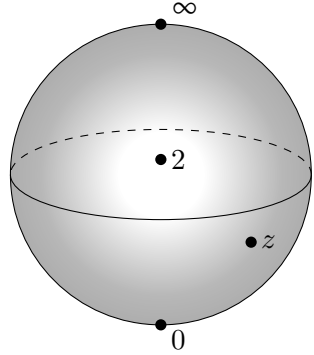

For $z\in \C$ with $|z|<1$, we consider the points on $\mathbb{S}^2=\hat{\C}$
\[ {\bf x}=(x_1,x_2,x_3,x_4)=(0,z,2,\infty)\]
for which we put weights $\bs{\alpha}=(\alpha_1,\alpha_2,\alpha_3,\alpha_4)$ with $\alpha_i<Q$ and 
$\alpha_1+\alpha_2-Q>0$ and $\alpha_3+\alpha_4-Q>0$. We view $\hat{\C}=\D\cup \D^*$ where $\D$ is the unit disk and $\D^*$ the exterior. We consider a metric $g=g(z)|dz|^2$ on $\mathbb{S}^2$, invariant by $z\mapsto 1/z$, satisfying $g(z)=1/|z|^2$ for $1/2<|z|<3/2$, $g(0)=1$ and $g(z)=1/|z|^4$ for $|z|\geq 2$, so that 
$(\D,g,\zeta)$ and $(\D^*,g,\zeta)$ are admissible surfaces with boundary if $\zeta(e^{i\theta})=e^{i\theta}$ is the parametrization of the boundary (incoming for $\D^*$, ougoing for $\D$). See Figure \ref{picsphere2}.

We can use the gluing Theorem \ref{Th:Segal_gluing} to write, with ${\bf x}_1=(0,z)$ and ${\bf x}_2=(2,\infty)$, and 
similarly $\bs{\alpha}_1=(\alpha_1,\alpha_2)$ and $\bs{\alpha}_2=(\alpha_3,\alpha_4)$
\begin{equation}\label{split_Sphere}
\cjg \prod_{i=1}^4 V_{\alpha_i}(x_i)\cjd_{\mathbb{S}^2,g}=\cjg \mc{A}_{\D,g,{\bf x}_1,\bs{\alpha}_1,\zeta},\mc{A}_{\D^*,g,{\bf x}_2,\bs{\alpha}_2,\zeta}\cjd_{\mc{H}}
\end{equation}
Here we view $\mc{A}_{\D^*,g,{\bf x}_2,\bs{\alpha}_2,\zeta}$ as an element in $\mc{H}$, but  we can equivalently consider it as an element in $\mc{H}^*$ (since the boundary is incoming), in which case the pairing above is the composition 
$\mc{A}_{\D^*,g_0,{\bf x}_2, \bs{\alpha}_2,\zeta}\circ \mc{A}_{\D,g,{\bf x}_1,\bs{\alpha}_1,\zeta}$.
The two amplitudes in the pairing are in $\mc{H}$ (or $\mc{H}^*$ for the incoming one) 
by the assumptions on the weights $\alpha_j$.
 We can then use the spectral decomposition  \eqref{fcompletevir}
\begin{align*}
&\cjg \prod_{i=1}^4 V_{\alpha_i}(x_i)\cjd_{\mathbb{S}^2,g}\\
 &=\frac{1}{2\pi}\int_0^\infty\sum_{\nu,\nu',\tilde{\nu},\tilde{\nu}'}
\cjg \mc{A}_{\D,g,{\bf x}_1,\bs{\alpha}_1,\zeta},\Psi_{Q+ip,\nu,\tilde{\nu}}\cjd_{\mc{H}}\cjg \Psi_{Q+ip,\nu',\tilde{\nu}'}, \mc{A}_{\D^*,g,{\bf x}_2,\bs{\alpha}_2,\zeta}\cjd_{\mc{H}}F_{Q+ip}^{-1}(\nu,\nu')F_{Q+ip}^{-1}(\tilde{\nu},\tilde{\nu}')dp.
\end{align*}
To compute each term, we use the factorisation \eqref{matrixcoefampl} obtained as a consequence of the Ward identity
\[\mc{A}_{\D^*,g,{\bf x}_2, \bs{\alpha},\zeta}(\Psi_{Q+ip,\nu',\tilde{\nu}'})=w(\Delta_{\alpha_3},\Delta_{\alpha_4},\Delta_{Q+ip},\nu',{\bf x}_2)
\overline{w(\Delta_{\alpha_3},\Delta_{\alpha_4},\Delta_{Q+ip},\tilde{\nu}',{\bf x}_2)}
\mc{A}_{\D^*,g,{\bf x}_2, \bs{\alpha}_2,\zeta}(\Psi_{Q+ip}).\]
The same factorization also holds for $\cjg \mc{A}_{\D,g,{\bf x}_1,\bs{\alpha}_1,\zeta},\Psi_{Q+ip,\nu,\tilde{\nu}}\cjd_{\mc{H}}$:
\[\cjg \mc{A}_{\D,g,{\bf x}_1,\bs{\alpha}_1,\zeta},\Psi_{Q+ip,\nu,\tilde{\nu}}\cjd_{\mc{H}}=
w(\Delta_{\alpha_1},\Delta_{\alpha_2},\Delta_{Q+ip},\nu,{\bf x}_1)
\overline{w(\Delta_{\alpha_1},\Delta_{\alpha_2},\Delta_{Q+ip},\tilde{\nu},{\bf x}_1)}
\cjg \mc{A}_{\D,g,{\bf x}_1,\bs{\alpha}_1,\zeta},\Psi_{Q+ip}\cjd_{\mc{H}}.\]
Moreover, we see that $\mc{A}_{\D^*,g,{\bf x}_2, \bs{\alpha}_2,\zeta}(\Psi_{Q+ip})$ is the analytic continuation in 
$\alpha$ of $\alpha\mapsto \mc{A}_{\D^*,g,{\bf x}_2, \bs{\alpha}_2,\zeta}(\Psi_{\alpha})$. Since the eigenstate 
\[\Psi_{\alpha}=\mc{A}_{\D,|dz|^2,0,\alpha,\zeta^0}=e^{-c_{\rm L}S_{\rm L}^0(\D,g,|dz|^2)}\mc{A}_{\D,g,0,\alpha,\zeta^0}\]
is the disk amplitude (see Theorem \ref{Theorem_spectral}), we get 
\[ \mc{A}_{\D^*,g,{\bf x}_2, \bs{\alpha}_2,\zeta}(\Psi_{\alpha})= 
e^{-c_{\rm L}S_{\rm L}^0(\D,g,|dz|^2)} \mc{A}_{\D^*,g,{\bf x}_2, \bs{\alpha}_2,\zeta}\circ \mc{A}_{\D,g,0,\alpha,\zeta^0}
\]
which by the gluing result Theorem  \ref{Th:Segal_gluing} becomes 
\[ \mc{A}_{\D^*,g,{\bf x}_2, \bs{\alpha}_2,\zeta}(\Psi_{\alpha})=e^{-c_{\rm L}S_{\rm L}^0(\D,g,|dz|^2)} \cjg V_{\alpha}(0)V_{\alpha_3}(2)V_{\alpha_4}(\infty)\cjd_{\mathbb{S}^2,g}\]
and by \eqref{3pointDOZZ}, one gets
\[ \mc{A}_{\D^*,g,{\bf x}_2, \bs{\alpha}_2,\zeta}(\Psi_{\alpha})=C(g)2^{-2(\Delta_{\alpha}+\Delta_{\alpha_3}+\Delta_{\alpha_4})}C_{\gamma,\mu}^{\rm DOZZ}(\alpha,\alpha_3,\alpha_4)\]
with $C(g):=C_0e^{c_{\rm L}(S_{\rm L}^0(\mathbb{S}^2,g_{\mathbb{S}^2},g)+S_{\rm L}^0(\D,|dz|^2,g))}$ (with $C_0$ in \eqref{C_0def}), and this extends analytically to $\alpha=Q+ip\in Q+i\R$. Similarly, one obtains 
\[\begin{split} 
\cjg \mc{A}_{\D,g,{\bf x}_1,\bs{\alpha}_1,\zeta},\Psi_{\alpha}\cjd_{\mc{H}}=&e^{-c_{\rm L}S_{\rm L}^0(\D,g,|dz|^2)} \cjg V_{\alpha_1}(0)V_{\alpha_2}(z)V_{\alpha}(\infty)\cjd_{\mathbb{S}^2,g}\\
=& C(g)C_{\gamma,\mu}^{\rm DOZZ}(\alpha,\alpha_1,\alpha_2)|z|^{2(\Delta_\alpha-\Delta_{\alpha_1}-\Delta_{\alpha_2})}g(z)^{-4\Delta_{\alpha_2}}
\end{split}\]
and this extends anti-holomorphically in $\alpha=Q+ip$. Gathering everything, we obtain
\begin{theorem}\cite{GKRV20_bootstrap}\label{ThBootSphere}
The $4$ point correlation function on the sphere can be expressed as 
\[\cjg \prod_{i=1}^4 V_{\alpha_i}(x_i)\cjd_{\mathbb{S}^2,g}=\frac{C(g)^2g(z)^{-4\Delta_{\alpha_2}}}{2\pi}\int_{\R} \bbar{C_{\gamma,\mu}^{\rm DOZZ}(Q+ip,\alpha_1,\alpha_2)}C_{\gamma,\mu}^{\rm DOZZ}(Q+ip,\alpha_3,\alpha_4)|\mc{F}_{p,\bs{\alpha}}(z)|^2dp
\]
Where $\mc{F}_{p,\bs{\alpha}}(z)$ is the \textbf{Conformal Block}, given by 
\[ \begin{split}
\mc{F}_{p,\bs{\alpha}}(z)=& 2^{-(\Delta_{Q+ip}+\Delta_{\alpha_3}+\Delta_{\alpha_4})}z^{\Delta_{Q+ip}-\Delta_{\alpha_1}-\Delta_{\alpha_2}}\\
& \times \sum_{\nu\in \mathcal{T}}F_{Q+ip}^{-1}(\nu,\nu')w(\Delta_{\alpha_1},\Delta_{\alpha_2},\Delta_{Q+ip},\nu,0,z)w(\Delta_{\alpha_3},\Delta_{\alpha_4},\Delta_{Q+ip},\nu',2,\infty)\end{split}\]
\end{theorem}
We notice that the series in the conformal block is holomorphic in $|z|<1$, while the term $z^{\Delta_{Q+ip}-\Delta_{\alpha_1}-\Delta_{\alpha_2}}$ is holomorphic but multivalued and has to be thought as a function on a universal 
cover of $\hat{\C}\setminus \{0,1,\infty\}$. The convergence of this type of series is very tricky in general, due to the large number of terms to sum. On the other hand, the convergence in $|z|<1$ for almost all $p\in \R$  follows from our proof of Theorem \ref{ThBootSphere} (one can prove the convergence of the series by using some Cauchy-Schwarz argument and comparing with the case $(\alpha_3,\alpha_4)=(\alpha_1,\alpha_2)$). As far as we know, there was no proof of the convergence of the conformal block in that case before, although the conformal blocks are widely used in the physics literature. 

\subsection{Genus $2$ surface}\label{boot_genus2}
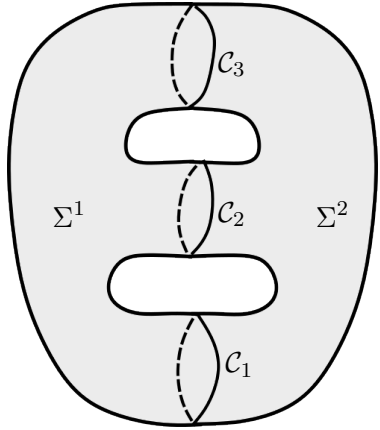
\begin{figure}
 \begin{tikzpicture}
 \node[inner sep=0pt] (pant) at (0,0)
{\includegraphics[width=0.3\textwidth]{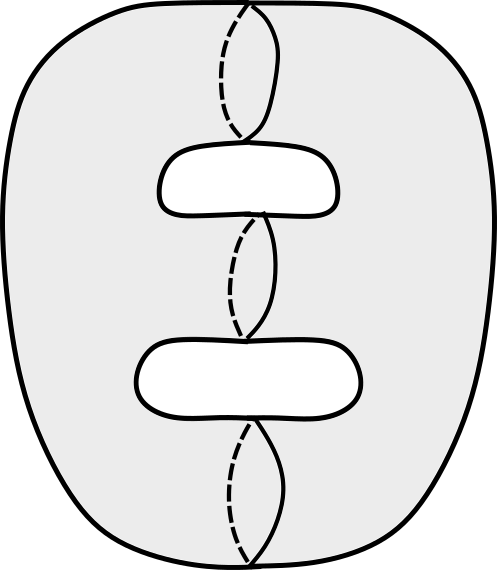}};
 \draw (-2,0) node[right,black]{$\Sigma^1$} ;
  \draw (1.5,0) node[right,black]{$\Sigma^2$} ; 
   \draw (0.2,0) node[right,black]{$\mc{C}_2$} ;
      \draw (0.2,2) node[right,black]{$\mc{C}_3$} ;
        \draw (0.3,-2) node[right,black]{$\mc{C}_1$} ;
\end{tikzpicture}
 \caption{A genus $2$ surface cut into two pairs of pants $\Sigma^1,\Sigma^2$ along $3$ curves 
        $\mc{C}=(\mc{C}_1,\mc{C}_2,\mc{C}_3)$.}\label{Genus2}
\end{figure}

We consider a surface $\Sigma$ of genus $2$ that we cut along three curves $\mc{C}=(\mc{C}_1,\mc{C}_2,\mc{C}_3)$, parametrized by $\bs{\zeta}=(\zeta_1,\zeta_2,\zeta_3)$, and we obtain two pairs of pants $\Sigma^1,\Sigma^2$ with parametrized boundary -- See Figure \ref{Genus2}. Assume that all the boundaries are incoming in $\Sigma^1$ and consider a metric $g$ on $\Sigma$ which is admissible when restricted on 
$(\Sigma^j,\bs{\zeta})$ for $j=1,2$.

We can use the gluing Theorem \ref{Th:Segal_gluing} to write
\begin{equation}\label{split_genus2}
Z_{\Sigma,g}=\mc{A}_{\Sigma,g}=\cjg \mc{A}_{\Sigma^1,g,\bs{\zeta}},\mc{A}_{\Sigma^2,g,\bs{\zeta}}\cjd_{\mc{H}^{\otimes 3}}
\end{equation}
where we view both  $\mc{A}_{\Sigma^j,g,\bs{\zeta}}$ as elements in $\otimes^3\mc{H}$ (or alternatively 
$\mc{A}_{\Sigma^1,g,\bs{\zeta}}\in \otimes^3 \mc{H}^*$ and the pairing above is a composition). For each $\mc{H}\times \mc{H}$ 
pairing above we use the spectral decomposition  \ref{fcompletevir} and get 
\[\begin{split}
Z_{\Sigma,g}=& \frac{1}{(2\pi)^3}\int_{\R_+^3}\sum_{\bs{\nu}\in\mc{T}^3,\tilde{\bs{\nu}}\in \mc{T}^3}\cjg \mc{A}_{\Sigma^1,g,\bs{\zeta}},\otimes_{j=1}^3\Psi_{Q+ip_j,\nu_j,\tilde{\nu}_j}\cjd_{\mc{H}^{\otimes 3}}\cjg\otimes_{j=1}^3\Psi_{Q+ip_j,\nu_j,\tilde{\nu}_j} ,\mc{A}_{\Sigma^1,g,\bs{\zeta}}\cjd_{\mc{H}^{\otimes 3}}\\
&  \qquad \qquad\quad  \times\prod_{j=1}^3F^{-1}_{Q+ip_j}(\nu_j,\nu_j')F^{-1}_{Q+ip_j}(\tilde{\nu}_j,\tilde{\nu}_j')dp_1dp_2dp_3
\end{split}\]
where $\bs{\nu}=(\nu_1,\nu_2,\nu_3)$ and $\tilde{\bs{\nu}}=(\tilde{\nu}_1,\tilde{\nu}_2,\tilde{\nu}_3)$.
Next we want to use the factorization \eqref{defp0} coming from the Ward identity (with choice ${\bf z}=(0,1,e^{i\pi/3})$ for example).
For this, we represent $(\Sigma^1,\bs{\zeta})$ under model form, as explained in Section \ref{holo_fact}: we get three biholomorphic maps $\psi^1_{j}:\D\to \psi^1_{j}(\D)\subset \C$ with $\psi^1_{1}(0)=0$, $\psi^1_{2}(0)=1$ and $\psi^1_3(0)=e^{i\pi/3}$ and 
$(\mc{P},\bs{\zeta})$ is biholomorphic to $\hat{\C}\setminus \bigcup_{j=1}^3\psi^1_j(\D^\circ)$ with parametriztions $\psi^1_j|_{\T}$ of each boundary. We get  from \eqref{defp0}
\begin{equation}\label{descendant_matrix}
\cjg \mc{A}_{\Sigma^1,g,\bs{\zeta}},\otimes_{j=1}^3\Psi_{Q+ip_j,\nu_j,\tilde{\nu}_j}\cjd_{\mc{H}^{\otimes 3}}=
C(\Sigma^1,g,\bs{\Delta}_{{\bf Q}+i{\bf p}})w(\Sigma^1,\bs{\Delta}_{{\bf Q}+i{\bf p}},\bs{\nu})\bbar{w(\Sigma^1,\bs{\Delta}_{{\bf Q}+i{\bf p}},\tilde{\bs{\nu}})}C_{\gamma,\mu}^{\rm DOZZ}({\bf Q}+i{\bf p})
\end{equation}
where ${\bf Q}+i{\bf p}:=(Q+ip_1,Q+ip_2,Q+ip_3)$, $\bs{\Delta}_{\bs{Q}+i{\bf p}}:=(\Delta_{Q+ip_1},\Delta_{Q+ip_2},\Delta_{Q+ip_3})$ and $w (\Sigma^2,\bs{\Delta}_{{\bf Q}+i{\bf p}},\bs{\nu})$ are coefficients depending in a holomorphic way on the model form 
biholomorphisms $(\psi^2_1,\psi^2_2,\psi^2_3)$ associated to $\Sigma^2$. We do the same with $\Sigma^2$, using the the amplitudes are real valued 
\[\cjg \otimes_{j=1}^3\Psi_{Q+ip_j,\nu_j,\tilde{\nu}_j},\mc{A}_{\Sigma^2,g,\bs{\zeta}}\cjd_{\mc{H}^{\otimes 3}}=
C(\Sigma^2,g,\bs{\Delta}_{{\bf Q}+i{\bf p}})w(\Sigma^2,\bs{\Delta}_{{\bf Q}+i{\bf p}},\bs{\nu})\bbar{w(\Sigma^2,\bs{\Delta}_{{\bf Q}+i{\bf p}},\tilde{\bs{\nu}})}C_{\gamma,\mu}^{\rm DOZZ}({\bf Q}-i{\bf p}).
\]
Using that $C^{\rm DOZZ}_{\gamma,\mu}(Q-ip_1,Q-ip_2,Q-ip_3)=\bbar{C^{\rm DOZZ}_{\gamma,\mu}(Q+ip_1,Q+ip_2,Q+ip_3)}$, and combining what we just discussed, we obtain the result:
\begin{theorem}\cite{GKRV21_Segal}
The partition function of the genus $2$ surface $(\Sigma,g)$ cut along the curves $\mc{C}$ parametrized by $\bs{\zeta}$ can be written under the form
\[Z_g= \frac{C'(\Sigma,\bs{\zeta},g)}{(2\pi)^3}\int_{\R_+^3}|C^{\rm DOZZ}_{\gamma,\mu}(Q+ip_1,Q+ip_2,Q+ip_3)|^2|\mc{F}_{{\bf p}}(\Sigma,\bs{\zeta})|^2dp_1dp_2dp_3 \]
where $\mc{F}_{\bf p}(\Sigma,\bs{\zeta})$ is the \textbf{Conformal Block} of $\Sigma$ defined by 
\[\mc{F}_{\bf p}(\Sigma,\bs{\zeta})=\prod_{j=1}^3((\psi^1_j)'(0)(\psi^2_j)'(0))^{\Delta_{Q+ip_j}}\sum_{\bs{\nu}\in \mc{T}^3}w(\Sigma^1,\bs{\Delta}_{{\bf Q}+i{\bf p}},\bs{\nu})
w(\Sigma^2,\bs{\Delta}_{{\bf Q}+i{\bf p}},\bs{\nu})\prod_{j=1}^3F^{-1}_{Q+ip_j}(\nu_j,\nu_j')\]
and $C'(\Sigma,\bs{\zeta},g)$ is a real valued explicit constant depending only on $\Sigma$, the parametrized curves $\bs{\zeta}$, the metric $g$.
\end{theorem}
The convergence of the conformal block for almost every ${\bf p}$ also follows from our proof, and was not proved before, as far as we know.
Let us now make two important remarks, that will be discussed further in next section:
\begin{enumerate}
\item The conformal block a priori depends on the choice of cutting curves $\bs{\zeta}$.
\item We need to choose a logarithm determination of the complex number $(\psi^1_j)'(0)(\psi^2_j)'(0)$ in order to define 
the conformal block. To make sense of it, we need to view $\Sigma$ as a \textbf{marked} Riemann surface, thus as an element in the Teichm\"uller space. Moreover, $\Sigma\mapsto \mc{F}_{\bf p}(\Sigma,\cdot)$ is holomorphic in an appropriate sense that we shall discuss in the next Section. 
\end{enumerate}

\section{Conformal blocks}\label{sec:block}
As explained above, the partition functions of a CFT can be viewed as functions on the moduli space $\mc{M}_\mathfrak{g}$ of Riemann surface of genus $\mathfrak{g}$ while the $m$-point correlation functions are functions on the moduli space $\mc{M}_{\mathfrak{g},m}$ of Riemann surfaces with $m$ marked points. These moduli spaces are complex manifolds of dimension $3\mathfrak{g}-3$ and $3\mathfrak{g}-3+m$.
The conformal blocks, in some sense, are the holomorphic content of these partition/correlation functions when we think of the dependence in the moduli parameter. It turns out that the Ward identities in Proposition \ref{propward} translate  in terms of identities on the conformal blocks, called Ward identities for conformal blocks, see \eqref{Ward_for_blocks}. In the Vertex Operator Algebra approach, conformal blocks are introduced  as a basis of solutions of these identities. Since these identities are only depending on the central charge (via the commutation relation of Virasoro algebra) and on the conformal weights, the solutions are universal to conformal field theories with a given central charge and family of conformal weights. Moreover, the Ward identities for conformal blocks are holomorphic in the moduli parameters (for example the marked points) except for some singularities when the marked points collide. The solutions are also holomorphic in these parameters but the singularities create some monodromy for the solutions, in a way similar to hypergeometric functions for instance.
 Since the correlation functions of a CFT also obey some Ward identities, one can show that they can be decomposed 
 using the basis of conformal blocks, and the coefficients in front of these blocks are (product of) the structure constants.
The special functions that we constructed from the probabilistic amplitudes in Section \ref{sec:CB}, which we called conformal blocks, are not yet proven to be solutions of the Ward identities and they are not a priori globally defined on the moduli space $\mc{M}_{\mathfrak{g},m}$, due to the fact that they depend on some choice of $3\mathfrak{g}-3+m$ cutting curves decomposing the surfaces into pairs of pants, annuli with $1$ point or disk with $2$ points.

As we shall explain, it turns out that the conformal blocks are functions on the Teichm\"uller spaces
$\mc{T}_\mathfrak{g}$ (and $\mc{T}_{\mathfrak{g},m}$ in case of marked points), which are the universal cover of $\mc{M}_{\mathfrak{g}}$ (and $\mc{M}_{\mathfrak{g},m}$): they have monodromies when applying a deck transformation, i.e. a diffeomorphism of the surface not isotopic to the identity. On the sphere with $4$ points $(0,1,z,\infty)$ in the example of Section \ref{sphere_boot}, 
the moduli parameter can be chosen to be $z$ and the monodromy of the conformal block comes from the term $z^{\Delta_{Q+ip}-\Delta_{\alpha_1}-\Delta_{\alpha_2}}$ in Theorem \ref{ThBootSphere}, which is not univalued on $\hat{\C}\setminus \{0,1,\infty\}$.
 We stress that this is not incompatible with the correlation functions being defined on the moduli space. 
We explain now how to define globally the conformal blocks, and why they are solutions of the Ward identities.

\subsection{Moduli and Teichm\"uller spaces} 
We say that a surface $(\Sigma,J,{\bf x},\bs{\zeta})$ is a Riemann surface of type $(\mathfrak{g},m,b^+,b^-)$ if $\Sigma$ is an oriented  compact surface  with boundary, $J$ is a complex structure, ${\bf x}=(x_1,\dots,x_m)$ is a collection of $m$ distinct points in $\Sigma^\circ$ and $\bs{\zeta}=(\zeta_1,\dots,\zeta_b)$ are analytic parametrizations of the boundary circles $\pl_1\Sigma,\dots,\pl_b\Sigma$ where $\zeta_j$ has outgoing orientation for $j=1,\dots b^+$ and incoming orientation for $j=b^++1,\dots,b=b^++b^-$. The \textbf{moduli space}
$\mc{M}_{\mathfrak{g},m,b^+,b^-}$ is the set of surfaces of type $(\mathfrak{g},m,b^+,b^-)$ quotiented out by the equivalence relation: 
$(\Sigma,J,{\bf x},\bs{\zeta})\sim (\Sigma',J',{\bf x}',\bs{\zeta}')$ iff there is a biholomorphism 
$\Phi:(\Sigma,J)\to (\Sigma',J')$ such that $x'_j=\Phi(x_j)$ and $\zeta'_j=\Phi\circ\zeta_j$. This is an infinite dimensional space if $b=b^++b^->0$, otherwise it has complex dimension $3\mathfrak{g}-3+m$. To define the \textbf{Teichm\"uller space} $\mc{T}_{\mathfrak{g} ,m,b^+,b^-}$, we fix a 
Riemann surface $(\Sigma^0,J^0,{\bf x}^0,\bs{\zeta}^0)$ of type $(\mathfrak{g},m,b^+,b^-)$ and we consider the set of equivalence classes of marked surfaces $(\Sigma,\Phi,J,{\bf x},\bs{\zeta})$ where $(\Sigma,J,{\bf x},\bs{\zeta})$ is of type $(\mathfrak{g},m,b^+,b^-)$, 
$\Phi:\Sigma^0\to \Sigma$ is an orientation preserving diffeomorphism (called a marking), analytic near $\pl \Sigma^0$, such that 
$\Phi\circ \zeta_j^0=\zeta_j$ for all $j\leq b$, $\Phi(x^0_j)=x_j$ for $j\leq m$, and  $(\Sigma^1,\Phi^1,J^1,{\bf x}^1,\bs{\zeta}^1)\sim (\Sigma^2,\Phi^2,J^2,{\bf x}^2,\bs{\zeta}^2)$ iff there is a biholomorphism $\Phi:(\Sigma^1,J^1)\to (\Sigma^2,J^2)$ such that 
$(\Phi^2)^{-1}\circ \Phi\circ \Phi^1$ is isotopic to the identity within the set of diffeomorphisms equal to the identity on ${\bf x}^0$ and on $\pl \Sigma^0$. It is an infinite dimensional space if $b^++b^->0$, otherwise it has complex dimension $3\mathfrak{g}-3+m$. When there is no boundary, this corresponds to the usual Teichm\"uller space (see \cite{Imayoshi-Taniguchi} for a reader friendly reference).

\subsection{Conformal block amplitude for building blocks}
For simplicity, we only explain the case when the building block is a pair of pants, i.e. a surface  of type $(0,0,b^+,b^-)$ with $b^++b^-=3$, following \cite{BGKR2}.
Let ${\rm BiHol}(\D)$ denote the set of biholomorphisms from $\D$ to a subset of $\C$, which extend holomorphically near $\D$.
Let us first recall the model form for a building block with $\pl_j\Sigma$ outgoing for $j\leq b^+$ and incoming for $j>b^+$, as discussed in Section \ref{holo_fact}: it is determined by three biholomorphisms 
$\bs{\psi}=(\psi_1,\psi_2,\psi_3)$ with $\psi_j^{\rm in}\in {\rm BiHol}(\D)$ for $j=1,2,3$ where $\psi_j^{\rm in}:=\psi_j\circ o$ (outgoing)
if $j\leq b^+$ and $\psi_j^{\rm in}:=\psi_j$ if $j>b^+$ (incoming), where $o(z)=1/z$.
The moduli space $\mc{M}_{0,0,b^+,b^-}$ then identifies with
\[\mc{M}_{0,0,b^+,b^-}\simeq \{ \bs{\psi}=(\psi_1,\psi_2,\psi_3)\,|\, \psi_j^{\rm in}\in \textrm{BiHol}(\D),  \psi_j^{\rm in}(0)=x_j^0\}\]
where ${\bf x}^0=(x_1^0,x_2^0,x_3^0)=(0,1,e^{i\pi/3})$.
One can show that the Teichm\"uller space of pairs of pants $\mc{T}_{0,b^+,b^-}:=\mc{T}_{0,0,b^+,b^-}$ is in one-to-one correspondence with 
\[\mc{T}_{0,b^+,b^-}\simeq \{ (\bs{\psi},\bs{\tau})=(\psi_1,\psi_2,\psi_3,\tau_1,\tau_2,\tau_3)\,|\, \tau_j\in\R,\,\psi_j^{\rm in}
\in \textrm{BiHol}(\D),  \psi_j^{\rm in}(0)=x_j^0, \psi_j'(0)=e^{i\tau_j} |\psi_j'(0)|\}.\]
In particular, we can define the map  for $\bs{\alpha}=(\alpha_1,\alpha_2,\alpha_3)\in \R^3$  
\[ \Gamma_{\bs{\alpha}}: \mc{T}_{0,b^+,b^-}\to \C ,\quad  \Gamma_{\bs{\alpha}}(\bs{\psi},\bs{\tau})=\prod_{j=1}^3|\psi'_j(0)|^{\Delta_{\alpha_j}}e^{i\Delta_{\alpha_j}\tau_j},\]    
which can be thought as a product of powers of complex conformal radii. Note that in small open sets $U\subset \mc{T}_{0,b^+,b^-}$, 
$\psi_j$ are holomorphic coordinates functions on $U$, $\psi'_j(0)$ are holomorphic functions and
$|\psi'_j(0)|^{\Delta_{\alpha_j}}e^{i\Delta_{\alpha_j}\tau_j}=(\psi'_j(0))^{\Delta_{\alpha_j}}$ 
where the determination of the log is defined by $\log(\psi_j'(0))=\log|\psi_j'(0)|+i\tau_j$. This means that $\Gamma_{\bs{\alpha}}$ is a holomorphic function on $\mc{T}_{0,b^+,b^-}$.

Now, we can consider the holomorphic part $\mc{H}^{1,0}$ 
of the Hilbert space $\mc{H}=L^2(H^{-s}(\T),\mu_0)$\footnote{Here the functions are assumed to be complex valued} by defining the Hilbert subspace and its orthogonal projection
\[ \mc{H}^{1,0}:=\{ F\in \mc{H}\,|\, \forall n>0, \tilde{{\bf L}}_n F=0\}, \quad \Pi^{1,0}:\mc{H}\to \mc{H}^{1,0}.\]
There is a slight issue with domains here since $\tilde{{\bf L}}_n$ are not bounded, but this can still be defined using the spectral resolution and $\mc{H}^{1,0}$ identifies with $L^2(\R_+\times \mc{T})$ by the map $\Pi_{\mc{V}}:\mc{H}^{1,0}\to L^2(\R_+\times \mc{T})$
\begin{equation}\label{PiV} 
(\Pi_{\mc{V}}u)(p,\nu):=\frac{1}{\sqrt{2\pi}}\cjg u, H_{Q+ip,\nu,0}\cjd_{\mc{H}}
\end{equation}
with $H_{Q+ip,\nu,0}:=\sum_{|\nu'|=|\nu|}F_{Q+{\rm i}p}^{-1/2}(\nu,\nu')\Psi_{Q+{\rm i}p,\nu',0}$ and $F_{Q+ip}^{-1/2}$ is viewed as a symmetric matrix; see \cite{BGKR2}. We can also define the adjoint $\Pi_{\mc{V}}^*: L^2(\R_+\times \mc{T})\to \mc{H}^{1,0}$ 
of $\Pi_{\mc{V}}$. We note that for $p$ fixed, $(\Pi_{\mc{V}}F)(p,\cdot)$ is an element of the Verma module generated by the primary field $\Psi_{Q+ip}$.

Now, for a marked pair of pants $(\bs{\psi},\bs{\tau})\in \mc{T}_{0,b^+,b^-}$ and 
$\mc{P}=\hat{\C}\setminus \cup_{j=1}^{3}\psi_j^{\rm in}(\D)$ a representative of this pant with boundary parametrization $\bs{\zeta}=(\psi_1|_{\T},\psi_2|_{\T},\psi_3|_{\T})$, we define its \textbf{conformal block amplitude} as the operator
 \begin{equation}\label{propertyB}
 \begin{split}
& \mc{B}_{(\bs{\psi},\bs{\tau})}: L^2_{\rm comp}(\R_+^{b^-};  L^2(\mc{T})^{\otimes b^-})\to L^2_{\rm loc}(\R_+^{b^+}; L^2(\mc{T})^{\otimes b^+})\\
&(\mc{B}_{(\bs{\psi},\bs{\tau})}{\bf u})(p_1,\dots,p_{b^+}):=\Big(\otimes^{b^+}\Pi_{\mc{V}}\circ \mc{A}_{\mc{P},g,\bs{\zeta}}\circ \otimes^{b^-}\Pi_{\mc{V}}^*\Big( \frac{{\bf u}\, \Gamma_{{\bf Q}+i{\bf p}}(\bs{\psi},\bs{\tau})}{C(\mc{P},g,\bs{\Delta}_{{\bf Q}+i{\bf p}})C^{\rm DOZZ}({\bf Q}+i{\bf p})}\Big)\Big)(p_1,\dots,p_{b^+}),
 \end{split}\end{equation}
where ${\bf p}=(p_1,p_2,p_3)$, $g$ is any admissible metric on $(\mc{P},\bs{\zeta})$, $C(\mc{P},g,\bs{\Delta}_{{\bf Q}+i{\bf p}})$ is the constant \eqref{constantC(P,g)} and $L^2_{\rm comp}$ (resp. $L^2_{\rm loc}$) denotes the space of $L^2$ functions with compact support in the variable ${\bf p}_-\in \R_+^{b^-}$ (resp. the space of measurable functions that are $L^2$ on  any compact set of $\R_+^{b^-}$).
A few comments are in order and to simplify let us assume $b^+=0$ and $b^-=3$. 
First, $\mc{B}_{(\bs{\psi},\bs{\tau})}$ in \eqref{propertyB} is independent of the choice of $g_{\mc{P}}$ thanks to the normalization by 
$1/C(\mc{P},g,\bs{\Delta}_{{\bf Q}+i{\bf p}})$ (the Weyl anomalies cancel out)
and it is essentially the projection of the Liouville amplitude on the holomorphic part $\mc{H}^{1,0}$ of the Hilbert space, up to the constants 
\[\frac{\Gamma_{{\bf Q}+i{\bf p}}(\bs{\psi},\bs{\tau})}{C(\mc{P},g,\bs{\Delta}_{{\bf Q}+i{\bf p}})C^{\rm DOZZ}({\bf Q}+i{\bf p})}=
(\bbar{\Gamma_{{\bf Q}+i{\bf p}}(\bs{\psi},\bs{\tau})}C'(\mc{P},g_\mc{P})C^{\rm DOZZ}({\bf Q}+i{\bf p}))^{-1}\]
where $C'(\mc{P},g_\mc{P})$ is a constant not depending on ${\bf p}$. The evaluation of the block amplitude on 
the primary states $\otimes_{j=1}^3\Psi_{Q+ip_j}$ gives
\[\mc{B}_{(\bs{\psi},\bs{\tau})}(\otimes_{j=1}^3\Psi_{Q+ip_j})=\mc{A}_{\mc{P},g_{\mc{P}},\bs{\zeta}}(\otimes_{j=1}^3\Psi_{Q+ip_j}))\frac{\Gamma_{{\bf Q}+i{\bf p}}(\bs{\psi},\bs{\tau})}{C(\mc{P},g,\bs{\Delta}_{{\bf Q}+i{\bf p}})C^{\rm DOZZ}({\bf Q}+i{\bf p})}= \Gamma_{{\bf Q}+i{\bf p}}(\bs{\psi},\bs{\tau})\]
which can be seen as a normalization constant (the value on the trivial Young diagram), holomorphic in the Teichm\"uller complex coordinates $\bs{\psi}$. The evaluation on descandents states $\Psi_{Q+ip_j,\nu_j,\emptyset}$ can be computed from the Ward identity as in \eqref{descendant_matrix} and we get for $\bs{\nu}(\nu_1,\nu_2,\nu_3)$
\[\mc{B}_{(\bs{\psi},\bs{\tau})}(\otimes_{j=1}^3\Psi_{Q+ip_j,\nu_j,\emptyset})=
w(\mc{P},\bs{\Delta}_{{\bf Q}+i{\bf p}},\bs{\nu})\Gamma_{{\bf Q}+i{\bf p}}(\bs{\psi},\bs{\tau})\]
where $w$ are the coefficients appearing in the factorization \eqref{pantclaim}, and are viewed as the matrix coefficients of the operator 
$\mc{B}_{(\bs{\psi},\bs{\tau})}$. This makes the connection with the bootstrap description above and the construction of the conformal block $\mc{F}_{\bf p}$ in  Section \ref{boot_genus2}.

\subsection{Gluing conformal block amplitudes}\label{sec:gluingblocks}
\begin{figure}
 \begin{tikzpicture}
\draw[->,>=stealth',draw=black,very thick] (2,0) arc[radius=1, start angle=0, end angle=360] node[left,red]{};
 \draw (0,0) node[right,black]{${\rm v}_1$} ;
  \draw (-2,-0.3) node[right,black]{${\rm v}_2$} ; 
   \draw[->,>=stealth',draw=black,very thick] (0,0) --node[midway,below]{$$} (-1,0) ;
      \draw[-,draw=black,very thick] (-1,0) --node[midway,below]{$$} (-2,0) ;
    \draw[->,>=stealth',draw=black,very thick] (-3.4,1.4) --node[midway,below]{$$} (-2.7,0.7)  ;
      \draw[-,draw=black,very thick] (-2.7,0.7) --node[midway,below]{$$} (-2,0)  ;
    \draw[->,>=stealth',draw=black,very thick] (-3.4,-1.4) --node[midway,below]{$$} (-2.7,-0.7)  ;
      \draw[-,draw=black,very thick] (-2.7,-0.7) --node[midway,below]{$$} (-2,0)  ;
   \node at (0,0) [circle,fill,inner sep=1.5pt]{};
      \node at (-2,0) [circle,fill,inner sep=1.5pt]{};
\end{tikzpicture}
 \caption{Graph of a a genus $1$ surface with $2$ boundaries cut along $2$ curves. The edge relating ${\rm v}_1$ to ${\rm v}_2$ is $((1,1),(2,3))$ and corresponds to $\mc{C}_2$, the edges linking ${\rm v}_1$ to itself is $((1,2),(1,3))$ and corresponds to $\mc{C}_1$.}\label{Genus1_graph}
\end{figure}
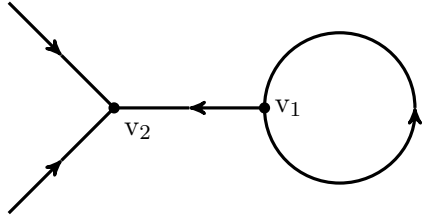

\begin{figure}
 \begin{tikzpicture}
 \node[inner sep=0pt] (pant) at (0,0)
{\includegraphics[width=0.4\textwidth]{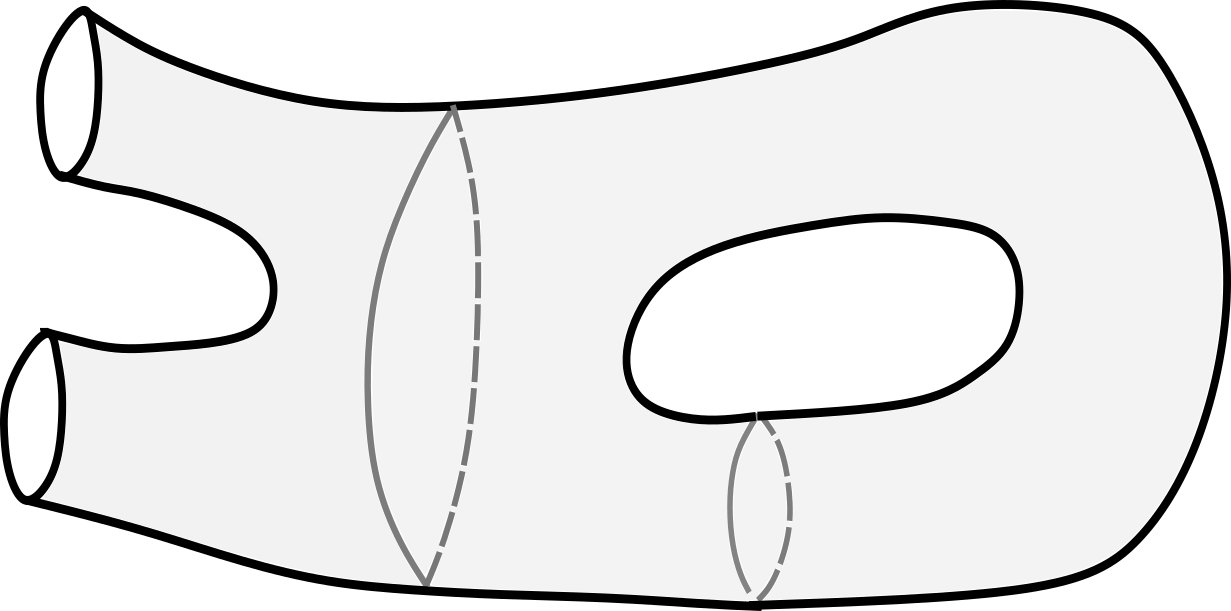}};
 \draw (-4,1.3) node[right,black]{$\pl_1\Sigma$} ;
  \draw (-4.2,-0.7) node[right,black]{$\pl_3\Sigma$} ; 
        \draw (1,-1) node[right,black]{$\mc{C}_1$} ;
          \draw (-1.3,0) node[right,black]{$\mc{C}_2$} ;
           \draw (2.5,0) node[right,black]{$\mc{P}_1$} ;
             \draw (-2.3,-1) node[right,black]{$\mc{P}_2$} ;
\end{tikzpicture}
 \caption{A genus $1$ surface cut along two curves into two pairs of pants $\mc{P}_1$ and $\mc{P}_2$.}\label{Genus1}
\end{figure}
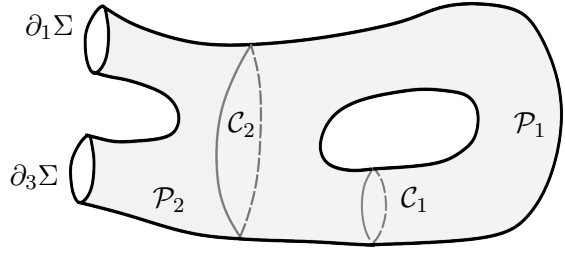

The conformal block amplitude of a general surface relies on its decomposition into geometric building blocks. Let us explain the case when there is no marked points, for simplicity.
First, we can decompose a genus $\mathfrak{g}$ surface $\Sigma$ with $b$ boundary components denoted $\pl_k\Sigma$  into $j_0:=2\mathfrak{g}-2+b$ pairs of pants $\mc{P}_j$, cut along $i_0:=3\mathfrak{g}-3+b$ curves $\mc{C}_i$. One can associate a graph 
$\mc{G}$ with vertices ${\rm v}_j$ corresponding to $\mc{P}_j$. At  each vertex we attach $3$ couples $(j,1),(j,2),(j,3)$, 
where each $(j,k)$ corresponds to a boundary component $\pl_k\mc{P}_j$. 
Now, if $\mc{P}_j$ is glued to $\mc{P}_{j'}$ in the surface $\Sigma$ by the relation 
$\pl_{k}\mc{P}_j=\pl_{k'}\mc{P}_{j'}=\mc{C}_i$, we glue the couples $(j,k)$ to $(j',k')$ to form an edge denoted 
${\rm e}_{i}=((j,k),(j',k'))$ relating ${\rm v}_j$ to ${\rm v}_{j'}$ and here we think of this edge as being 
oriented from ${\rm v}_j$ to ${\rm v}_{j'}$. We then take the convention that $\pl_k\mc{P}_j$ is outgoing and 
$\pl_{k'}\mc{P}_{j'}$ is incoming. Note that it is possible to take $j=j'$ and have edges of the form $((j,k),(j,k'))$ (self-gluing).
For the remaining couples (not linking two vertices) we fix an orientation, either $((j,k),\emptyset)$ (outgoing) or $(\emptyset,(j,k))$ (incoming) and notice that they correspond to boundaries of $\Sigma$. All this is purely topological so far. See Figure \ref{Genus1_graph} for the graph associated to the surface of Figure \ref{Genus1}.

Denote $b^+_j$ and $b^-_j$ the number of outgoing/incoming boundary components of $\mc{P}_j$.
 It can be shown that there is a natural gluing map 
\[ {\rm Gl}_{\mc{G}}: \mc{T}_{0,b^+_1,b^-_1}\times \dots \times \mc{T}_{0,b^+_{j_0},b^-_{j_0}}\to \mc{T}_{\mathfrak{g},b^+,b^-}\]
where for simplicity we have noted $\mc{T}_{0,b_j^+,b_j^-}$ and $\mc{T}_{\mathfrak{g},b^+,b^-}$ instead of 
$\mc{T}_{0,0,b_j^+,b_j^-}$ and $\mc{T}_{\mathfrak{g},0,b^+,b^-}$ (we forget the marked points as we only discuss the case with no marked points here). Each $(\bs{\psi}_j,\bs{\tau}_j)\in\mc{T}_{0,b_j^+,b_j^-}$ is represented by a pairs of pant 
$(\mc{P}_j,\bs{\zeta}_j)$  (with parametrized boundary) under model form together with a marking diffeomorphism $\Phi_j:\mc{P}_0\to \mc{P}_j$ where $\mc{P}_0$ is a fixed pant, say $\mc{P}_0=\hat{\C}\setminus \cup_{j=1}^3 D(x_j,1/4)$ with $(x_1,x_2,x_3)=(0,1,e^{i\pi/3})$. Gluing these pairs of pants along their boundary as prescribed by the graph $\mc{G}$ and gluing the marking diffeomorphisms $\Phi_j$ together produces the map ${\rm Gl}_{\mc{G}}$. This map is surjective but not injective, the defect of injectivity corresponding exactly to two different pair of pant decompositions of a surface $(\Sigma,J,\bs{\zeta})$ along two families of curves $\mc{C}_i^1$ and $\mc{C}_i^2$, being isotopic. This means that, to a collection $(\bs{\psi},\bs{\tau})=(\bs{\psi}_1,\bs{\tau}_1,\dots,\bs{\psi}_{j_0},\bs{\tau}_{j_0})$ in 
$\mc{T}_{0,b^+_1,b^-_1}\times \dots \times \mc{T}_{0,b^+_{j_0},b^-_{j_0}}$, one can associate a marked Riemann 
surface $(\Phi,\Sigma,J,\bs{\zeta})$ with $b$ parametrized boundary circles, together with $i_0$ interior parametrized curves 
$\xi_i:\T\to \mc{C}_i$
cutting the pants, and this correspondence is one-to-one. We can then define the conformal block amplitude of 
$(\bs{\psi},\bs{\tau})$, or equivalently $(\Phi,\Sigma,J,\bs{\zeta},\bs{\xi})$ with $\bs{\xi}=(\xi_1,\dots,\xi_{j_0})$, by composing the block amplitudes using partial traces in a way similar to \eqref{partial_trace}, except that we work in the $L^2(\mc{T})$ space over Young diagrams, and freezing the $p$ variables associated to the cutting curves $\mc{C}_i$ to be $p_i$. 
The conformal block $\mc{B}_{{\rm Gl}(\bs{\psi},\bs{\tau})}({\bf p})$ of the glued surface is, for ${\bf p}=(p_1,\dots,p_{i_0})$,  a continuous map 
\[\mc{B}_{{\rm Gl}(\bs{\psi},\bs{\tau})}({\bf p}): L^2_{\rm comp}(\R_+^{b^-},L^2(\mc{T})^{\otimes b^-})\to L^2_{\rm loc}(\R_+^{b^+},L^2(\mc{T})^{\otimes b^+}) \]
whose dependence in ${\bf p}$ is $L^2_{\rm loc}(\R_+^{i_0})$.
Instead of writing a general formula, which involves more notations, let us give a concrete example:
take the surface $(\Sigma,\bs{\zeta})$ in Figure \ref{Genus1}, represented by the model form 
building blocks $(\bs{\psi}_1,\bs{\tau}_1)$ and $(\bs{\psi}_1,\bs{\tau}_2)$, and assume that the boundary parametrizations $\bs{\zeta}=(\zeta_1,\zeta_2)$ are incoming: this means that $\pl_j\mc{P}_2$ are all incoming for $j=1,2,3$ (thus $b_2^+=0$ and $b^-_2=3$) 
and $\pl_1\mc{P}_1,\pl_2\mc{P}_1$ are outgoing while $\pl_3\mc{P}_1$ is incoming (thus $b_1^+=2$ and $b^-_1=1$).
The conformal block is defined, for $(\bs{\psi},\bs{\tau})=(\bs{\psi}_1,\bs{\tau}_1,\bs{\psi}_2,\bs{\tau}_2)$ and 
${\bf p}=(p_1,p_2)$ the parameters associated to the cutting curves $\mc{C}_1,\mc{C}_2$, by
\begin{align*} 
&\mc{B}_{{\rm Gl}_{\mc{G}}(\bs{\psi},\bs{\tau})}({\bf p})u=\\
& \int_{\R_+^2}\sum_{\nu'_1,\nu_2',\nu_1,\nu_2\in \mc{T}}\mc{B}_{\bs{\psi}_2,\bs{\tau}_2}(s_1,s_2,p_2;\nu'_1,\nu'_2,\nu_2)\mc{B}_{\bs{\psi}_1,\bs{\tau}_1}(p_2,p_1,p_1;\nu_2,\nu_1,\nu_1)u(s_1,s_2;\nu'_1,\nu'_2)ds_1ds_2
\end{align*}
for $u\in  L^2_{\rm comp}(\R_+^{2};  L^2(\mc{T})^{\otimes 2})$, where here $\mc{B}_{\bs{\psi}_j,\bs{\tau}_j}(\cdot\, ;\cdot)$ denotes the integral kernel of the operator $\mc{B}_{\bs{\psi}_j,\bs{\tau}_j}$ of \eqref{propertyB}.

\begin{figure}
 \begin{tikzpicture}
 \node[inner sep=0pt] (pant) at (0,0)
{\includegraphics[width=0.4\textwidth]{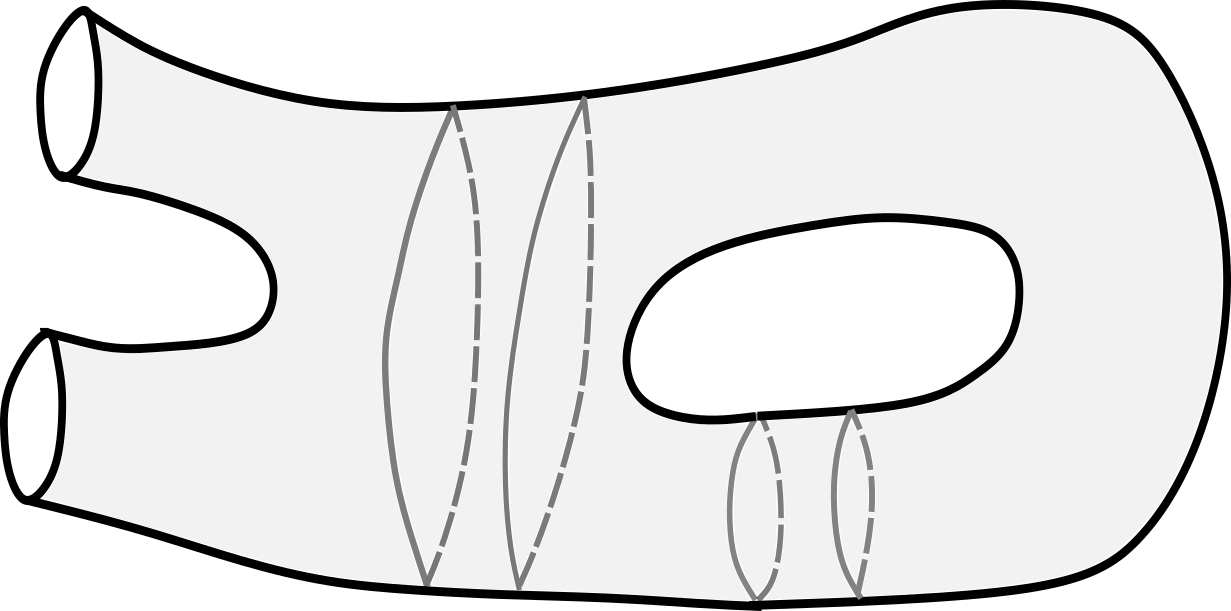}};
 \draw (-4,1.3) node[right,black]{$\pl_1\Sigma$} ;
  \draw (-4.2,-0.7) node[right,black]{$\pl_3\Sigma$} ; 
        \draw (1.3,-1) node[right,black]{$\mc{C}^2_1$} ;
         \draw (0,-1) node[right,black]{$\mc{C}^1_1$} ;
          \draw (-1.8,0) node[right,black]{$\mc{C}^1_2$} ;
           \draw (-0.2,0.7) node[right,black]{$\mc{C}^2_2$} ;
\end{tikzpicture}
 \caption{Two different cutting of the surface $\Sigma$ with same  graph $\mc{G}$}\label{Genus1cut}
\end{figure}
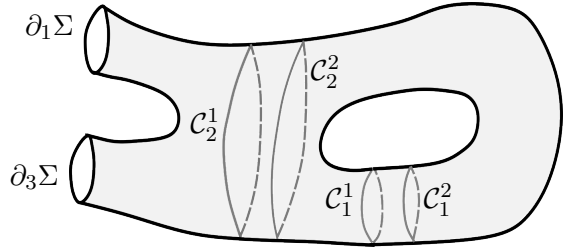

As said above, two collections $(\bs{\psi}^1,\bs{\tau}^1)$ and  $(\bs{\psi}^2,\bs{\tau}^2)$ in $\mc{T}_{0,2,1}\times \mc{T}_{0,0,3}$ can lead to the same marked surface ${\rm Gl}_{\mc{G}}(\bs{\psi}^1,\bs{\tau}^1)={\rm Gl}_{\mc{G}}(\bs{\psi}^2,\bs{\tau}^2)$ in 
$\mc{T}_{1,0,2}$, which can be viewed as the same marked surface $\Sigma$ with two different parametrized cutting curves $\bs{\xi}^1=(\xi^1_1,\xi^1_2)$ and $\bs{\xi}^2=(\xi^2_1,\xi^2_2)$ with images $(\mc{C}_1^1,\mc{C}^1_2)$ and $(\mc{C}_1^2,\mc{C}^2_2)$ respectively, see Figure \ref{Genus1cut}.
In order to compute the dependence of the conformal block on the choice of parametrized cutting curves $\xi_1:\T\to \mc{C}_1$ and $\xi_2:\T\to \mc{C}_2$, we can choose a $1$-parameter family $\bs{\xi}^t=(\xi_1^t,\xi_2^t)$ for $t\in [1,2]$, producing a family of model forms $(\bs{\psi}^t,\bs{\tau}^t)=(\bs{\psi}_1^t,\bs{\tau}_1^t,\bs{\psi}_2^t,\bs{\tau}_2^t)\in \mc{T}_{0,2,1}\times \mc{T}_{0,0,3}$ such that ${\rm Gl}(\bs{\psi}^t,\bs{\tau}^t)=(\bs{\psi}^1,\bs{\tau}^1)$ is fixed. Then, it is shown in \cite{BGKR2} that:
\begin{lemma}\label{variationblock} 
Under the assumption described above, the conformal block has a variation given by
\[\begin{split} 
\pl_t \mc{B}_{{\rm Gl}_{\mc{G}}(\bs{\psi}^t,\bs{\tau}^t)}({\bf p})=& \Big(\frac{c_{\rm L}}{24\pi i}\int_{\T}v_2^t({\rm S}(\psi^t_{11})-{\rm S}(\psi^t_{23}))+\frac{c_{\rm L}}{24\pi i}\int_{\T}v_1^t({\rm S}(\psi^t_{12})-{\rm S}(\psi^t_{13}))\Big)\mc{B}_{{\rm Gl}_{\mc{G}}(\bs{\psi}^t,\bs{\tau}^t)}({\bf p})
\end{split}\]
where $\bs{\psi}^t_j=(\psi_{j1}^t,\psi_{j2}^t,\psi_{j3}^t)$, ${\rm S}(f)=(f''/f')'-\frac{1}{2}(f''/f')^2$ is the Schwarzian derivative of $f$, and 
\[v_1^t=\pl_s ((\xi_1^t)^{-1}\circ \xi_1^{t+s})|_{s=0} , \quad v_2^t=\pl_s ((\xi_2^t)^{-1}\circ \xi_2^{t+s})|_{s=0}\]
are the variations of the cutting curves $\xi_1^t$ and $\xi_2^t$.
\end{lemma}
Using this variation expression, we can define a cocycle by integrating this variation: for two sets of parametrized  cutting curves $\bs{\xi}^1$ and $\bs{\xi}^2$ as above  we set
\[ \Upsilon_{\Sigma}(\bs{\xi}^1,\bs{\xi}^2)=\int_1^2 \Theta_{\bs{\xi}^t}(v^t)dt\]
where $t\in [0,1]\mapsto \bs{\xi}^t$ is any $C^1$-family of analytic parametrizations of the cutting curves with endpoints 
$\bs{\xi}^t|_{t=1}=\bs{\xi}^1$ and $\bs{\xi}^t|_{t=2}=\bs{\xi}^2$, $v^t=(\pl_s ((\xi_1^t)^{-1}\circ \xi_1^{t+s})|_{s=0},\pl_s ((\xi_2^t)^{-1}\circ \xi_2^{t+s})|_{s=0})$ and 
\begin{equation}\label{ThetaDef} 
 \Theta_{\bs{\xi}^t}(v^t):=\frac{1}{2\pi i}\int_{\T}v_2^t({\rm S}(\psi^t_{11})-{\rm S}(\psi^t_{23}))dz+\frac{1}{2\pi i}\int_{\T}v_1^t({\rm S}(\psi^t_{12})-{\rm S}(\psi^t_{13}))dz.
 \end{equation}
The function  $\Upsilon_{\Sigma}$ is a cocycle in the sense that, for three sets of cutting curves,
\[  \Upsilon_{\Sigma}(\bs{\xi}^1,\bs{\xi}^2)+ \Upsilon_{\Sigma}(\bs{\xi}^2,\bs{\xi}^3)= \Upsilon_{\Sigma}(\bs{\xi}^1,\bs{\xi}^3).\]
When the surface $\Sigma$ is closed with genus $\mathfrak{g}$, the cocycle $ \Upsilon_{\Sigma}$ (defined exactly as above) allows to define a holomorphic line bundle $\mc{L}$ over Teichm\"uller space $\mc{T}_\mathfrak{g}$: for $\mc{G}$ a gluing graph representing a decomposition into $j_0=2\mathfrak{g}-2$ pairs of pants and $i_0=3\mathfrak{g}-3$ cutting curves $\mc{C}_i$, the line $\mc{L}_X$ over a marked complex surface 
$X:=(\Phi,\Sigma,J)$ viewed as an element in $\mc{T}_\mathfrak{g}$ is defined as follows: 
if ${\rm Cut}_{\mc{G}}$ denotes the set of analytic parametrized curves $\bs{\xi}=(\xi_1,\dots,\xi_{i_0})$ with $\xi_i(\T)$ isotopic to $\mc{C}_i$ and $\bs{\xi}^0\in {\rm Cut}_{\mc{G}}$ is a fixed element, then set
\[ \mc{L}_X:=\{ F: {\rm Cut}_{\mc{G}}\to \C\,|\, \forall \bs{\xi}^0\in {\rm Cut}_{\mc{G}}, \,  F(\bs{\xi})= F(\bs{\xi}^0)e^{\frac{c_{\rm L}}{12}\Upsilon_{\Sigma}(\bs{\xi}^0,\bs{\xi})} \}.\]
Moreover, choosing a ($C^1$-regular) family of 
analytic negatively curved Riemannian metric $g_X$ for each element $X\in \mc{T}_\mathfrak{g}$, for example the hyperbolic metric, allows to put a metric $\|\cdot\|_{\mc{L}}$ on the line bundle $\mc{\mc{L}}$ by evaluating\footnote{In fact one adds an explicit 
multiplicative factor involving the Weyl anomaly $S_{\rm L}^0$, see \cite{BGKR2}.} the modulus of an element $F\in \mc{L}_X$ on the (unique) 
geodesic $\gamma_i$ of $g_X$ in the homotopy class of the curves $\mc{C}_i$, parametrized  by arclength by the circle $\T$.
We show:
\begin{theorem}\cite{BGKR2}
Let $\Sigma$ be a closed surface of genus $\mathfrak{g}$ and $\mc{G}$ a gluing graph representing a decomposition into $2\mathfrak{g}-2$ pairs of pants. Then the conformal block  $X\mapsto \mc{F}_{{\bf p}}(X)=\mc{F}_{\mc{G},{\bf p}}(X)$ defined for ${\bf p}\in \R_+^{i_0}$ by 
\[ \mc{F}_{\bf p}(X): \bs{\xi}\in  {\rm Cut}_{\mc{G}}\mapsto \mc{B}_{{\rm Gl}_{\mc{G}}(\bs{\psi}(\bs{\xi}),\bs{\tau}(\bs{\xi}))}({\bf p})\]
is a global holomorphic section of $\mc{L}$ over Teichm\"uller space $\mc{T}_{\mathfrak{g}}$, where 
\[(\bs{\psi}(\bs{\xi}),\bs{\tau}(\bs{\xi}))=(\bs{\psi}_1(\bs{\xi}),\bs{\tau}_1(\bs{\xi}),\dots,\bs{\psi}_{j_0}(\bs{\xi}),\bs{\tau}_{j_0}(\bs{\xi}))\] 
are the model form building blocks associated to the choice of set of cutting curves $\bs{\xi}$. Finally, the partition function decomposes as 
\[ Z_{\Sigma,g_X}=C_0 \int_{\R_+^{i_0}}\bs{\rho}({\bf p})\|\mc{F}_{{\bf p}}(X)\|^2_{\mc{L}} {\rm d} {{\bf p}} \]
for some $C_0$ depending only the genus and $\rho({\bf p})$ is an explicit product of $j_0$ structure constants $C^{\rm DOZZ}(\cdot)$.
\end{theorem}
In the case with marked points, a similar result holds for the first part of this theorem. We remark that the family 
$\mc{F}_{\bf p}(X)$ as $X$ covers $\mc{T}_{\mathfrak{g}}$ generate a vector space of 
$L^2(\R_+^{i_0})$ (in the ${\bf p}$ variable) where the $L^2$ product is defined to be weighted by the $\rho({\bf p})$ function. 
We call this space the vector space of conformal blocks. 

\subsection{Ward identity for conformal blocks}
The definition of conformal blocks given in the langage of Vertex Operator Algebra can be summarized as follows.
For $x_1,\dots,x_m$ some disjoint points on a closed Riemann surface $(\Sigma,J)$, we associate some vector spaces 
$\mc{V}_{\alpha_1},\dots,\mc{V}_{\alpha_m}$ with a representation of the Virasoro algebra ${\bf L}_n$, where $\mc{V}_{\alpha}$ are Verma module, i.e. 
\[ \mc{V}_{\alpha}={\rm span}\{ {\bf L}_{-\nu}\Psi_{\alpha}\,|\, \nu \in \mc{T} \}\]
using the notation of Section \ref{rep_virasoro}, with $\Psi_{\alpha}$ being a highest weight vector, killed by all ${\bf L}_n$ if $n>0$.
Fix an atlas of coordinate charts $w_j:U_j\to \hat{\C}$ for $j\in \mc{J}$  such that the change of coordinates  $\omega_k\circ \omega_j^{-1}$ are elements in ${\rm PSL}_2(\C)$. Such a family is called a \textbf{complex projective structure}. Without loss of generality, we can assume that for each marked point $x_j$ there is a coordinate chart $\psi_j^{-1}=\omega_j$ that is part of the projective atlas and
 $\psi_j(0)=x_j$.
As defined in \cite[Section 12.1]{Teschner_Vartanov}, a conformal block on $(\Sigma,J)$  is a linear form 
\begin{equation}\label{Ward_for_blocks} 
\mc{F}: \bigotimes_{j=1}^m \mc{V}_{\alpha_j} \to \C
\end{equation}
satisfying the following invariance property, called \textbf{Ward identity}:  for all meromorphic vector field $v$ on $\Sigma$ with poles contained in ${\bf x}=(x_1,\dots,x_m)$, then for all $u\in \otimes_{j=1}^r\mc{V}_{\alpha_j}$
\begin{equation}\label{WardVOA}
 \mc{F}( {\bf L}_{v_j}^{(j)}u)=0
 \end{equation}
where ${\bf L}_{v_j}^{(j)}(u_1\otimes \dots \otimes u_{m}):=u_1\otimes \dots \otimes {\bf L}_{v_j}u_j\otimes \dots \otimes u_m$ and 
\begin{equation}\label{Lvj} 
v_j:=\psi_j^*v \quad v_j(z)=\sum_{n\in \Z} v_{jn}z^{n+1}\pl_z, \quad {\bf L}_{v_j}:=\sum_{n\in \Z}v_{jn}{\bf L}_n.
\end{equation}

In our setting of Liouville CFT, we can show the following version of the Ward identity (here we only state the case of incoming boundaries and no extra marked point for simplicity in order to connect with the discussion above). 
\begin{theorem}\cite{BGKR2}\label{Ward_blocks}
Consider a surface $(\Sigma,J,\bs{\zeta})$ of type $(\mathfrak{g},0,0,b)$ and let $\hat{\Sigma}$ be the closed surface with $b$ marked points $x_1,\dots,x_b$ obtained by gluing a disk $\D$ at each boundary $\pl_j\Sigma$ using the parametrization $\zeta_j$, which extends as a holomorphic embedding denoted $\psi_j:\D \to \psi_j(\D)\subset \hat{\Sigma}$.
Let $v$ be a meromorphic vector field on $\hat{\Sigma}$ with poles contained in ${\bf x}=(x_1,\dots,x_b)$, or more generally a holomorphic vector field defined on a neighborhood of $\Sigma$. Take a graph $\mc{G}$ representing a decomposition of $\Sigma$ into pairs of pants with $i_0$ analytic cutting curves $\xi_i:\T \to \Sigma^\circ$ forming $\bs{\xi}\in {\rm Cut}_{\mc{G}}$.
 Then the following holds true for the conformal block $\mc{B}_{{\rm Gl}_{\mc{G}}(\bs{\psi}(\bs{\xi}),\bs{\tau}(\bs{\xi}))}({\bf p})$ with ${\bf p}\in \R_+^{i_0}$
\[ \begin{split}
\mc{B}_{{\rm Gl}_{\mc{G}}(\bs{\psi}(\bs{\xi}),\bs{\tau}(\bs{\xi}))}({\bf p})(\sum_{j=1}^b {\bf L}_{v_j}^{(j)}u) +\Big(-\sum_{j=1}^b \frac{c_{\rm L}}{24\pi i}\int_{\T}{\rm S}(\psi_{j})v_jdz +\frac{c_{\rm L}}{12}\Theta_{\bs{\xi}}(v^{\rm gl})dz\Big) \mc{B}_{{\rm Gl}_{\mc{G}}(\bs{\psi}(\bs{\xi}),\bs{\tau}(\bs{\xi}))}({\bf p})u=0
\end{split} \]
for all $u\in L^2_{\rm comp}(\R_+^{b},((1+|\nu|)^{-1/2}L^2(\mc{T}))^{\otimes b})$, where $\Theta_{\bs{\xi}}$ is the defect $1$-form 
defined as in \eqref{ThetaDef} and $v^{\rm gl}=(\xi_1^*v,\dots,\xi_{i_0}^*v)$ is the vector field $v$ restricted on the cutting curves, and ${\bf L}_{v_j}$ are defined as in \eqref{Lvj}.
\end{theorem}
The action of ${\bf L}_{v_j}$ on  $L^2_{\rm comp}(\R_+^{b},((1+|\nu|)^{-1/2}L^2(\mc{T}))^{\otimes b})$ is only on the $\mc{T}$ 
variables and is understood via the conjugation by the map $\Pi_{\mc{V}}$ of \eqref{PiV}, and to make connection with the discussion above, what happens is that, roughly speaking, for $p$ fixed the space $\mc{V}_p:={\rm span}\{\Psi_{Q+ip,\nu,0}\,|\, \nu \in \mc{T}\}$ is mapped to $L^2(\mc{T})$ by $\Pi_{\mc{V}}$ (the drawback is that, due to continuous spectrum, this only makes sense when integrated in $p$ on some interval if one thinks of $\mc{V}_p$ as a subspace of  the Hilbert space $\mc{H}$). 
The weight $(1+|\nu|)^{-1/2}$ is used since ${\bf L}_{n}$ are unbounded and we need to restrict to a domain where they act reasonably well.
We notice that here, the Schwarzian derivatives terms come from the fact we have not chosen a complex projective structure on the surface. The Schwarzian derivatives appearing in the $\Theta_{\bs{\xi}}(v^{\rm gl})$ term (recall \eqref{ThetaDef}) can be written in terms of Schwarzian of changes of coordinates (of the form ${\rm S}(\psi_{jk}\circ \psi_{j'k'}^{-1})$ if $((j,k),j',k')$ is an edge of the graph and $\psi_{jk},\psi_{j'k'}$ the model form holomorphic maps that are glued together). In particular, choosing the model 
building blocks where $\psi_j$ are in ${\rm PSL}_2(\C)$, one gets a projective structure and the Schwarzian derivatives in Theorem \ref{Ward_blocks} disapear and we recover the Ward identity of \eqref{Ward_blocks}.

The proof of Theorem \ref{Ward_blocks} is related to Proposition \ref{diff_amplitudes} and consists in differentiating the block amplitude along the variation given by the flow of the vector field $v$ for small time (which is a biholomorphism moving slightly $\Sigma$ in $\hat{\Sigma}$).
 
In conclusion, we get:\\

\textbf{Summary:} For each gluing graph $\mc{G}$, there exists a family of conformal blocks 
$\mc{F}_{\mc{G},{\bf p}}$ on closed surfaces with marked points that are holomorphic sections of a line bundle $\mc{L}$ on the Teichm\"uller space $\mc{T}_{\mathfrak{g},m}$, 
and the conformal blocks $\mc{B}_{{\rm Gl}_{\mc{G}}(\bs{\psi}(\bs{\xi}),\bs{\tau}(\bs{\xi}))}({\bf p})$ on surfaces with $b$ boundary components satisfy the Ward identity. 

\bibliographystyle{alpha}
\bibliography{virasoro}

\newcommand{\etalchar}[1]{$^{#1}$}
\begin{thebibliography}{BGKR24b}

\bibitem[AGT10]{Alday_2010}
L.~F. Alday, D.~Gaiotto, and Y.~Tachikawa.
\newblock Liouville correlation functions from four-dimensional gauge theories.
\newblock {\em Letters in Mathematical Physics}, 91(2):167--197, January 2010.

\bibitem[AHS23]{nina}
M.~Ang, N.~Holden, and X.~Sun.
\newblock Integrability of {SLE} via conformal welding of random surfaces.
\newblock {\em Communications on Pure and Applied Mathematics}, 10 2023.

\bibitem[ARS21]{AngRemySun21_FZZ}
M.~{Ang}, G.~{Remy}, and X.~{Sun}.
\newblock {FZZ formula of boundary Liouville CFT via conformal welding}.
\newblock {\em arXiv e-prints}, page arXiv:2104.09478, April 2021.

\bibitem[ARS22]{Xin}
M.~Ang, G.~Remy, and X.~Sun.
\newblock The moduli of annuli in random conformal geometry.
\newblock {\em arXiv e-prints}, page
  \href{https://arxiv.org/abs/2203.12398}{arXiv:2203.12398}, 2022.

\bibitem[ARSZ23]{ang2023derivation}
M.~Ang, G.~Remy, X.~Sun, and T.~Zhu.
\newblock Derivation of all structure constants for boundary liouville cft,
  2023.

\bibitem[AS21]{AngSun21_CLE}
M.~{Ang} and X.~{Sun}.
\newblock {Integrability of the conformal loop ensemble}.
\newblock page arXiv:2107.01788, July 2021.

\bibitem[BGK{\etalchar{+}}22]{BGKRV}
G.~{Baverez}, C.~{Guillarmou}, A.~{Kupiainen}, R.~{Rhodes}, and V.~{Vargas}.
\newblock {The Virasoro structure and the scattering matrix for Liouville
  conformal field theory}.
\newblock {\em Probability and Mathematical Physics}, (To appear), 2022.

\bibitem[BGKR24a]{BGKR2}
G.~Baverez, C.~Guillarmou, A.~Kupiainen, and R.~Rhodes.
\newblock The conformal blocks of {L}iouville {CFT}.
\newblock {\em preprint}, 2024.

\bibitem[BGKR24b]{BGKR1}
G.~Baverez, C.~Guillarmou, A.~Kupiainen, and R.~Rhodes.
\newblock {Semigroup of annuli in Liouville CFT}.
\newblock {\em arXiv:2403.10914}, 2024.

\bibitem[Bor86]{Borcherds}
R.~Borcherds.
\newblock Vertex algebras, {K}ac-{M}oody algebras, and the monster.
\newblock {\em Proceedings of the National Academy of Sciences of the United
  States of America}, 83:3068--3071, 1986.

\bibitem[BPZ84]{BPZ84}
A.A. Belavin, A.M. Polyakov, and A.B. Zamolodchikov.
\newblock Infinite conformal symmetry in two-dimensional quantum field theory.
\newblock {\em Nuclear Physics B}, 241(2):333--380, 1984.

\bibitem[Cer22]{cercle}
B.~Cercl\'e.
\newblock Three-point correlation functions in the ${\rm sl}_3$ {T}oda theory
  ii: the {F}ateev-{L}itvinov formula.
\newblock {\em arXiv:2208.12085}, 2022.

\bibitem[CRV23]{Cercle-Rhodes-Vargas}
B.~Cercl\'e, R.~Rhodes, and V.~Vargas.
\newblock Probabilistic construction of {T}oda conformal field theories.
\newblock {\em Annales Henri Lebesgue}, 6:31--64, 2023.

\bibitem[CS12]{CS10}
D.~Chelkak and S.~Smirnov.
\newblock Universality in the 2d {Ising} model and conformal invariance of
  fermionic observables.
\newblock {\em Invent. Math.}, 189(3):515--580, 2012.

\bibitem[Dav88]{David}
F.~David.
\newblock Conformal field theories coupled to 2d gravity in the conformal
  gauge.
\newblock {\em Modern Physics Letters A}, 3(17):1651--1656, 1988.

\bibitem[DK89]{Distler_Kawai}
J.~Distler and H.~Kawai.
\newblock Conformal field theory and 2-d quantum gravity or who's afraid of
  {J}oseph {L}iouville?
\newblock {\em Nuclear Physics B}, 321:509--517, 1989.

\bibitem[DKRV16]{DKRV16}
F.~David, A.~Kupiainen, R.~Rhodes, and V.~Vargas.
\newblock Liouville quantum gravity on the {R}iemann sphere.
\newblock {\em Comm. Math. Phys.}, 342(3):869--907, 2016.

\bibitem[DMS21]{MatingOfTrees}
B.~{Duplantier}, J.~{Miller}, and S.~{Sheffield}.
\newblock {Liouville quantum gravity as a mating of trees}.
\newblock {\em Ast\'erisque}, 427, 2021.

\bibitem[DO94]{DornOtto94}
H.~Dorn and H.-J. Otto.
\newblock Two- and three-point functions in {L}iouville theory.
\newblock {\em Nuclear Phys. B}, 429(2):375--388, 1994.

\bibitem[DRV16]{DRV16_tori}
F.~David, R.~Rhodes, and V.~Vargas.
\newblock Liouville quantum gravity on complex tori.
\newblock {\em J. Math. Phys.}, 57(2):022302, 25, 2016.

\bibitem[FB08]{Fyodorov_2008}
Y.V. Fyodorov and J-P. Bouchaud.
\newblock Freezing and extreme-value statistics in a random energy model with
  logarithmically correlated potential.
\newblock {\em Journal of Physics A: Mathematical and Theoretical},
  41(37):372001, August 2008.

\bibitem[FF84]{FeiginFuchs}
B.L. Feigin and D.B. Fuchs.
\newblock Verma modules over the virasoro algebra,.
\newblock In {\em Topology : general and algebraic topology, and applications :
  proceedings of the international topological conference held in Leningrad,
  August 23-27, 1982}, Lecture notes in mathematics (Springer-Verlag) ; 1060,
  1984.

\bibitem[FGG73]{Ferrara}
S.~Ferrara, A.~F. Grillo, and R.~Gatto.
\newblock {Tensor representations of conformal algebra and conformally
  covariant operator product expansion}.
\newblock {\em Annals Phys.}, 76:161--188, 1973.

\bibitem[FLDR09]{Fyodorov_2009}
Y.V. Fyodorov, P.~Le~Doussal, and A.~Rosso.
\newblock Statistical mechanics of logarithmic rem: duality, freezing and
  extreme value statistics of 1/fnoises generated by gaussian free fields.
\newblock {\em Journal of Statistical Mechanics: Theory and Experiment},
  2009(10):P10005, October 2009.

\bibitem[FLM88]{Frenkel:1988xz}
I.~Frenkel, J.~Lepowsky, and A.~Meurman.
\newblock {\em {Vertex Operator Algebras and the monster}}.
\newblock 1988.

\bibitem[Fre07]{Frenkel07}
E.~Frenkel.
\newblock Lectures on the {L}anglands program and conformal field theory.
\newblock In {\em Frontiers in number theory, physics, and geometry. {II}},
  pages 387--533. Springer, Berlin, 2007.

\bibitem[FS87]{FriedanShenker87}
D.~Friedan and S.~Shenker.
\newblock The analytic geometry of two-dimensional conformal field theory.
\newblock {\em Nuclear Phys. B}, 281(3-4):509--545, 1987.

\bibitem[Gaw96]{Gawedzki96_CFT}
K.~Gawedzki.
\newblock Lectures on conformal field theory.
\newblock {\em Nucl. Phys. B}, 328:733--752, 1996.

\bibitem[GKR23]{CILT}
C.~Guillarmou, A.~Kupiainen, and R.~Rhodes.
\newblock Compactified imaginary {L}iouville theory.
\newblock {\em arXiv e-prints}, page
  \href{https://arxiv.org/abs/2310.18226}{arXiv:2310.18226}, 2023.

\bibitem[GKRV20]{GKRV20_bootstrap}
C.~{Guillarmou}, A.~{Kupiainen}, R.~{Rhodes}, and V.~{Vargas}.
\newblock Conformal bootstrap in {L}iouville theory.
\newblock {\em Acta Mathematica, to appear}, 2020.

\bibitem[GKRV21]{GKRV21_Segal}
C.~{Guillarmou}, A.~{Kupiainen}, R.~{Rhodes}, and V.~{Vargas}.
\newblock {Segal's axioms and bootstrap for Liouville Theory}.
\newblock {\em arXiv e-prints}, page
  \href{https://arxiv.org/abs/2112.14859}{arXiv:2112.14859}, December 2021.

\bibitem[GRSS24]{blocguillaume}
P.~Ghosal, G.~Remy, X.~Sun, and Y.~Sun.
\newblock Probabilistic conformal blocks for {L}iouville cft on the torus.
\newblock {\em arXiv e-prints}, page
  \href{https://arxiv.org/abs/2003.03802}{arXiv:2310.18226}, 2024.

\bibitem[GRV19]{GRV2019}
C.~Guillarmou, R.~Rhodes, and V.~Vargas.
\newblock Polyakov's formulation of {$2d$} bosonic string theory.
\newblock {\em Publ. Math. Inst. Hautes \'{E}tudes Sci.}, 130:111--185, 2019.

\bibitem[HS19]{XinCardy}
N.~Holden and X.~Sun.
\newblock Convergence of uniform triangulations under the {C}ardy embedding.
\newblock {\em Acta Mathematica}, 230(1):93--203, 2019.

\bibitem[Hua97]{Huang1997TwoDimensionalCG}
Y-Z. Huang.
\newblock Two-dimensional conformal geometry and vertex operator algebras.
\newblock 1997.

\bibitem[IT92]{Imayoshi-Taniguchi}
Y.~Imayoshi and M.~Taniguchi.
\newblock {\em An introduction to Teichm\"uller spaces}.
\newblock Springer Verlag, 1992.

\bibitem[Jaf18]{Jaffe}
A.~Jaffe.
\newblock Reflection positivity then and now.
\newblock {\em arXiv e-prints}, page
  \href{https://arxiv.org/pdf/1802.07880.pdf}{arXiv:1802.07880}, 2018.

\bibitem[Kah85]{Kahane85}
J-P. Kahane.
\newblock Sur le chaos multiplicatif.
\newblock {\em Ann. Sci. Math. Qu\'{e}bec}, 9(2):105--150, 1985.

\bibitem[Kle91]{Klebanov}
I.R. Klebanov.
\newblock String theory in two dimensions.
\newblock {\em arXiv:9108019}, 1991.

\bibitem[Kos15]{kostov}
I.~Kostov.
\newblock {\em Two-dimensional quantum gravity, in The Oxford Handbook of
  Random Matrix Theory}.
\newblock Oxford University Press, 09 2015.

\bibitem[KPZ88]{Knizhnik:1988ak}
V.~G. Knizhnik, A.~Polyakov, and A.~B. Zamolodchikov.
\newblock {Fractal Structure of 2D Quantum Gravity}.
\newblock {\em Mod. Phys. Lett. A}, 3:819, 1988.

\bibitem[KRV20]{KRV_DOZZ}
A.~Kupiainen, R.~Rhodes, and V.~Vargas.
\newblock Integrability of {L}iouville theory: proof of the {DOZZ} formula.
\newblock {\em Ann. of Math. (2)}, 191(1):81--166, 2020.

\bibitem[Mac77a]{mack2}
G.~Mack.
\newblock {Convergence of Operator Product Expansions on the Vacuum in
  Conformal Invariant Quantum Field Theory}.
\newblock {\em Commun. Math. Phys.}, 53:155, 1977.

\bibitem[Mac77b]{Mack1}
G.~Mack.
\newblock Duality in quantum field theory.
\newblock {\em Nucl. Phys.}, B118:445--457, 1977.

\bibitem[MO19]{Davesh_MAULIK_2019}
D.~Maulik and A.~Okunkov.
\newblock Quantum groups and quantum cohomology.
\newblock {\em Ast{\'e}risque}, 408:1--212, 2019.

\bibitem[Nak04]{Nakayama}
Y.~Nakayama.
\newblock Liouville field theory -- a decade after the revolution.
\newblock {\em UT-04-02}, 2004.

\bibitem[Nek03]{Nek}
N.A. Nekrasov.
\newblock {Seiberg-Witten Prepotential from Instanton Counting}.
\newblock {\em Advances in Theoretical and Mathematical Physics}, 7(5):831 --
  864, 2003.

\bibitem[NQSZ23]{nolin}
P.~Nolin, W.~Qian, X.~Sun, and Z.~Zhuang.
\newblock {Backbone exponent for two-dimensional percolation}.
\newblock {\em arXiv e-prints}, arXiv:2309.05050, 2023.

\bibitem[Oik21]{Oikarinen}
J.~Oikarinen.
\newblock Stress-energy in {L}iouville conformal field theory on compact
  riemann surfaces.
\newblock {\em arXiv:2108.06767}, 2021.

\bibitem[OS73]{osterwalder-schrader}
K.~Osterwalder and R.~Schrader.
\newblock Axioms for euclidean {G}reen's functions.
\newblock {\em Comm. Math. Phys.}, 31(2):83--112, 1973.

\bibitem[Ost09]{Ostrovsky_2009}
D.~Ostrovsky.
\newblock Mellin transform of the limit lognormal distribution.
\newblock {\em Communications in Mathematical Physics}, 288(1):287--310, March
  2009.

\bibitem[Ost18]{Ostrovsky_2018}
D.~Ostrovsky.
\newblock A review of conjectured laws of total mass of bacry--muzy gmc
  measures on the interval and circle and their applications.
\newblock {\em Reviews in Mathematical Physics}, 30(10):1830003, October 2018.

\bibitem[Pol70]{Pol}
A.~M. Polyakov.
\newblock Conformal symmetry of critical fluctuations.
\newblock {\em JETP Lett.}, 12:381--383, 1970.

\bibitem[Pol74]{Pol74}
A.~M. Polyakov.
\newblock {Nonhamiltonian approach to conformal quantum field theory}.
\newblock {\em Zh. Eksp. Teor. Fiz.}, 66:23--42, 1974.

\bibitem[Pol08]{Pol08}
A.~Polyakov.
\newblock From quarks to strings.
\newblock {\em arXiv:0812.0183}, 2008.

\bibitem[PRV19]{slava}
D.~Poland, S.~Rychkov, and A.~Vichi.
\newblock The conformal bootstrap: Theory, numerical techniques, and
  applications,.
\newblock {\em Reviews of Modern Physics}, 91:015002, 2019.

\bibitem[Rem20]{Remy20}
G.~Remy.
\newblock The {F}yodorov-{B}ouchaud formula and {L}iouville conformal field
  theory.
\newblock {\em Duke Math. J.}, 169(1):177--211, 2020.

\bibitem[{Rib}14]{Ribault14}
S.~{Ribault}.
\newblock {Conformal field theory on the plane}.
\newblock {\em arXiv e-prints}, page
  \href{https://arxiv.org/abs/1406.4290}{arXiv:1406.4290}, June 2014.

\bibitem[RS71]{Ray-Singer}
D.~Ray and I.~Singer.
\newblock R-torsion and the laplacian on riemannian manifolds.
\newblock {\em Advances in Mathematics}, 7(2):145--210., 1971.

\bibitem[RV14]{rhodes2014_gmcReview}
R.~Rhodes and V.~Vargas.
\newblock Gaussian multiplicative chaos and applications: a review.
\newblock {\em Probab. Surv.}, 11:315--392, 2014.

\bibitem[RZ20]{Remy_2020}
G.~Remy and T.~Zhu.
\newblock The distribution of gaussian multiplicative chaos on the unit
  interval.
\newblock {\em The Annals of Probability}, 48(2), March 2020.

\bibitem[Seg88]{Segal87}
G.~B. Segal.
\newblock The definition of conformal field theory.
\newblock In {\em Differential geometrical methods in theoretical physics
  ({C}omo, 1987)}, volume 250 of {\em NATO Adv. Sci. Inst. Ser. C Math. Phys.
  Sci.}, pages 165--171. Kluwer Acad. Publ., Dordrecht, 1988.

\bibitem[Sei90]{SeibergRev}
N.~Seiberg.
\newblock {Notes on quantum Liouville theory and quantum gravity}.
\newblock {\em Prog. Theor. Phys. Suppl.}, 102:319--349, 1990.

\bibitem[Smi01]{Smirnov_2001}
S.~Smirnov.
\newblock Critical percolation in the plane: conformal invariance, {C}ardy's
  formula, scaling limits.
\newblock {\em Comptes Rendus de l'Acad{\'e}mie des Sciences - Series I -
  Mathematics}, 333(3):239--244, August 2001.

\bibitem[Smi10]{Sm10}
S.~Smirnov.
\newblock Conformal invariance in random cluster models. i. holomorphic
  fermions in the ising model.
\newblock {\em Annals of Mathematics}, 172(2):1435--1467, 2010.

\bibitem[SV13]{Schiffmann_2013}
O.~Schiffmann and E.~Vasserot.
\newblock Cherednik algebras, {W}-algebras and the equivariant cohomology of
  the moduli space of instantons on $a^2$.
\newblock {\em Publications math{\'e}matiques de l'IH{\'E}S}, 118(1):213--342,
  May 2013.

\bibitem[Tes95]{Tesc}
J.~Teschner.
\newblock On the {L}iouville three point function.
\newblock {\em Phys. Lett. B}, 363:65--70, 1995.

\bibitem[Tes01]{Teschner_revisited}
J.~Teschner.
\newblock Liouville theory revisited.
\newblock {\em Classical and Quantum Gravity}, 18(23):R153--R222, nov 2001.

\bibitem[Tro91]{Troyanov}
M.~Troyanov.
\newblock Prescribing curvature on compact surfaces with conical singularities.
\newblock {\em Transactions of the American Mathematical Society},
  324(2):793--821, 1991.

\bibitem[TV15]{Teschner_Vartanov}
J.~Teschner and G.S. Vartanov.
\newblock Supersymmetric gauge theories, quantization of $m_{\rm flat}$, and
  conformal field theory.
\newblock {\em Adv. Theor. Math. Phys.}, 19(1):1--135, 2015.

\bibitem[Var17]{Vargas}
V.~Vargas.
\newblock Lecture notes on liouville theory and the {DOZZ} formula.
\newblock {\em Arxiv.1712.00829}, 2017.

\bibitem[Wil69]{Wilson:1969}
K.G. Wilson.
\newblock Non-lagrangian models of current algebra.
\newblock {\em Physical Review}, 179(5):1499--1512, 1969.

\bibitem[Wu22]{Wu}
B.~Wu.
\newblock Conformal bootstrap on the annulus in {L}iouville {CFT}.
\newblock {\em arXiv e-prints}, page
  \href{https://arxiv.org/abs/2203.11830}{arXiv:2203.11830}, 2022.

\bibitem[ZZ96]{Zamolodchikov96}
A.~Zamolodchikov and Al. Zamolodchikov.
\newblock Conformal bootstrap in {L}iouville field theory.
\newblock {\em Nuclear Phys. B}, 477(2):577--605, 1996.

\end{thebibliography}

\end{document}